\documentclass[aps,prb,twocolumn,floats,epsfig,pdflatex]{revtex4-2}
\usepackage{amsmath}
\usepackage{amsfonts}
\usepackage{booktabs}
\usepackage{graphicx}
\usepackage{tikz}
\usetikzlibrary{shapes.geometric, positioning}
\usetikzlibrary{arrows.meta}
\usepackage{braket}
\usetikzlibrary{matrix,fit,positioning}
\usepackage{amssymb}
\usepackage{amsbsy}
\usepackage{comment}
\usepackage{xcolor}
\usepackage{ulem}
\usepackage[colorlinks=true, citecolor=red, urlcolor=blue, linkcolor=cyan, pdfencoding=auto, psdextra]{hyperref}
\usepackage{hyperref}
\usepackage{bbm}
\usepackage{orcidlink}

\newcommand{\beq}{\begin{equation}}
\newcommand{\eeq}{\end{equation}}
\newcommand{\bea}{\begin{eqnarray}}
\newcommand{\eea}{\end{eqnarray}}

\begin{document}

%\title{Universal quasienergy distributions in Floquet systems with protected nullspaces}
\title{Exponentially many protected zero modes and spectral universality in Floquet many-body systems}
%\title{Protected Floquet zero modes: spectral isolation and quasienergy folding}
\author{Saptadip Roy\,\orcidlink{0009-0006-6066-5719}}
\email{spssr3123@iacs.res.in}
\author{Arnab Sen\,\orcidlink{0000-0001-5342-8332}}
\email{tpars@iacs.res.in}
\affiliation{School of Physical Sciences, Indian Association for the Cultivation of Science, Kolkata 700032, India}
\date{\today}

\begin{abstract}
We investigate how chiral symmetry shapes quasienergy statistics in periodically driven PXP-type chains and a transverse-field Ising chain with three-spin interactions, using two-step square-pulse drives. In all the models, the Floquet unitary \(U\) hosts exponentially many (in system size) protected zero-quasienergy modes due to an interplay between chirality and a lattice symmetry, while a protected imbalance at \(\pi\) is forbidden. In one class of drives, \(U\) follows the chiral circular orthogonal ensemble, \(\mathrm{ch\text{-}COE}(\nu_0,0)\), where \(\nu_0\) and \(\nu_\pi\) denote the chiral imbalances at quasienergies \(0\) and \(\pi\), respectively. Its eigenphase density depends only on the Hilbert-space dimension and the protected-mode count, and has a spectral edge separating the zero modes from the rest of the spectrum followed by a square-root onset that approaches a constant in the bulk. A different structure emerges when a dynamical symmetry combines half-period time translation with a unitary operation $G$ that exchanges the two drive steps. In the symmetry sector of interest, the resulting half-period unitary \(W=GU_1=U_2G\) satisfies \(U=W^2\), with any phase in \(G^2\) absorbed into the definition of \(W\). In such Floquet models, \(W\) follows a generalized \(\mathrm{ch\text{-}COE}(\nu_0,\nu_\pi)\), allowing nonzero imbalances at both \(0\) and \(\pi\). \(W\) can either exchange symmetry blocks of \(U\) or preserve a block even within the same model. In the former case, \(U\) retains the \(\mathrm{ch\text{-}COE}(\nu_0,0)\) description. In the latter, squaring \(W\) wraps its eigenphases and produces spectral features near zero quasienergy in \(U \) that have no static counterpart. When \(W\) has protected modes at both \(0\) and \(\pi\), the density of \(U\) has a spectral edge followed by a second square-root onset. When only one of these imbalances is nonzero, the spectral edge is replaced by a plateau at roughly half the bulk eigenphase density followed by a square-root onset to the bulk value. We derive both density profiles, confirm them numerically, and show that overlapping spectral branches have level statistics consistent with a superposition of two GOE spectra. Finally, we show how $U=W^n$ with $2<n<\infty$, where $W$ belongs to ch-COE, can also be realized in a concrete Floquet setup.
\end{abstract}

\maketitle

%----------------------------------------------------------------------------------------------------------------------------------------------------------------------------------------------------------
\section{Introduction} 
\label{sec_introduction} 

Random matrix theory (RMT) provides a powerful framework~\cite{Brody1981, Beenakker1997, Guhr1998, DAlessio2016} for describing the statistical properties of complex quantum systems whose exact microscopic features are too intricate to solve. Originally introduced by Wigner to model the energy levels of heavy atomic nuclei~\cite{Wigner1955}, RMT posits that the statistical properties of a highly complex system can be accurately captured by an ensemble of matrices with random elements constrained only by certain fundamental symmetries. For example, the presence or absence of time-reversal symmetry and spin-rotation symmetry classifies chaotic or complex quantum systems into $3$ classical Gaussian ensembles~\cite{Guhr1998, Zirnbauerbook} (also called Wigner-Dyson ensembles): the Gaussian orthogonal ensemble (GOE), the Gaussian unitary ensemble (GUE), and the Gaussian symplectic ensemble (GSE), whose entries are, respectively, real, complex, or quaternionic random variables. While the Berry-Tabor conjecture~\cite{Berry1977} states that the quantum energy levels of a system that is classically integrable (regular) obey Poisson statistics, the Bohigas–Giannoni–Schmit conjecture~\cite{Bohigas1984} states that the quantum energy levels of a classically chaotic system show the same statistical properties as the eigenvalues of an appropriate Gaussian ensemble. Crucially, quantum many-body systems with no semiclassical limit, including spin-$1/2$ chains, also display the same statistical features in their energy levels~\cite{DAlessio2016, Hsu1993, Luitz2015}.

For many-body lattice models with local Hamiltonians, quantum chaos is commonly diagnosed by comparing energy-level statistics, such as the level-spacing distribution~\cite{DAlessio2016} or adjacent-gap-ratio distribution~\cite{Oganesyan2007, Atas2013}, with the predictions of the appropriate Gaussian ensemble. In such models, because the local density of states, $\rho(E)$, depends on energy $E$, spectral unfolding is required for the level-spacing distribution to match RMT predictions, while the adjacent-gap-ratio distribution bypasses this. Furthermore, these models often display additional lattice and/or internal symmetries; the Hamiltonian must be further block-diagonalized using the model's commuting unitary symmetries. In the chaotic regime, short-range spectral correlations in the bulk of the spectrum (i.e., far from the energy band edges) are expected to follow the appropriate random-matrix ensemble, with Dyson index $\beta=1$ (GOE), $\beta=2$ (GUE), and $\beta=4$ (GSE), within each sufficiently large symmetry-resolved block~\cite{DAlessio2016, Santos2010}.

For many-body systems with an additional $E$ to $-E$ symmetry (also called spectral mirror symmetry), one must move beyond the original Wigner-Dyson ensembles to describe the universal properties of such eigenspectra. While the energy levels show standard GOE/GUE/GSE correlations in the bulk of the spectrum, they display new universal features near $E=0$~\cite{Verbaarschot2000, Altland1997} that differ across the $7$ symmetry classes that emerge beyond the Wigner-Dyson ensembles. Within this $10$-fold classification of random matrices, often called the Altland–Zirnbauer (AZ) classification~\cite{Altland1997}, there is a one-to-one correspondence between each ensemble and symmetric spaces in Cartan's classification~\cite{Zirnbauer1996, Heinzner2005}. In particular, the $3$ chiral ensembles, denoted by ch-GOE, ch-GUE, and ch-GSE (also known as Gaussian BDI,  Gaussian AIII, and Gaussian CII, respectively), describe situations where a chiral symmetry, represented by a unitary operator $\mathcal{C}$, anticommutes with the Hamiltonian $H$~\cite{Verbaarschot2000}. Consequently, it does not split the eigenstates of the Hamiltonian into independent chiral sectors, but pairs states at energies \(E\) and \(-E\), while allowing unpaired zero modes at $E=0$. A unique characteristic of chiral RMT classes is an index-protected zero-mode multiplicity set by the imbalance between the two chiral subspaces. When the number of zero modes is large, level repulsion pushes the $E \neq 0$ levels away from the origin ($E=0$), and the spectral density approaches the so-called Marchenko-Pastur distribution~\cite{Marchenko1967}, which controls both the typical value of the smallest $E \neq 0$ eigenvalue as well as the characteristic power-law behavior of the density of states away from it.

A particularly interesting many-body setting where these features appear is when intertwining a chiral and a point-group symmetry leads to the ``protection" of an exponentially large number of exact zero modes (in system size) in interacting lattice models~\cite{Turner2018pxp1, Turner2018pxp2, Schecter2018}. For a Hamiltonian $H$ with a chiral symmetry $\mathcal{C}$ such that $\{H,\mathcal{C}\}=0$, the signed zero-mode imbalance in a symmetry sector $S$ of $H$ that also commutes with $\mathcal{C}$ is 
\begin{equation}
\label{eq:zeroimbalance}
  \nu_{0S}=\mathrm{Tr}(P_S\mathcal{C})=\mathcal{N}_{+,S}-\mathcal{N}_{-,S}  
\end{equation} 
where $P_S$ is the sector projector, and $\mathcal{N}_{+,S}$ ($\mathcal{N}_{-,S}$) refer to the number of zero modes with positive (negative) chirality in $S$. Nonzero-energy pairs
make no contribution to the trace. In interacting spin chains, resolving symmetries can make
$|\mathrm{Tr}(P_S\mathcal{C})|$ exponentially large in system size even when the trace over the full Hilbert space is small~\cite{Schecter2018}. While several works have focused on finding quantum many-body scars in such zero-mode manifolds~\cite{Banerjee2021, Karle2021, Biswas2022, Udupa2023, Sau2024, Budde2024, Pal2025, Ivanov2025, Shashikanta2025, Gupta2026, Eloi2026, ivanov2025exactarealawscareigenstates, Mukherjee2026}, two recent works~\cite{jonay2026, kawabata2026} have shown that such protected zero modes can coexist with many-body quantum
chaos and that chiral level repulsion can separate them from the remaining states on a scale much
larger than the local mean level spacing. 

Periodic driving~\cite{DAlessio2016, Bukov2015, Eckardt2017, Sen2021} changes the geometry of this problem because the spectrum lies on a circle. In driven systems, the Hamiltonian spectrum is replaced by the quasienergy spectrum of the Floquet unitary. For a time-dependent Hamiltonian
$H(t + T) = H(t)$, with $T$ being the period of the drive protocol, the Floquet unitary is
\begin{equation}
\label{eq:defofU}
    U \equiv U(T,0)=\mathcal{T}\exp \left[ -i\int_0^T H(t) dt\right]
\end{equation}
with eigenvalues $\exp(i\theta)$ (here $\mathcal{T}$ denotes time-ordering). Generic Floquet
systems with many interacting degrees of freedom are expected to exhibit circular-ensemble statistics~\cite{DAlessio2014}, with the appropriate ensemble
determined by their unitary and antiunitary symmetries within each irreducible sector~\cite{Regnault2016}. Unlike time-independent local Hamiltonians, Floquet systems do not conserve energy. In a generic Floquet-ergodic regime~\cite{DAlessio2014, Lazarides2014, Ponte2015}, the quasienergy density of states, $\rho(\theta)$, is expected to become completely flat across the spectrum for large systems, even for a local $H(t)$.

Given a unitary chiral operator $\mathcal{C}$, if $U(T,0)$ satisfies $\mathcal{C} U(T,0) \mathcal{C}^{-1} =U(T,0)^{\dagger}$, then chiral symmetry pairs $\theta$ with $-\theta$ and the ch-GOE, ch-GUE, and ch-GSE are replaced by appropriate chiral circular ensembles~\cite{Zirnbauer_1996, Nishigaki_2003, Zirnbauerbook, Altland1997} denoted as ch-COE, ch-CUE, and ch-CSE (where C stands for circular). Such ensembles are characterized by $2$ chiral imbalances~\cite{Asboth2014, Nathan_2015} since both $\theta=0$ and $\theta=\pi$ are fixed points under $\theta \rightarrow -\theta$. The chiral imbalance at both $\theta=0$ and $\theta=\pi$ in any symmetry sector $S$ of $U(T,0)$ such that the symmetry commutes with $\mathcal{C}$, which we denote by $\nu_{0S}$ and $\nu_{\pi S}$ respectively, are defined as 
\begin{eqnarray}
\label{eq:zeropi_im}
    \nu_{0S}=\frac{1}{2} [\mathrm{Tr}(P_S\mathcal{C})+\mathrm{Tr}(P_S\mathcal{C}U(T,0))] \nonumber\\
    \nu_{\pi S}=\frac{1}{2} [\mathrm{Tr}(P_S\mathcal{C})-\mathrm{Tr}(P_S\mathcal{C}U(T,0))]. 
\end{eqnarray}
to give the corresponding imbalances in that sector. These imbalances are protected as long as the symmetries of $U(T,0)$ stay unchanged upon varying some parameter. Since $U(T,0) \rightarrow \mathbbm{1}$ as $T \rightarrow 0$ (Eq.~\eqref{eq:defofU}), where $\mathbbm{1}$ denotes the identity matrix, this immediately restricts one chiral imbalance. Given the topological nature of $\nu_{0S}$ and $\nu_{\pi S}$, these quantities do not change as the driving period $T$ varies. This implies $\nu_{\pi S}=0$, so {\it{any}} non-trivial spectral features appear only near $\theta=0$ if $|\nu_{0S}|$ is large. We specifically consider interacting Floquet problems, where both $|\nu_{0S}|$ and $\mathcal{D}_S$, where the latter denotes the Hilbert space dimension of $S$, grow exponentially with system size~\cite{PXP_driven1, PXP_driven2}, but $1 \ll |\nu_{0S}| \ll \mathcal{D}_S$ for large lattices. 
The ordinary chiral circular orthogonal
ensemble, ch-COE $(\nu_{0S},0)$, where the second argument is zero since there is no chiral imbalance at $\pi$, then provides the simplest benchmark for a circular BDI block. 
We ask when this benchmark describes the spectrum of $U(T,0)$ and when the structure of the drive produces a
different density of eigenphases near $\theta=0$ within the same symmetry class. 

In a nutshell, we see that $U(T,0)$ satisfies ch-COE $(\nu_{0S},0)$ if the two halves of the drive are not related to each other by a unitary symmetry. On the other hand, a {\it{dynamical symmetry}} in the Floquet setup allows for richer possibilities. The mechanism considered here is a symmetry that exchanges the two halves of the drive in square-pulse protocols, where $U(T,0)=U_2 U_1$, with $U_1=U(T/2,0)$ and $ U_2=U (T,T/2)$. If a unitary $G$ satisfies 
\begin{equation}
\label{eq:mechanismG}
    H(t+T/2) = G H(t) G^{-1},
\end{equation}
 then $U_2=G U_1 G^{-1}$ and $G^2$ commutes with the half-period evolutions. It follows that
\begin{equation}
\label{eq:symmG}
  W(T,0) \equiv G U_1,\  W(T,0)^2 = G^2U(T,0).  
\end{equation}
In a sector where $G^2 = \mathbbm{1}$, this gives $U(T,0) = W(T,0)^2$. More generally, when $G^2$ acts as a phase in a
sector preserved by $W(T,0)$, a choice of phase for $W(T,0)$ gives the same relation. The combined
action of a spatial translation or spin inversion and half-period evolution is closely related to the construction used for
time-glide symmetries~\cite{PhysRevB.95.195155} where $G$ corresponds to a spatial reflection.

The same $W(T,0)$ also appears in the symmetry algebra of $U(T,0)$ in our models. Each chiral operation
or time-reversal operation conjugates $U(T,0)$ to $U(T,0)^\dagger$. The product of two operations of either kind commutes
with $U(T,0)$ and coincides with $W(T,0)$ up to a phase. We further show that the half-step symmetry relations
also establish that $W(T,0)$ satisfies circular BDI symmetry. As $T \rightarrow 0$, $W(T,0) \rightarrow G$, which allows a nonzero protected index at either or both $\phi=0$ and $\phi=\pi$ from Eq.~\eqref{eq:zeropi_im} (where the eigenvalues of $W(T,0)$ are $\exp(i \phi)$), and thus $W(T,0)$ belongs to ch-COE($\nu_0,\nu_\pi)$ in general. $W(T,0)$ can either exchange symmetry blocks of $U(T,0)$ or preserve a
symmetry block of $U(T,0)$ within the same model. For the former case, $U(T,0)$ continues to be described by a ch-COE($\nu_0,0$). When $W(T,0)$ preserves a symmetry-resolved block $S$ of $U(T,0)$, the map $\theta = 2 \phi$ (mod $2\pi$) folds two parts of its spectrum onto that block. The density of the eigenphases, $\rho_U(\theta)$ and $\rho_W(\phi)$, are then related by 
\begin{equation}
\label{eq:foldWtoU}
    \rho_U(\theta)=\frac{1}{2} \left[\rho_W(\theta/2)+\rho_W(\theta/2+\pi)\right]\ ,
\end{equation}
with arguments understood modulo $2\pi$. Protected eigenphases at both $\phi=0$ and $\phi=\pi$ contribute to
the zero-quasienergy manifold of $U(T,0)$ from Eq.~\eqref{eq:foldWtoU}. Each gap in the continuous spectrum of $W(T,0)$ contributes its own
threshold to the folded density. Two unequal gaps give a first spectral edge when one branch starts contributing
and a second square-root onset when the other enters. When only one branch is gapped, the other
contributes all the way to zero, producing a plateau at approximately half the bulk density. The wrapping ($\theta=2\phi$ mod $2\pi$) further changes the adjacent gap-ratio distribution, which is then described by a superposition of two GOE spectra~\cite{Alet_paper} rather than a single GOE. 

We illustrate these possibilities with Floquet versions of three non-integrable spin-chain models on periodic lattices using square-pulse drives. The uniformly driven PXP chain (model I)~\cite{PXP_driven1, PXP_driven2} gives the ordinary ch-COE benchmark, since the periodic driving has no unitary symmetry $G$ that exchanges the two halves of the drive. In the staggered PXP chain
(model II)~\cite{Chen2021, Mukherjee2022}, a one-site translation exchanges the two drive steps, and we find  $W(T,0)$, which has protected imbalances at both
fixed eigenphases. Using two large symmetry-resolved sectors of $U(T,0)$, we show that while its eigenspectrum satisfies ordinary ch-COE in one sector, it cannot be captured by ch-COE in the other sector. In the periodic three-spin Ising chain (model III)~\cite{Udupa2023}, a global spin operation exchanges
the two drive steps, and the sectors displayed have a protected imbalance at only one fixed eigenphase of $W(T,0)$. We also discuss the spectral features of the corresponding static versions for models I--III, which are described by ch-GOE. Since these models are non-integrable, the eigenspectrum of both $H$ in the static case and $U(T,0)$ in the Floquet case for models I--III is generated using exact diagonalization (ED) techniques for finite $L$. To compare the numerical data with RMT predictions, we work in the large irreducible sectors of each model, using the maximum number of mutually commuting symmetries to maximize the numerically accessible chain length $L$. For the Floquet versions, we choose the drive period $T$ and other parameters in $H(t)$ to be comparable to each other to be away from any perturbative regime so that $\rho(\theta)$ is constant in the bulk of the spectrum (sufficiently away from $\theta=0$).

The rest of the paper is organized as follows. In Sec.~\ref{sec_circular_BDI} we review the density of states $\rho(E)$ for ch-GOE, relevant for the static problems, as well as the density of eigenphases $\rho(\theta)$, relevant for the Floquet versions, from the circular ensemble with BDI symmetry. The latter is determined solely by $|\nu_0S|$ and $\mathcal{D}_S$ and has no other fitting parameters in the Floquet regime where the bulk value of $\rho(\theta)$ is constant. We also work out the generalized $\rho(\phi)$, which is relevant to $W(T,0)$ since it can have protected modes at both $\phi=0$ and $\phi=\pi$, unlike $U(T,0)$. In Sec.~\ref{sec_models}, we introduce the PXP model (model I), the staggered PXP model (model II), and the three-spin Ising model (model III) under periodic boundary conditions (PBC), in both their static and periodically driven versions. For each model, we identify the relevant symmetries and the chiral imbalances $\nu_{0,\pi}$, construct the operator $W(T,0)$ with $W^2=U(T,0)$ where it exists, and compare the predicted densities of eigenphases and the level statistics from RMT with ED results on finite chains. We conclude and discuss some open directions in Sec.~\ref{sec_discussions_conclusions}. We discuss how to numerically extract the mean level spacing for the static problems near $E=0$, derive the density of states $\rho(\theta)$ using a saddle-point method, and provide technical remarks on ED in appendices~\ref{app_delta_fit}--\ref{app_tech_remarks}. Appendix~\ref{app_W_n} generalizes the Floquet construction of model III to $n$-step drives with $U(T,0)=W(T,0)^n$ for $n \geq 3$, and explicitly discusses numerical results for $n=3$.

%----------------------------------------------------------------------------------------------------------------------------------------------------------------------------------------------------------
\section{Brief review of chiral GOE and chiral COE}
\label{sec_circular_BDI}

\subsection{Chiral GOE}
\label{subsec:chiralGOE}
We begin by revisiting relevant results from the chiral Gaussian orthogonal ensemble (ch-GOE) (also called Gaussian BDI), where a canonical representative $H_s$ that anticommutes with a unitary chiral operator $\mathcal{C}_s$ can be cast into a block off-diagonal form as
\begin{equation} \label{eq_H_block}
H_s=\begin{pmatrix}0&M\\M^{T}&0\end{pmatrix} \ ,
\end{equation}
where $M$ is a real rectangular $p \times q$ matrix whose elements are independent Gaussian random variables with zero mean and variance $\sigma^2$. In addition to chiral symmetry $H_s$ has (an anti-unitary) time-reversal symmetry $\mathcal{T}_s$ such that $\mathcal{T}_s^2=\mathbbm{1}$, the $\mathcal{D} \times \mathcal{D}$ identity operator and $\rm{Tr}(\mathcal{C}_s):= \nu = p-q$ is the chiral imbalance of the sector with size $\mathcal{D}=p+q$; we take $p\geq q$. There exists a basis in which $\mathcal{T}_s$ can be taken as the complex conjugation operator $\mathcal{K}$. We list the symmetry relations for the aforementioned operators below. 
\begin{equation} \label{eq_Gaussian_BDI_action}
    \begin{aligned}
        \mathcal{C}_s H \mathcal{C}_s^{-1} = -H \ &, \quad \mathcal{C}_s^2 = \mathbbm{1} ; \\
        \mathcal{T}_s H \mathcal{T}_s^{-1} = H \ &, \quad \mathcal{T}_s^2 = \mathbbm{1} ; \\
        \mathcal{P}_s H \mathcal{P}_s^{-1} = -H \ &, \quad \mathcal{P}_s^2 = \mathbbm{1} .
    \end{aligned}
\end{equation}
where $\mathcal{P}_s$ is another anti-unitary operator, called particle-hole symmetry, which can be constructed as $\mathcal{P}_s := \mathcal{C}_s\mathcal{T}_s$. These three operators mutually commute. The requirement $\mathcal{P}_s^2=\mathbbm{1}$ enforces commutation between $\mathcal{C}_s$ and $\mathcal{T}_s$; the other two commutation relations, namely $\left[\mathcal{P}_s, \mathcal{C}_s\right]=0 =\left[\mathcal{P}_s, \mathcal{T}_s\right]$ then follow immediately from the definition of $\mathcal{P}_s$.

Due to chirality, the spectrum of $H_s$ is symmetric about $E=0$, and the non-zero eigenvalues of $H_s^2$ are those of the $p \times p$ Wishart matrix $MM^{T}$, which has the following joint probability distribution.
\begin{equation} \label{eq_Wishart_dist}
\begin{aligned} 
    & P(\lambda_1, \lambda_2, \cdots , \lambda_q) \\ \quad &\propto \left( \prod_{i=1}^{q} \lambda_i^{(\nu-1)/2} e^{- \lambda_i/(2 \sigma^2)} \right) \prod_{i<j} |\lambda_i - \lambda_j| \ ,
\end{aligned}
\end{equation}
where $\lambda_i = E_i^2>0$ for $i=1,\cdots,q$ and the proportionality constant follows from the normalization condition. $H_s$ has exactly $\nu$ zero modes provided $M$ has the maximum rank $q$ and its non-zero levels form $q$ number of $\pm E_i$ pairs. For $\mathcal{D} \to \infty $, the density of states, $\rho(E)$, normalized to unity, follows the Marchenko–Pastur law~\cite {Marchenko1967}, which can be obtained from a saddle-point analysis of Eq.~\eqref{eq_Wishart_dist}:
\begin{equation}  \label{eq_MP_law}
\begin{aligned} 
    \rho(E) = \frac{\nu}{\mathcal{D}} \delta(E)  &+ \frac{1}{\pi \sigma^2 \mathcal{D} |E|} \sqrt{(E_+^2 - E^2) (E^2 - E_-^2)}\\ & \quad  \times \Theta(|E| - E_-) \ \Theta(E_+ - |E|) \ , 
\end{aligned}
\end{equation}
\begin{equation} \label{eq_E_pm}
\begin{aligned} 
      & E_{\pm} = \sigma\sqrt{(\mathcal{D}/2)} \left[\sqrt{1 + (\nu/\mathcal{D})} \pm \sqrt{1 - (\nu/\mathcal{D})} \ \right] \ .
\end{aligned}
\end{equation}
The continuous spectrum therefore occupies the interval $[-E_+,-E_-]\cup[E_-,E_+]$, while the fraction $\nu/\mathcal D$ of the states remains as exact zero modes at $E=0$. Thus, for nonzero $\nu/\mathcal D$, the zero modes are separated from the continuous bulk by the hard spectral gap $E_{-}$.~\cite{jonay2026}. 

Setting $\nu=0$ in Eq.~\eqref{eq_MP_law} in $\rho(E)$, we obtain the Wigner semicircle law 
%{\color{blue}question: to avoid confusion, should we denote $\rho_W(E)$ as $\rho_{\rm {Wg}}(E)$ and accordingly $\delta_W$ as $\delta_{\rm{Wg}}$?}
\begin{equation}
\label{eq:Wignersemicircle}
    \rho_{\rm {Wg}}(E)=\frac{1}{\pi \sigma^2 \mathcal{D}} \sqrt{2\mathcal{D}\sigma^2-E^2} \times\Theta(\sqrt{2\mathcal{D}} \sigma -|E|)
\end{equation}
from which we see that $\rho_{\rm {Wg}}(E)$ is maximized at $E=0$. Defining the level spacing $\delta_{\rm{Wg}}$ in the neighborhood of $E=0$, we then get 
\begin{equation} \label{eq_typ_spacing}
\begin{aligned}
    \frac{1}{\delta_{\rm{Wg}}} &:= \mathcal{D} \rho_{\rm {Wg}}(E)  \\
    \Rightarrow \delta_{\rm{Wg}} &= \frac{\pi \sigma}{\sqrt{2 \mathcal{D}}} \ .
\end{aligned}
\end{equation}
So in terms of $\delta_{\rm{Wg}}$, one gets the gap separating the zero modes $E_- = (\delta_{\rm{Wg}} \mathcal{D}/\pi) \left[\sqrt{1 + (\nu/\mathcal{D})} - \sqrt{1 - (\nu/\mathcal{D})} \right]$, which approximately equals 
\begin{equation}
\label{eq:gapenh}
    E_{-}=(\nu \delta_{\rm{Wg}})/\pi
\end{equation}
when $1\ll \nu \ll \mathcal{D}$, which is the regime of interest for this study. Eq.~\eqref{eq:gapenh} shows that a spectral gap, much larger than the typical level spacing, arises when many ($\nu \gg 1$) protected zero modes exist at $E=0$.

For a local microscopic Hamiltonian, RMT predictions for the spectral gap and its neighborhood can be tested faithfully around the middle of the spectrum. We therefore introduce a system-specific cut-off $W_{\rm cut}$ such that the density of states, apart from a delta function at $E=0$, can be fitted as (see Appendix~\ref{app_delta_fit} for more details)
\begin{align} \label{eq_modified_MP}
    \rho_{\rm fit}(E) \approx 
    \begin{cases}
        0,  \ |E|<\Delta \\
        \frac{1}{\mathcal{D}\delta_{\rm fit}} \sqrt{1- \left( \frac{\Delta}{E}\right)^2}, \ \Delta \leq|E| \leq W_{\rm cut} \ ;
    \end{cases}
\end{align}
where $\Delta$ is the first non-zero positive eigenvalue and $\delta_{\mathrm{fit}}$ is a fitting parameter in Eq.~\eqref{eq_modified_MP}. The RMT prediction is that 
\begin{equation}
\label{eq:staticRMT_prediction}
    \Delta/\delta_{\rm fit} \approx \nu / \pi
\end{equation}
in the limit $1 \ll \nu \ll \mathcal{D}$. We will test this prediction in the static versions of models I--III later in the paper. $\delta_{\mathrm{fit}}$ can also be extracted using another method discussed in Ref.~\cite{kawabata2026} and Appendix~\ref{app_delta_fit}, which gives results quite similar to the fitting procedure explained above. Eq.~\eqref{eq_modified_MP} also shows that $\rho(E)$ exhibits a characteristic square-root onset at the spectral gap. 

\subsection{Chiral COE}
\label{subsec:chiralCOE}

We now consider a unitary matrix $U$ that has BDI symmetry (in accordance with A-Z classification~\cite{PhysRevB.55.1142}), defined by chiral symmetry $\mathcal{C_F}$, time-reversal $\mathcal{T_F}$, and particle-hole $\mathcal{P_F}$, which satisfy the following relations.
\begin{equation} \label{eq_circular_BDI_action}
    \begin{aligned}
        \mathcal{C_F} U \mathcal{C_F}^{-1} = U^{\dagger} \ &, \quad \mathcal{C_F}^2 = \mathbbm{1} ; \\
        \mathcal{T_F} U \mathcal{T_F}^{-1} = U^{\dagger} \ &, \quad \mathcal{T_F}^2 = \mathbbm{1} ; \\
        \mathcal{P_F} U \mathcal{P_F}^{-1} = U \ &, \quad \mathcal{P_F}^2 = \mathbbm{1} .
    \end{aligned}
\end{equation}
Here alike the static case, $\mathcal{P_F}:= \mathcal{C_F} \mathcal{T_F}$ and all the three operators, namely $\mathcal{C_F},\mathcal{T_F},\mathcal{P_F}$ are mutually commuting. The spectrum of $U$ is given by the eigenphases $\{e^{i\theta}\}$ that lies on a unit circle and $\theta=0, \pi$ are the fixed points of spectral reflection $\{e^{i \theta}\} \leftrightarrow \{e^{-i \theta}\}$ due to chiral symmetry. As is clear from Eq.~\eqref{eq:zeropi_im}, consequently, there are two chiral imbalances, denoted by $\nu_0$ and $\nu_\pi$. We denote by \(\mathrm{ch\text{-}COE}(\nu_0,\nu_\pi)\) the chiral circular orthogonal ensemble at fixed zero- and \(\pi\)-quasienergy imbalances. The pair \((\nu_0,\nu_\pi)\) labels members of the BDI ensemble family, rather than distinct symmetry classes; the eigenphase density depends on their magnitudes. The standard chiral BDI random matrix ensemble has $\nu_\pi=0$~\cite{Edelman2022Cartan}. For such an RMT ensemble, in terms of the variables $y_i := \sin^2(\theta_i/2)$, the joint probability distribution function (JPDF) for the eigenphases of $U$ reads~\cite{Edelman2022Cartan, Nishigaki_2003}
\begin{equation} \label{eq_circ_BDI_JPDF}
\begin{aligned} 
    & P(y_1,y_2, \cdots , y_q) \\ & \quad \propto \left(\prod_i y_i^{(|\nu_0| -1)/2} \ (1-y_i)^{-1/2} \right) \ \prod_{i<j} |y_i - y_j|\ ,
\end{aligned}
\end{equation}
where $\nu_0$ and $\nu_{\pi}=0$ are the imbalances in $\theta=0,\pi$ subspaces respectively. From the expression in Eq.~\eqref{eq_circ_BDI_JPDF}, we derive the density of states, normalized to unity, for the eigenphases $\theta \in (-\pi, \pi]$ of $U$ using a saddle-point approach (see Appendix~\ref{app_derivation})~\cite{Ramli_2012, Marino_2016, tricomi1985integral}
\begin{equation} \label{eq_circ_BDI_dos}
\begin{aligned}
    \rho (\theta) &= \frac{|\nu_0|}{\mathcal{D}} \delta(\theta) + \frac{(\mathcal{D}-1)}{2 \pi \mathcal{D}} \sqrt{1 - \frac{a(\theta_0)}{\sin^2(\theta/2)}} \ \Theta(|\theta| - \theta_0) \ , \\ 
    \rm{with} & \quad a(\theta_0) := \sin^2(\theta_0/2) = \left(\frac{|\nu_0| -1}{\mathcal{D}-1} \right)^2 \ .
\end{aligned}
\end{equation}
Thus, the spectrum has sharp edges at $\theta=\pm \theta_0 = \pm 2 \arcsin{(\sqrt{a})}$ where the density of states becomes zero from the usual uniform distribution $\approx 1/(2\pi)$ (observed throughout the entire range of $\theta$ in Wigner-Dyson classes) at the bulk. We note that the gap that separates the nullspace depends on the scaling of $|\nu_0|$ with $\mathcal{D}$ and may remain $\mathcal{O}(1)$ in the thermodynamic limit if the ratio $(|\nu_0|/\mathcal{D})$ is fixed. As observed from Eq.~\eqref{eq_circ_BDI_dos}, there is no repulsion from $\theta = \pi$ modes. We emphasize here that, unlike ch-GOE, the density of states in Eq.~\eqref{eq_circ_BDI_dos} and the gap $\theta_0$ do not depend on any model-specific parameter(s). $\rho(\theta)$ again exhibits a square-root onset at the spectral edges $\theta=\pm \theta_0$ that approaches the bulk value of $1/2\pi$. 

To understand the eigenphase density for $W(T,0)$, we need to generalize this analysis to the case where both chiral imbalances $\nu_0$ and $\nu_{\pi}$ can be nonzero, and there are spectral edges both at $\phi=0$ and $\phi=\pi$. Modifying Eq.~\eqref{eq_circ_BDI_JPDF} as~\cite{Holcomb2012, Edelman2022Cartan}
\begin{equation} \label{eq_circ_BDI_JPDF_generic}
\begin{aligned} 
    & P(y_1,y_2, \cdots , y_q) \\ & \quad \propto \left(\prod_i y_i^{(|\nu_0| -1)/2} \ (1-y_i)^{(|\nu_{\pi}| -1)/2} \right) \ \prod_{i<j} |y_i - y_j|\ ,
\end{aligned}
\end{equation}
We obtain a more complicated density of states for the eigenphases $\phi \in (-\pi,\pi]$ of $W$ as follows (See Appendix~\ref{app_derivation_additional}).~\cite{Ramli_2012, Marino_2016, tricomi1985integral}
\begin{equation} \label{eq_W_dos}
\begin{aligned}
\rho_W(\phi) &= \frac{|\nu_0|}{\mathcal D}\,\delta(\phi) + \frac{|\nu_\pi|}{\mathcal D}\,\delta(\phi-\pi) \\
&\quad + \left(\frac{\mathcal{D}-2}{2\pi\mathcal D}\right) \sqrt{1-\frac{A^2}{\sin^2(\phi/2)}-\frac{B^2}{\cos^2(\phi/2)}} \\
&\qquad \qquad \times\  \Theta\!\left(|\phi|-\phi_0\right)\, \Theta\!\left(\phi_\pi-|\phi|\right) \ , \\
A&=\frac{|\nu_0|-1}{\mathcal D-2}\ , \quad B=\frac{|\nu_\pi|-1}{\mathcal D-2} \ , \\
\phi_0 &= \arcsin(A+B)+\arcsin(A-B) \ , \\
\phi_\pi &= \pi-\arcsin(A+B)+\arcsin(A-B) \ .
\end{aligned}
\end{equation}
As apparent from Eq.~\eqref{eq_W_dos}, $\rho_W$ has sharp spectral edges at $\pm \phi_0,\pm \phi_{\pi}$. Again, this density of states only depend on $|\nu_0|$, $|\nu_\pi|$ and $\mathcal{D}$, and not on any other microscopic details.

In the limit $1\ll|\nu_0|,|\nu_\pi|\ll\mathcal{D}$, Eq.~\eqref{eq_W_dos} essentially reduces to Eq.~\eqref{eq_circ_BDI_dos} in the neighborhood of $\phi=0$ and $\phi=\pi$. In this regime, we have $A,B\ll1$ and $(\mathcal{D}-2)/\mathcal{D}\simeq(\mathcal{D}-1)/\mathcal{D}\simeq1$. Using the expansion $\arcsin x=x+x^3/6+\mathcal{O}(x^5)$,
\begin{equation}
\begin{aligned}
\phi_0 &= 2A+\tfrac{1}{3}A^3+AB^2+\dots\simeq2A\ ,\\ 
\pi-\phi_\pi &= 2B+A^2B+\tfrac{1}{3}B^3+\dots\simeq 2B \ .
\end{aligned}
\end{equation}
Near $\phi=0$ ($\phi\ll1$) we have $\sin(\phi/2)\simeq\phi/2$ and $\cos(\phi/2)\simeq1$, so that the square root in Eq.~\eqref{eq_W_dos} becomes
\begin{equation}
\begin{aligned}
f^2(\phi)&=1-\frac{A^2}{\sin^2(\phi/2)}-\frac{B^2}{\cos^2(\phi/2)} \\
&\simeq1-\frac{4A^2}{\phi^2}-\Big(\frac{A^2}{3}+B^2\Big),
\end{aligned}
\end{equation}
which is the square root of Eq.~\eqref{eq_circ_BDI_dos} with $\theta=\phi$ and $a(\theta_0)\simeq A^2$, up to corrections of relative order $A^2,B^2$. The second imbalance $\nu_\pi$ thus only shifts the edge slightly. At the edge, $f^2(\phi_0)=0$ and $(f^2)'(\phi_0)=8A^2/\phi_0^3\simeq1/A$, so that for $\phi=\phi_0+\epsilon$ with $0<\epsilon\ll A$, $f^2\simeq\epsilon/A$ and
\begin{equation} \label{eq_W_near_zero}
\rho_W(\phi_0+\epsilon)\simeq\frac{1}{2\pi}\sqrt{\frac{\epsilon}{A}} .
\end{equation}
In Eq.~\eqref{eq_circ_BDI_dos}, $1-a/\sin^2(\theta/2)$ vanishes at $\theta_0$, and $\sqrt{a}=(|\nu_0|-1)/(\mathcal{D}-1)\simeq A$, so that $\rho(\theta_0+\epsilon)\simeq\frac{1}{2\pi}\sqrt{\epsilon/A}$, which is the same as Eq.~\eqref{eq_W_near_zero}. Near $\phi=\pi$ let $\phi=\pi-\chi$ with $0<\chi\ll1$. Then $\sin(\phi/2)=\cos(\chi/2)\simeq1$ and $\cos(\phi/2)=\sin(\chi/2)\simeq\chi/2$, so that
\begin{equation}
\begin{aligned}
f^2&\simeq1-\frac{4B^2}{\chi^2}-\left(A^2 + \frac{B^2}{3}\right)\ ,\\
\rho_W&\simeq\frac{1}{2\pi}\sqrt{1-\frac{4B^2}{\chi^2}}\ \Theta(\chi-2B)\ ,
\end{aligned}
\end{equation}
i.e.\ Eq.~\eqref{eq_circ_BDI_dos} with $\nu_0\to\nu_\pi$ and $\theta\to\pi-\phi$ (the same BDI density with the protected modes at $\pi$). At the edge, $(f^2)'=8B^2/\chi^3\simeq1/B$, so that $\rho_W(\phi_\pi-\epsilon)\simeq\frac{1}{2\pi}\sqrt{\epsilon/B}$ for $0<\epsilon\ll B$: the mirror image of the edge at $\phi_0$ with $A$ replaced by $B$.

%Before closing this section, we note that a microscopic Hamiltonian \(H\), Floquet unitary \(U(T,0)\), or auxiliary unitary \(W(T,0)\) may admit multiple chiral or time-reversal operators satisfying the corresponding BDI relations [Eqs.~\eqref{eq_Gaussian_BDI_action} and \eqref{eq_circular_BDI_action}]. The product of two chiral operators, or of two time-reversal operators, is a unitary symmetry that commutes with the corresponding \(H\), \(U(T,0)\), or \(W(T,0)\). Resolving this symmetry can produce smaller blocks; a block has BDI symmetry, however, only if the relevant chiral and time-reversal operators preserve that block.

\section{Models, their symmetries, and numerical results}
\label{sec_models}

In this section, we introduce the static and Floquet versions of models I--III and discuss their symmetries in both cases. We restrict our numerical analysis to irreducible sectors with the maximum number of mutually commuting symmetries to enable ED to reach reasonably large values of $L$ in each case. In these irreducible sectors, we show that both the static and Floquet versions of models I-III possess BDI symmetry, defined in Eq.~\eqref{eq_Gaussian_BDI_action} and Eq.~\eqref{eq_circular_BDI_action}, respectively. In particular, the Floquet versions of models II and III have additional features that we will elaborate on. We briefly discuss other irreducible sectors of these models as well (in particular, model II), but we have not analyzed them numerically in any detail in this work. We will label the chiral, time-reversal, and particle-hole operators with a subscript $s$ ($\mathcal{F}$) for the static (Floquet) case.    

We test the RMT predictions for the density of eigenvalues discussed in Sec.~\ref{sec_circular_BDI} against ED data extracted from the models introduced here. We compute all eigenspectra in the appropriate symmetry-resolved sectors, using the chiral block structure inherent to the problems considered here and the Chebyshev propagation method described in Appendix~\ref{app_tech_remarks}. This allows us to obtain spectra of irreducible sectors as large as $\mathcal{D} \sim 1.5 \times 10^5$.
For each sector, the dimension of the sector $\mathcal{D}$, the chiral imbalances $\nu_0$ for the static case as well as $\nu_{0,\pi}$ for the Floquet case are obtained exactly from the basis states, as the traces $\mathrm{Tr}(\mathcal{C})=\nu_0+\nu_\pi$ and $\mathrm{Tr}(\mathcal{C} \mathcal{U})=\nu_0-\nu_\pi$ (Eq.~\eqref{eq:zeropi_im}), where $\mathcal{U}=\mathbbm{1}$ for $U(T,0)$ and $\mathcal{U}=G$ for $W(T,0)$ in the limit $T \rightarrow 0$, given the invariance of these quantities with $T$. The saddle-point prediction from RMT for the density of eigenvalues is free of any other model-specific parameters in the Floquet case, and requires an additional fitting parameter for the bulk level spacing $\delta_{\rm fit}$ near $E=0$ extracted from the model's data for the static case (Eq.~\eqref{eq_modified_MP}). Unless stated otherwise, histograms of the density of states are normalized to unity over the whole spectrum, so the protected $0/\pi$ modes appear as $\delta$-functions of weight $|\nu_{0/\pi}|/\mathcal{D}$, which we indicate with a vertical line in the numerical plots (we truncate the $\delta$-peak to a finite height in each case for visual clarity and indicate it with a red double oblique). We also monitor the adjacent gap-ratio distribution, $P(r)$, for $r \in [0,1]$ to investigate the statistics of the bulk eigenstates. For this, we monitor the level-spacing ratios~\cite{Alet_paper, PhysRevE.80.065201} defined as 
\begin{align} 
\label{eq_lsr}
    r_n := \frac{{\rm min} (E_n - E_{n-1}, E_{n+1} - E_{n})}{{\rm max}(E_{n} - E_{n-1}, E_{n+1} - E_{n})} \ , \quad 0\leq r_n \leq 1 
\end{align}
for a strictly ordered spectrum of eigenvalues $\{E\}$ (likewise, eigenphases $\{\theta\}$ for the driven problem) lying in the bulk of positive energies (positive phases) for all three aforementioned systems, thereby avoiding the correlations between the levels at $\pm E$ ($\pm\theta$) that are induced by the chiral symmetry. We compare the obtained $P(r)$ with the Wigner--Dyson distribution of the GOE~\cite{PhysRevLett.110.084101},
\begin{align} \label{eq_Pr_beta}
    P_{\mathrm{GOE}}(r) &= \frac{27}{4} \frac{(r+r^2)} {\left(1+r+r^2\right)^{\frac{5}{2}}}\ ,
\end{align}
which gives the mean level-spacing ratio $\langle r\rangle\simeq0.5359$, and with the Poisson distribution, for which $\langle r\rangle\simeq0.3863$, where the former is expected for cases described by ch-GOE and ch-COE($\nu_0,\nu_\pi)$ (in the regime where $1 \ll |\nu_{0,\pi}| \ll \mathcal{D}$) while the latter describes integrable models. For the cases where $U(T,0)=W(T,0)^2$ and $W(T,0)$ preserves a symmetry-resolved block of $U(T,0)$, $P(r)$ for $U(T,0)$ is expected to follow a superposition of two independent GOE spectra~\cite{Alet_paper} with an intermediate value for $\langle r \rangle \simeq 0.4234$.

%-----------------------------------------------------------------------------------------------------------------------------------------------------------------------------------------
\subsection{Model I: PXP model under PBC}
\label{subsec_model_I}
We consider the spin-$1/2$ PXP model~\cite{Turner2018pxp1, Turner2018pxp2, Sachdev_pxp, Fendley_pxp, Lesanovsky_pxp} with $L$ (even) spins on a ring, where the Hamiltonian reads
\begin{align} \label{eq_PXP_Ham}
    H =  \sum_{j=1}^{L} P_{j-1} \sigma_j^x P_{j+1} \ , 
\end{align}
with $\sigma_k^{\alpha}$ for $\alpha \in \{x,y,z\}$ and $P_k = (\mathbbm{1} - \sigma_k^z)/2$ are the Pauli matrices and the down-spin projector at site $k$, respectively. This quantum many-body model is defined in a constrained Hilbert space~\cite{sil2026} in which adjacent spin-up excitations are forbidden, capturing the dynamics of Rydberg atom chains in the so-called `Rydberg blockade' regime~\cite{Bernien2017}. Its non-integrable nature and the emergence of quantum many-body scars~\cite{Turner2018pxp1, Turner2018pxp2, Serbyn_2021} make it a canonical setting for studying weak ergodicity breaking and anomalous thermalization under its time evolution quenched from the N\'eel state. 

%--------------------------------------------------------------# Fig. 1 #--------------------------------------------------------------------------------------------------------------
\begin{figure*} 
    \centering
    \includegraphics[width=0.49\linewidth]{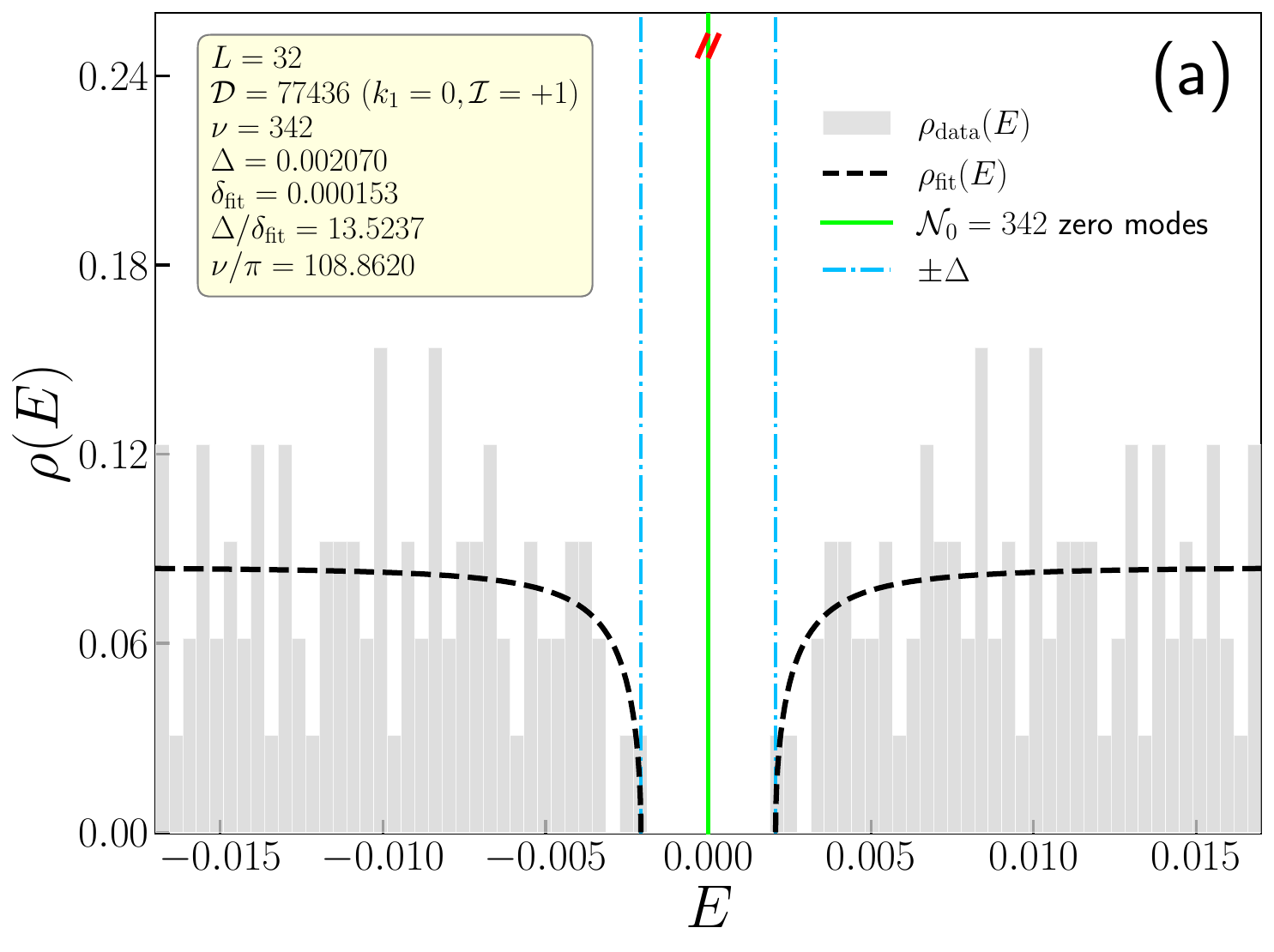}
    \includegraphics[width=0.49\linewidth]{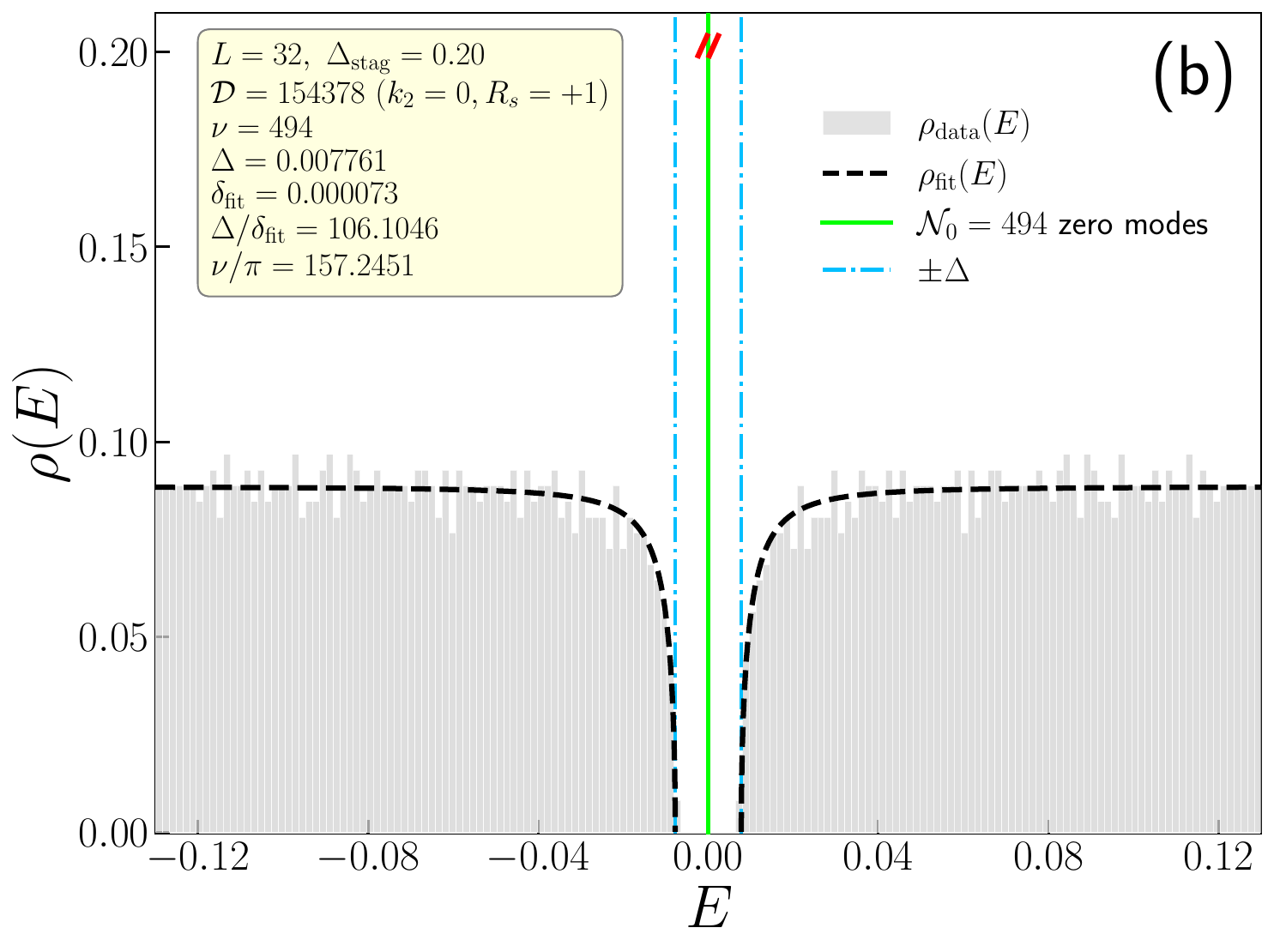}
    \includegraphics[width=0.49\linewidth]{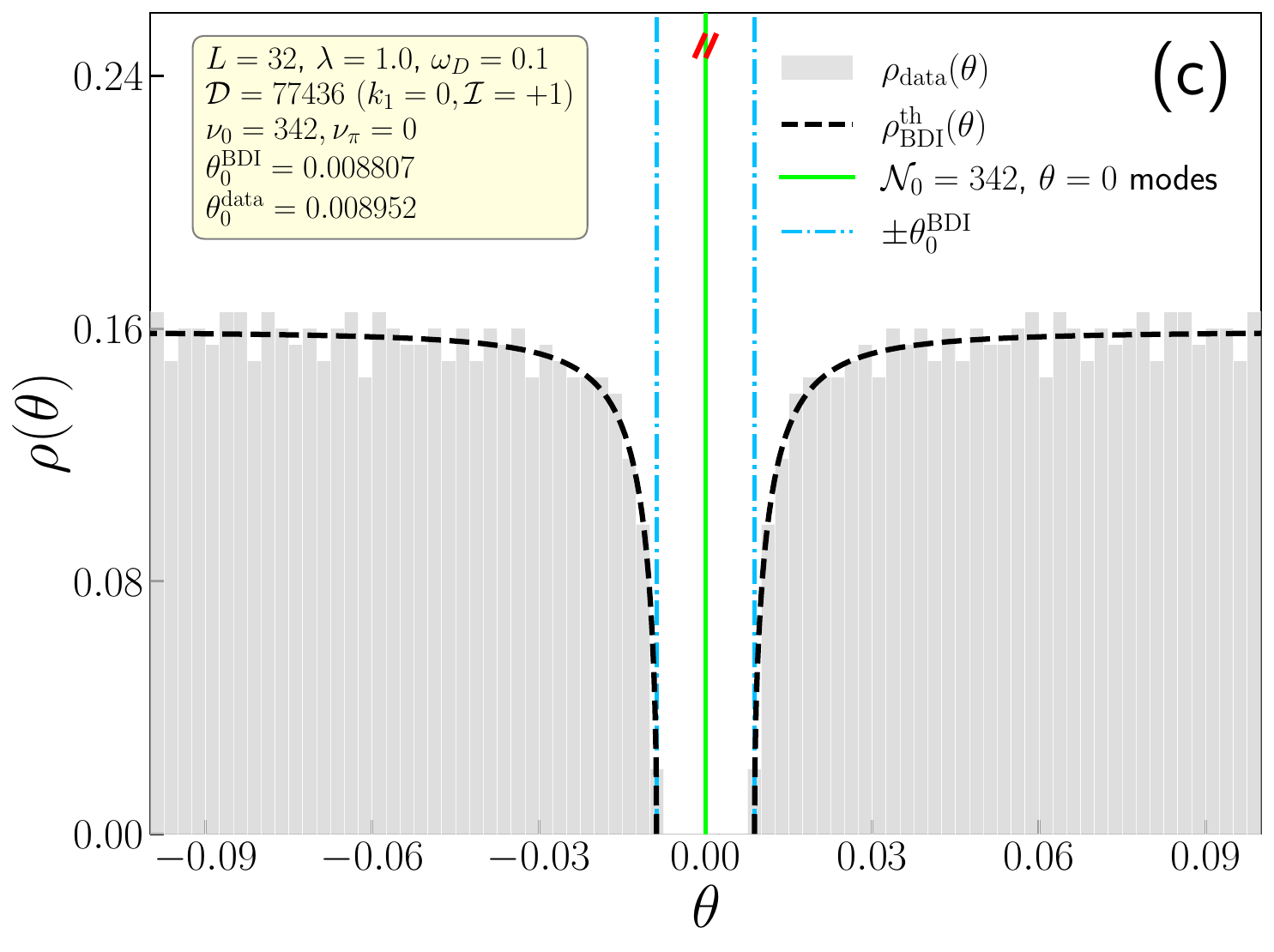}
    \includegraphics[width=0.49\linewidth]{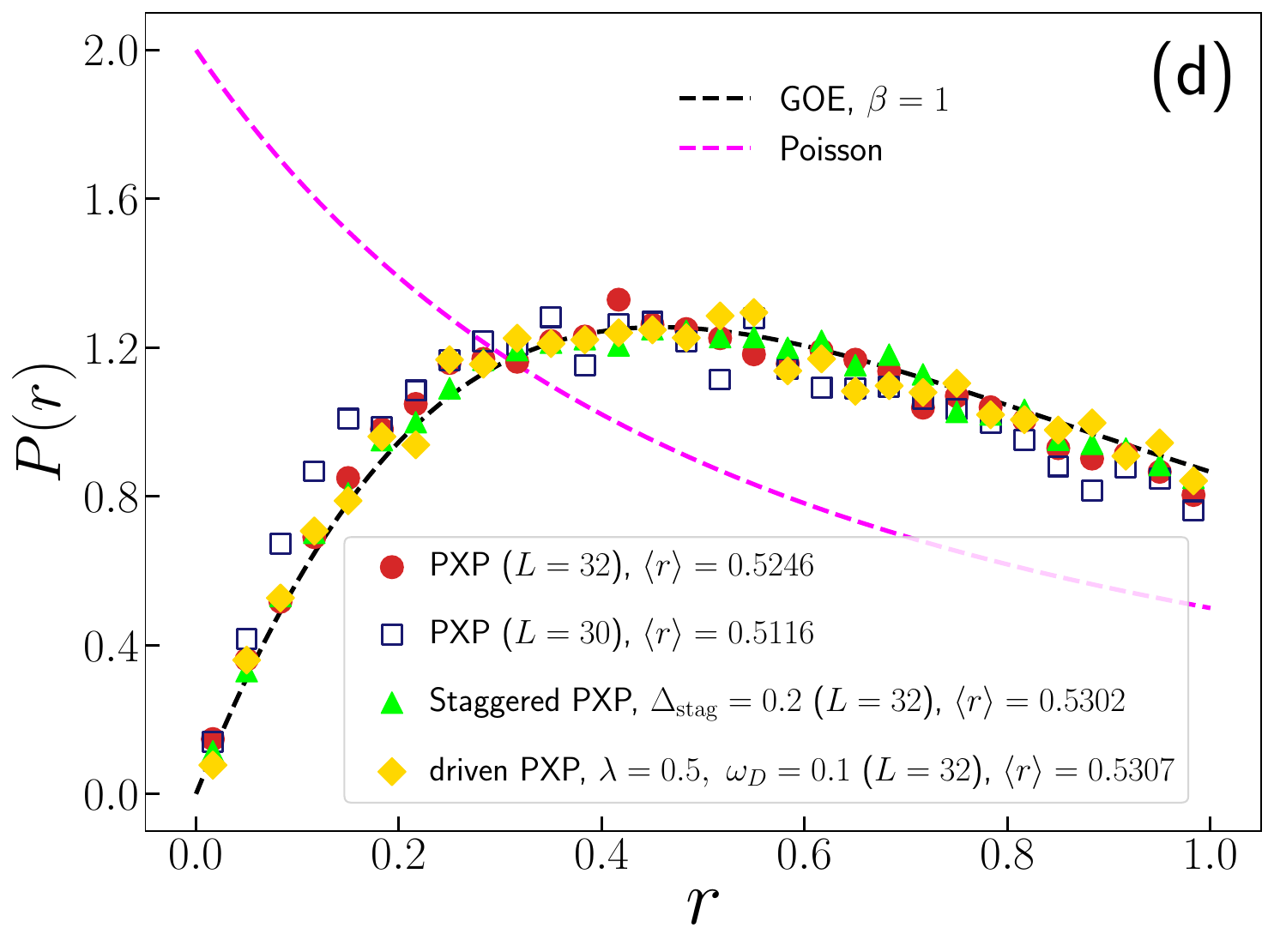}
    \caption{Density of states (a) near zero modes for the PXP Hamiltonian, (b) near the zero modes for the staggered PXP model with staggering strength $\Delta_{\rm stag}=0.2$, and (c) near $\theta=0$ modes for the eigenphases of $U(T,0)$ for the driven PXP model with $\lambda=1,  \omega_D=0.1$, and (d) the comparison of $P(r)$ vs $r$ statistics among the three problems. The plots (a) and (c) correspond to the system size $L=32$ in the $k_1=0,\mathcal{I}=+1$ sector for the static and the driven PXP model, respectively; while the irreducible sector is $k_2=0, R_s=+1$ with system size $L=32$ for the staggered PXP problem (b). In panel (d), we also show data for the static PXP model with $L=30$ (same sector as $L=32$), and we inscribe the means, $\langle r\rangle$, of the level-spacing ratios for each case.}
    \label{fig_1}
\end{figure*}

%---------------------------------------------------------------------------------------------------------------------------------------------------------------------------------------

%--------------------------------------------------------------# Fig. 2 #--------------------------------------------------------------------------------------------------------------
\begin{figure}
    \centering
    \includegraphics[width=\linewidth]{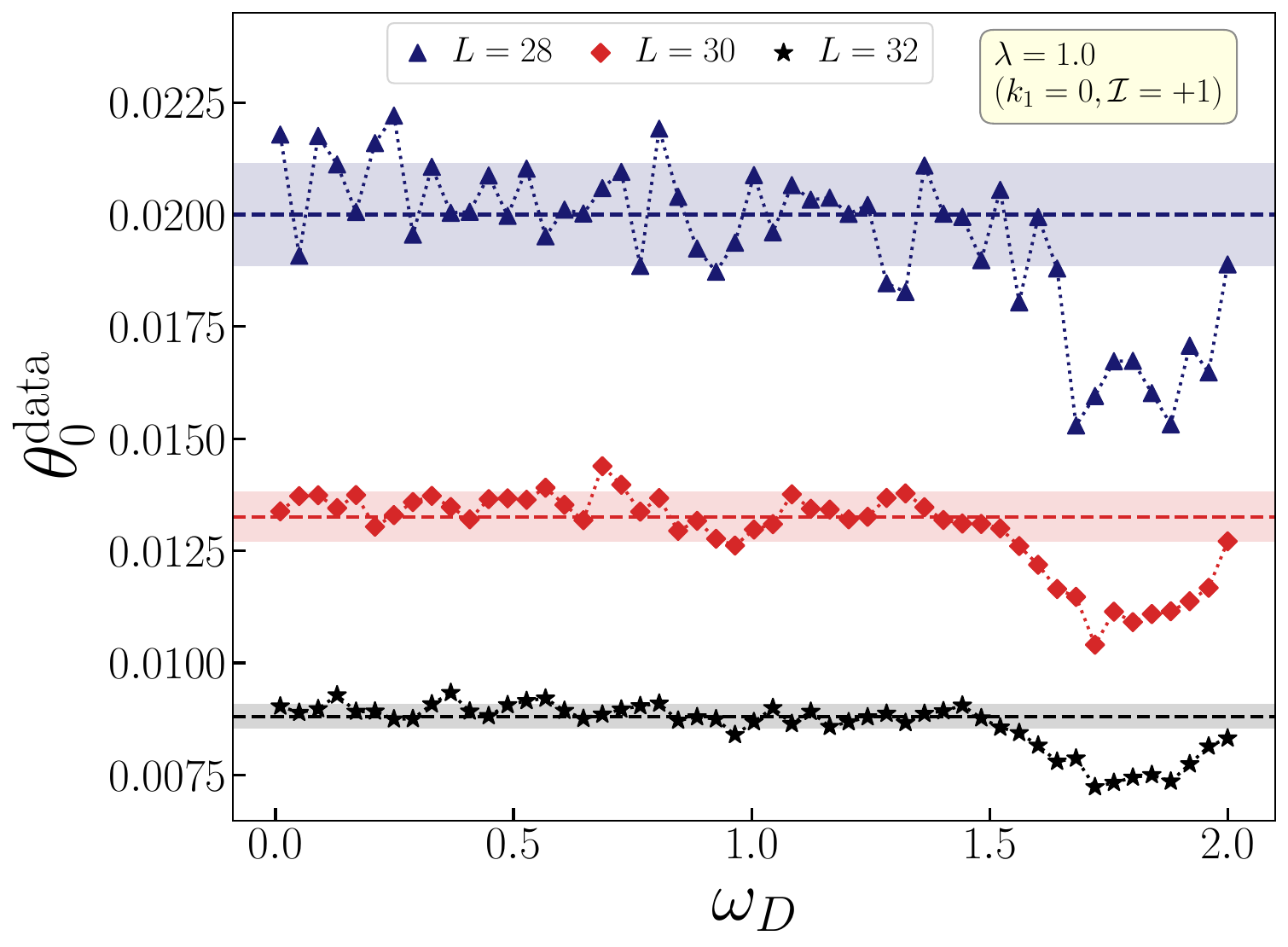}
    \caption{Numerically observed gap $\theta_0^{\rm data}$ as a function of drive frequency $\omega_D$ for the driven PXP model with $\lambda=1.0$ for system sizes $L=28,30$ and $32$ in the $k_1=0, \mathcal{I}=+1$ symmetry resolved sector. The corresponding theoretically predicted gaps $\theta_0^{\rm BDI}=2\arcsin(\frac{|\nu_0|-1}{\mathcal{D}-1})$ are shown by horizontal dashed lines, and the shaded regions correspond to $\theta_0^{\rm BDI} \pm 3\left(\nu_0^{1/3}/\mathcal{D}\right)$.}
    \label{fig_2}
\end{figure}

%---------------------------------------------------------------------------------------------------------------------------------------------------------------------------------------

The static model (Eq.~\eqref{eq_PXP_Ham}) has two symmetries: translation by one lattice site, \(T_1=e^{ik_1}\) (with the lattice spacing set to unity), and spatial inversion \(\mathcal{I}\), which maps site \(i\) to site \(L-i+1\). Since inversion maps \(k_1\) to \(-k_1\), the momentum sectors \(k_1=0\) and \(k_1=\pi\) can be further resolved into sectors with \(\mathcal{I}=\pm1\). The largest symmetry sector obtained by resolving all commuting symmetries is \(k_1=0,\mathcal{I}=+1\). In this sector, the model in Eq.~\eqref{eq_PXP_Ham} is compatible with the BDI symmetry class [Eq.~\eqref{eq_Gaussian_BDI_action}], with
\begin{equation}
\label{eq:BDIstaticmodelI}
\mathcal{C}_s=\prod_j \sigma_j^z,\qquad
\mathcal{T}_s=\mathcal{K},\qquad
\mathcal{P}_s=\mathcal{C}_s\mathcal{K},
\end{equation}
where these operators implement chiral, time-reversal, and particle-hole symmetry, respectively. In the \(\sigma^z\) basis, \(H\) is real and symmetric, so complex conjugation \(\mathcal{K}\) serves as the time-reversal operator. We numerically find that \(\mathrm{Tr}(\mathcal{C}_s)\) in this sector is \(21, 32, 53, 82, 133, 212,\) and \(342\) for \(L=20,22,\ldots,32\), respectively. These values indicate that the chiral imbalance \(\nu_0\) grows approximately as \(\varphi^{L/2}\), where \(\varphi\approx1.618\) is the golden ratio (see also, Refs.~\cite{Turner2018pxp1, Turner2018pxp2, Buijsman2022}). Accordingly, we expect the eigenvalue statistics of \(H\) in the \(k_1=0,\mathcal{I}=+1\) sector to follow those of the ch-GOE [Eq.~\eqref{eq_Gaussian_BDI_action}] with an enhanced spectral gap around $E=0$ dictated by Eq.~\eqref{eq:staticRMT_prediction} due to the large degeneracy of zero modes.

We show the behavior of $\rho(E)$ for the static version of model I in Fig.~\ref{fig_1} (a) in the vicinity of $E=0$ in the $k_1=0, \mathcal{I}=+1$ sector using $L=32$, which gives a large $\mathcal{D}=77436$ for the corresponding symmetry sector and $\nu_0=\mathrm{Tr}(\mathcal{C}_s)=342$. The density of states around $E=0$ shows a $\delta$-function with the weight $\nu/\mathcal{D}$ predicted from the trace, followed by a hard gap, significantly larger than the typical level spacing, and a flat bulk density. The fitted form from Eq.~\eqref{eq_modified_MP} using the range $|E|\leq W_{\rm cut}$ (Fig.~\ref{fig_1}(a); see Appendix~\ref{app_delta_fit}) is also shown using a dashed curve. The first nonzero positive eigenvalue $\Delta$ and the bulk level spacing $\delta_{\rm fit}$ are listed inside Fig.~\ref{fig_1}(a). The ratio $\Delta/\delta_{\rm fit}=13.5$ is, however, much smaller than $\nu/\pi=108.9$. For  a smaller chain of $L=30$,  $\Delta/\delta_{\rm fit}=4.97$ and $\nu/\pi=67.48$ in the same symmetry sector. We suspect that the large discrepancy compared to the RMT prediction (Eq.~\eqref{eq:staticRMT_prediction}) is tied to the depletion of states near $E=0$ (Fig.~\ref{fig_1}(a)) in the PXP model, although we currently do not have a full understanding of this unusual finite-size effect. We examine the extraction of $\delta_{\rm fit}$ in Appendix~\ref{app_delta_fit}, where we compare the curve-fit value with the level-counting estimate in Ref.~\cite{kawabata2026}: the two agree qualitatively, and our method is more stable (Table~\ref{tab:app_spacing_comparison}). The adjacent level-spacing distribution, $P(r)$, for the full spectrum of the static PXP model is close to $P_{\mathrm{GOE}}(r)$ and clearly different from the Poisson one (Fig.~\ref{fig_1}(d)), as expected from ch-GOE. The measured mean values $\langle r\rangle=0.5246$ ($L=32$) and $0.5116$ ($L=30$) for the static model are close to, and slightly below, the GOE value. We notice that the deviation of $\Delta/\delta_{\rm fit}$ from $\nu/\pi$ decreases (see Appendix~\ref{app_delta_fit}) and the bulk statistics qualitatively improve ($\langle r\rangle$ approaches the GOE value) as $L$ is increased from $30$ to $32$ for the static PXP model. The latter observation has been reported earlier in the literature~\cite{Turner2018pxp1}. However, as our observations show, deviations from ch-COE are more dramatic near $E=0$ than for quantities like $\langle r \rangle$, which depend on the full spectrum.

%The Floquet predictions below do not involve a fitted scale, since in Eq.~\eqref{eq_circ_BDI_dos} the gap and the density are fixed by $\nu_0$ and $\mathcal{D}$ alone.

For the Floquet version of model I, we add a magnetic field along the $z$-axis and periodically drive it using a square-pulse protocol~\cite{PXP_driven1, PXP_driven2} with a $T=2\pi/\omega_D$, where $\omega_D$ is the drive frequency,
\begin{equation}  \label{eq_PXP_driven}
\begin{aligned} 
    H(t) &=  \sum_{j=1}^{L} P_{j-1} \sigma_j^x P_{j+1}  + \lambda(t) \sum_{j=1}^L \sigma_j^z \ , \\
    \lambda(t) &= \begin{cases}
        \lambda , \ 0\leq t < \frac{T}{2} , \\
        - \lambda , \ \frac{T}{2} \leq t <T  \ .
    \end{cases}
\end{aligned}
\end{equation}
The stroboscopic time evolution operator, or the Floquet unitary, $U(T,0) = \exp(-i TH(-\lambda)/2) \, \exp(-i TH(\lambda)/2)$ (in the units $\hbar=1$) also commutes with $T_1$ and $\mathcal{I}$, and we diagonalize $U(T,0)$ in the largest symmetry sector $k_1=0, \mathcal{I}=+1$.
In this sector, $U(T,0)$ is compatible with the BDI symmetry class [(Eq.~\eqref{eq_circular_BDI_action})], with
\begin{equation}
\label{eq:BDIFloquetmodelI}
\begin{aligned}
\mathcal{C_F}=\prod_j \sigma_j^z &,\ \mathcal{T_F}=U(T,0)^{-1/2}\exp(-i TH(-\lambda)/2) \, \mathcal{K}, \\
&\mathcal{P_F}=\mathcal{C_F} \mathcal{T_F},
\end{aligned}
\end{equation}
where these operators implement chiral, time-reversal, and particle-hole symmetry, respectively. Eq.~\eqref{eq:BDIFloquetmodelI} can be verified by using $\mathcal{K} H(\pm \lambda) \mathcal{K}=H(\pm \lambda)$ and $\mathcal{C_F} H(\pm \lambda) \mathcal{C_F}=-H(\mp \lambda)$, where $H(\pm \lambda)$ refer to the Hamiltonians with the appropriate sign of $\lambda$ in Eq.~\eqref{eq_PXP_driven} for the two half-cycles. Here \(U(T,0)^{-1/2}\) is taken on the principal branch $(-\pi,\pi)$, assuming \(U(T,0)\) has no eigenvalue at \(-1\) (quasienergy \(\pi\)). While $\nu_\pi=0$ does not rule out accidental $\pi$-modes at fine-tuned parameter values $(\lambda, T)$, we have not encountered this for any parameter values explored in this work (see also Ref.~\cite{PXP_driven2}). Since $\mathcal{C_F}=\mathcal{C}_s$, the chiral imbalance for $U(T,0)$ at $\theta=0$ is identical to that of the static case at $E=0$ for a given $L$, and thus grows as $\varphi^{L/2}$. For the Floquet problem, varying $T$ and $\lambda$ in the regime where the bulk $\rho(\theta)$ is constant gives {{\it different realizations}} of ch-COE($\mathcal{\nu}_0,0$) since $U(T,0)$ changes with $T$ and $\lambda$.  

For the Floquet unitary of the driven PXP model, Eq.~\eqref{eq_PXP_driven}, in the sector $k_1=0$, $\mathcal{I}=+1$, $\nu_\pi=0$ and $\nu_0=\mathrm{Tr}\,(\mathcal{C_F})$, the eigenphase density $\rho(\theta)$ is given by Eq.~\eqref{eq_circ_BDI_dos} without invoking any additional model-specific microscopic parameters. Fig.~\ref{fig_1}(c) shows $\rho(\theta)$ for $L=32$ ($\mathcal{D}=77436$, $\nu_0=342$) at $\lambda=1$ and $\omega_D=0.1$. The histogram follows the curve in Eq.~\eqref{eq_circ_BDI_dos} everywhere, without any adjustable parameters, including the sharp edges at $\theta=\pm\theta_0$. The match to ch-COE($\nu_0,0)$ is significantly better than the one obtained from ch-GOE for the static model (Fig.~\ref{fig_1}(a)).  The predicted saddle-point gap $\theta_0^{\rm BDI}=2\arcsin[(|\nu_0|-1)/(\mathcal{D}-1)]=8.807\times10^{-3}$ should be compared with the smallest positive eigenphase of the data, $\theta_0^{\rm data}=8.952\times10^{-3}$, which is only about $1.6\%$ larger. While the saddle-point gap does not depend on the drive parameters, as expected from Eq.~\eqref{eq_circ_BDI_dos}, fluctuations beyond the saddle-point result persist for any finite $\mathcal{D}$. This is tested in Fig.~\ref{fig_2}, which shows $\theta_0^{\rm data}$ as a function of $\omega_D\in(0,2]$ for $L=28$, $30$ and $32$, together with the saddle point predictions $\theta_0=2\arcsin[(|\nu_0|-1)/(\mathcal{D}-1)]$ ($0.0200$, $0.0133$ and $0.0088$, respectively) as horizontal lines. For $\omega_D\lesssim1.5$, the numerical values fluctuate around the predictions without any systematic dependence on $\omega_D$. In particular most of the fluctuations lie within a region $\left(|\nu_0|^{1/3}/\mathcal{D}\right)$ times an $\mathcal{O}(1)$ number~\cite{Holcomb2012, johnstone2008multivariate}), where the scale of fluctuations around $\theta^{\mathrm{BDI}}_0$ can be obtained from a more careful treatment~\cite{Verbaarschot1994, Majumdar2009} of  Eq.~\eqref{eq_circ_BDI_JPDF}, which yields 
\begin{equation}
\label{eq:fluctuation_theta0}
    \delta \theta_0/\theta_0^{\mathrm{BDI}} \sim |\nu_0|^{-2/3}
\end{equation}
where $\delta \theta_0$ represents the scale of fluctuations in $\theta_0^{\mathrm{data}}$ around $\theta_0^{\mathrm{BDI}}$. For $\omega_D\gtrsim1.5$ the smallest eigenphase falls below the prediction for all three sizes, i.e.\ the Floquet unitary deviates from the ch-COE ensemble at large drive frequency. This is expected since in the limit $\omega_D \to \infty$, $U(T,0)$ approaches $\mathbbm{1}$ for which all the $\mathcal{D}$ eigenvalues pile up at $\theta=0$. $P(r)$ for the Floquet unitary $U(T,0)$ of the driven PXP model is again close to the $P_{\mathrm{GOE}}(r)$ distribution, as expected from ch-COE, and clearly different from the Poisson one (Fig.~\ref{fig_1}(d)). The measured average $\langle r \rangle = 0.5307$ for the driven model is only marginally below the GOE value for $L=32$. 
%----------------------------------------------------------------------------------------------------------------------------------------------------------------------------------------------
\subsection{Model II: Staggered PXP model under PBC}
\label{subsec_model_II}

A more intricate problem from the RMT perspective is obtained by considering the PXP Hamiltonian (Eq.~\eqref{eq_PXP_Ham}) in the same constrained Hilbert space, but with an additional staggered detuning term~\cite{Chen2021, Mukherjee2022},
\begin{align} \label{eq_PXP_stag}
    H_{\rm stag} &= \sum_{j=1}^L P_{j-1} \sigma_j^x P_{j+1} + \Delta_{\rm stag} \sum_{j=1}^L (-1)^j \sigma_j^z \ .
\end{align}
For non-zero $\Delta_{\rm stag}$ and even $L$, $H_{\rm stag}$ commutes with the two-site translation $T_2 = T_1^2 = e^{ik_2}$ and with the site-resolved reflection $R_s$ that interchanges sites $j$ and $L-j+2$ modulo $L$. Since $R_s T_1 R_s^{-1} = T_1^{-1}$, it exchanges $k_2$ with $-k_2$, and can be resolved only at $k_2=0$ and $k_2=\pi$ through $R_s=\pm1$. We first state the BDI structure in a generic momentum sector $k_2$, and then restrict to the two sectors of interest, $k_2=0,\, R_s=+1$ and $k_2=\pi,\, R_s=+1$ (the latter exists for $L$ divisible by $4$), for the static and the Floquet cases. We use the notation $\Gamma=\prod_j\sigma_j^z$ and the complex conjugation $\mathcal{K}$ in the $\sigma^z$ basis, where $H_{\rm stag}$ is real.

%---------------------------------------------#  Table 1  #------------------------------------------------------------------------------------------------------------------------------
\begin{table}
\centering
\begin{tabular}{|c|c|c|c|}
\hline \hline
$k_2$ & $R_s$ & $\mathcal{D}$ & $\rm{Tr}(\mathcal{C}_{s,k_2})$\\
\hline \hline
$0$ & $+1$ & $154378$ & $494$\\
$0$ & $-1$ & $150197$ & $-493$\\
\hline \hline
$\pm\pi/8,\,\pm3\pi/8,\,\pm5\pi/8,\,\pm7\pi/8$ & -- & $304290$ & $0$\\
$\pm\pi/4,\,\pm3\pi/4$ & -- & $304560$ & $0$\\
$\pm\pi/2$ & -- & $304570$ & $0$\\
\hline \hline
$\pi$ & $(+1)\oplus(-1)$ & $304572$ & $0$\\
\hline
\end{tabular}
\caption{Chiral imbalance $\nu=\rm{Tr}(\mathcal{C}_{s,k_2})$ (see Eq.~\eqref{eq:BDIstaticmodelII}) of the static version of model II (Eq.~\eqref{eq_PXP_stag}) at $L=32$. The $k_2=\pi$ entry refers to the direct sum of the two reflection sectors, since the chiral operator exchanges them. The $k_2=0$ reflection-unresolved imbalance is the sum $494-493=1$ of the two resolved entries. For paired momenta $\pm k_2$, the dimension and imbalance shown are those of each momentum block separately.}
\label{tab:static_stag_imbalances}
\end{table}
%---------------------------------------------------------------------------------------------------------------------------------------------------------------------------------------

The operator $T_1\Gamma$ anticommutes with $H_{\rm stag}$ and squares to $T_2$. Together with $R_s \mathcal{K}$ that preserves $k_2$, the BDI operators [Eq.~\eqref{eq_Gaussian_BDI_action}] for a fixed momentum sector $k_2$ are
\begin{equation} \label{eq:BDIstaticmodelII}
\begin{aligned}
    \mathcal{C}_{s,k_2} = e^{-ik_2/2}\,T_1\Gamma , \  \mathcal{T}_{s,k_2} = R_s\mathcal{K} , \
    \mathcal{P}_{s,k_2} = \mathcal{C}_{s,k_2}\mathcal{T}_{s,k_2}\ .
\end{aligned}
\end{equation}
The relation $R_s T_1\Gamma R_s^{-1} = T_2^{-1}\,T_1\Gamma$ implies that $\mathcal{T}_{s,k_2}$ commutes with $\mathcal{C}_{s,k_2}$, so that all three operators square to $\mathbbm{1}$ and commute mutually. Therefore, each fixed-$k_2$ block has a BDI symmetry representation before any further reflection resolution.

At $k_2=0$, $\mathcal{C}_s = T_1\Gamma$ preserves each reflection sector. In the sector $R_s=+1$ the operators of Eq.~\eqref{eq:BDIstaticmodelII} reduce to
\begin{equation} \label{eq:BDIstaticmodelII_zero_momentum}
    \mathcal{C}_s = T_1\Gamma , \qquad \mathcal{T}_s = \mathcal{K} , \qquad \mathcal{P}_s = \mathcal{C}_s\mathcal{K}\ .
\end{equation}

At $k_2=\pi$, $R_s\mathcal{C}_sR_s^{-1}=-\mathcal{C}_s$, so $\mathcal{C}_s$ exchanges the sectors $R_s=\pm1$ and pairs an energy $E$ in one sector with $-E$ in the other. The BDI relations therefore hold only on the direct sum of the two reflection sectors. Within a single block, $k_2=\pi$ and $R_s=+1$ (or $R_s=-1$), there is no chiral symmetry and the only antiunitary symmetry is the time reversal $\mathcal{K}$ with $\mathcal{K}^2=+1$. Each block thus belongs to the Altland--Zirnbauer class AI~\cite{Altland1997}, so that its level statistics are those of the GOE. The spectrum of the block $R_s=-1$ is the exact mirror image $E\to-E$ of that of the $R_s=+1$ block.

The chiral imbalance is defined as $\nu=\rm{Tr}(\mathcal{C}_{s,k_2})$. For $L=32$ and $k_2=0$, resolving reflection yields the two large, oppositely signed imbalances listed in Table~\ref{tab:static_stag_imbalances}. Their sum is the small imbalance $\rm{Tr}_{k_2=0}(\mathcal C_s)=1$ in the reflection-unresolved sector. In all $k \neq 0$ momentum sectors, the trace vanishes. At $k_2=\pi$, the trace is evaluated on the direct sum of the two reflection sectors and is also zero because $\{R_s, \mathcal{C}_{s, \pi}\}=0$. Thus, we focus on the symmetry-resolved sector with $k_2=0, R_s=+1$ to maximize the chiral imbalance for the static version of model II. For model II (Eq.~\eqref{eq_PXP_stag}) at $L=32$, in the sector $k_2=0$, $R_s=+1$, the Hilbert-space dimension is $\mathcal D=154378$ and the chiral imbalance is $\nu=494$. At $\Delta_{\rm stag}=0.20$, the density of states, $\rho(E)$, near $E=0$ displays the predicted delta function with weight $\nu/\mathcal D$, followed by a hard gap and then a square-root onset to an approximately flat bulk density (see Fig.~\ref{fig_1} (b)). It is clear from a visual inspection that the fitted form of Eq.~\eqref{eq_modified_MP} describes the histogram in Fig.~\ref{fig_1}(b) in a much better manner than Fig.~\ref{fig_1}(a) (static version of model I). This is further supported by the ratio $\Delta/\delta_{\rm fit}=106.1$, which is reasonably close to the RMT estimate $\nu/\pi=157.2$ compared to the values obtained for model I (for $L=30$, the corresponding values are $53.5$ and $112.0$). The adjacent level-spacing gap ratio is $\langle r\rangle=0.5302$, close to the GOE value [Fig.~\ref{fig_1}(d)], as expected for ch-GOE.

%---------------------------------------------------------------#  Table 2  #------------------------------------------------------------------------------------------------------------
\begin{table}
\centering
\begin{tabular}{|c|c|c|c|c|c|c|}
\hline \hline
$k_2$ & $R_s$ & $\mathcal{D}$ & $\mathrm{Tr}(\widetilde{\mathcal{C}}_{\mathcal{F}})$& $\mathrm{Tr}(\mathcal{C}_{\mathcal{F},k_2})$ & $\nu_0$ & $\nu_\pi$ \\
\hline \hline 
$0$ & -- & 304575 & 147 & 1 & 74 & 73 \\
$\pm\pi/8$ & -- & 304290 & -138 & 0 & -69 & -69 \\
$\pm\pi/4$ & -- & 304560 & 132 & 0 & 66 & 66 \\
$\pm3\pi/8$ & -- & 304290 & -138 & 0 & -69 & -69 \\
$\pm\pi/2$ & -- & 304570 & 142 & 0 & 71 & 71 \\
$\pm5\pi/8$ & -- & 304290 & -138 & 0 & -69 & -69 \\
$\pm3\pi/4$ & -- & 304560 & 132 & 0 & 66 & 66 \\
$\pm7\pi/8$ & -- & 304290 & -138 & 0 & -69 & -69 \\
$\pi$ & -- & 304572 & 144 & 0 & 72 & 72 \\
\hline  \hline
$0$ & $+1$ & 154378 & 190 & 494 & 342 & -152 \\
$0$ & $-1$ & 150197 & -43 & -493 & -268 & 225 \\
$\pi$ & $+1$ & 152286 & 72 & -- & 72 & 0 \\ 
$\pi$ & $-1$ & 152286 & 72 & -- & 72 & 0 \\
\hline
\end{tabular}
\caption{Chiral imbalances of the staggered driven PXP chain with $L=32$ in all symmetry sectors. The first block lists the momentum sectors $k_2=2\pi m/16$ (eigenvalue $e^{ik_2}$ of $T_2$) without resolving the reflection $R_s$; the sectors $k_2$ and $-k_2$ have identical entries and are listed once. The second block lists the reflection-resolved sectors at $k_2=0$. $\mathcal{D}$ is the dimension of the sector, $\mathcal{C}_{\mathcal{F},k_2}=e^{-ik_2/2}U_2 T_1 \widetilde{\mathcal{C}}_{\mathcal{F}} =W _{k_2}(T,0)\,\widetilde{\mathcal{C}}_{\mathcal{F}}$ with $\mathcal{\widetilde{C}_{\mathcal{F}}}=\prod_j\sigma^z_j$, and $\nu_{0,\pi}=\left(\mathrm{Tr}(\widetilde{\mathcal{C}}_{\mathcal{F}})\pm \mathrm{Tr}(\mathcal{C}_{\mathcal{F},k_2})\right)/2$ are the chiral imbalances in the eigenspaces of eigenphase $0$ and $\pi$. For $k_2=\pi$, $\mathcal{C}_{\mathcal{F},k_2}$ maps $R_s=+1$ onto $R_s=-1$, so $\mathrm{Tr}(\mathcal{C}_{\mathcal{F},k_2})$ vanishes in $k_2=\pi$ sector and $\mathcal{C}_{\mathcal{F},k_2}$ is not defined in $R_s=\pm1$ blocks of $k_2=\pi$ sector.}
\label{tab:Floquet_stag_imbalances}
\end{table}

%----------------------------------------------------------------------------------------------------------------------------------------------------------------------------------------

For the Floquet version of model II, we drive the staggering detuning strength $\Delta_{\rm stag}$ periodically, again using a square-pulse protocol:
\begin{equation}  \label{eq_PXP_staggerd_driven}
\begin{aligned} 
    H_{\rm stag}(t) &=  \sum_{j=1}^{L} P_{j-1} \sigma_j^x P_{j+1}  + \Delta_{\rm stag}(t) \sum_{j=1}^L (-1)^j \sigma_j^z \ , \\
    & \Delta_{\rm stag}(t) = \begin{cases}
        \Delta_{\rm stag} , \ 0\leq t < \frac{T}{2} , \\
        - \Delta_{\rm stag} , \ \frac{T}{2} \leq t <T  \ .
    \end{cases}
\end{aligned}
\end{equation}
The unitary time evolution operator in stroboscopic time steps, i.e., the Floquet unitary $U(T,0)$, is given by
\begin{equation} \label{eq_time_evolution_op}
\begin{aligned} 
&U(T,0) = U_2 U_1 \\ 
&= \exp \left(- i \frac{T}{2} H_{\rm stag}(-\Delta_{\rm stag})\right) \exp \left(- i \frac{T}{2} H_{\rm stag}(\Delta_{\rm stag})\right) \ .
\end{aligned}
\end{equation}

Since $\mathcal{C}_s$ anticommutes with both half-period Hamiltonians and the Hamiltonians are real in the $\sigma^z$ basis,
\begin{equation}
    \mathcal C_s U_{1,2}\,\mathcal C_s^{-1}=U_{1,2}^{-1}, \qquad \mathcal K U_{1,2}\,\mathcal K=U_{1,2}^{-1}\ .
\end{equation}
In a fixed momentum sector $k_2$, we define
\begin{equation} \label{eq:BDIFloquetmodelIIset1}
\begin{aligned}
\mathcal C_{\mathcal{F},k_2} &=e^{-ik_2/2}U_2\mathcal{C}_s,\ \mathcal T_{\mathcal{F},k_2}=U_2 R_s\mathcal{K},\\
\mathcal P_{\mathcal F,k_2} &=\mathcal C_{\mathcal{F},k_2}\mathcal T_{\mathcal{F},k_2} = e^{-ik_2/2}\mathcal{C}_s R_s \mathcal{K}\ .
\end{aligned}
\end{equation}
The operators in Eq.~\eqref{eq:BDIFloquetmodelIIset1} mutually commute, and they satisfy the BDI relations in Eq.~\eqref{eq_circular_BDI_action}. Thus $U(T,0)$ has BDI symmetry in every fixed $k_2$ block before any further reflection resolution. In particular, at $k_2=0$, $R_s$ commutes with all three BDI operators, so it must be resolved to obtain irreducible BDI blocks. At $k_2=\pi$, $\mathcal{C}_{\mathcal{F},\pi}$ exchanges the two reflection sectors and is therefore not a chiral operator within either $ R_s=\pm 1$ block. However, $U(T,0)$ in the symmetry blocks with $k_2=\pi$ and $R_s=\pm 1$ is still BDI compatible due to the presence of another chiral operator, as we will see below.

We note that $\widetilde{\mathcal C}_{\mathcal F}=\Gamma=\prod_j \sigma_j^z$ is another chiral operator for $U(T,0)$ irrespective of the sectors since
\begin{equation}
\begin{aligned}
\widetilde{\mathcal C}_{\mathcal F} H_{\rm stag}(\pm\Delta_{\rm stag})\, \widetilde{\mathcal C}_{\mathcal F}^{-1} &=-H_{\rm stag}(\mp\Delta_{\rm stag}), \\
\widetilde{\mathcal C}_{\mathcal F} U(T,0)\, \widetilde{\mathcal C}_{\mathcal F}^{-1}&=U(T,0)^{-1}\ .
\end{aligned}
\end{equation}
However, the time-reversal operator $\mathcal{T}_{\mathcal{F},k_2}$ does not commute with $\widetilde{\mathcal C}_{\mathcal F}$; rather,
\begin{equation}
    \mathcal T_{\mathcal F,k_2}\widetilde{\mathcal C}_{\mathcal F}
    \mathcal T_{\mathcal F,k_2}^{-1}=U(T,0)\widetilde{\mathcal C}_{\mathcal F} \ .
\end{equation}
We can construct a time-reversal operator that commutes with $\Gamma$ by dressing, as in the Floquet version of model I; this yields another set of operators that obey the BDI relations in Eq.~\eqref{eq_circular_BDI_action} as follows.
\begin{equation} \label{eq:BDIFloquetmodelIIset2}
\begin{aligned}
\widetilde{\mathcal C}_{\mathcal F} = \Gamma,& \quad \widetilde{\mathcal T}_{\mathcal F,k_2} =U(T,0)^{-1/2}\mathcal T_{\mathcal F,k_2}\ , \\
\widetilde{\mathcal P}_{\mathcal{F},k_2} &=\widetilde{\mathcal C}_{\mathcal F} \widetilde{\mathcal T}_{\mathcal{F},k_2} = \mathcal{P}_{\mathcal{F},k_2} \ .
\end{aligned}
\end{equation}
Here, $U(T,0)^{-1/2}$ is defined using the principal branch, with eigenphases in the interval $(-\pi, \pi)$. In the parameter regime considered, we have not encountered any $\pi$ modes, so that $\widetilde{\mathcal T}_{\mathcal F,k_2}$ is unambiguously defined. The three operators in Eq.~\eqref{eq:BDIFloquetmodelIIset2} commute with each other, and each squares to $\mathbbm{1}$. Thus, Eq.~\eqref{eq:BDIFloquetmodelIIset2} provides a second BDI realization of $U(T,0)$ in addition to that in Eq.~\eqref{eq:BDIFloquetmodelIIset1}. Before resolving reflection (and at $k_2=0$ after resolving $R_s$), both realizations are defined in each fixed-$k_2$ sector. At $k_2=\pi$, the chiral operator in Eq.~\eqref{eq:BDIFloquetmodelIIset1} exchanges the $R_s=\pm 1$ sectors, whereas the operators in Eq.~\eqref{eq:BDIFloquetmodelIIset2} act within each sector. Thus, only one set of BDI operators (Eq.~\eqref{eq:BDIFloquetmodelIIset2}) is defined for $k_2=\pi$ and $R_s=\pm 1$.

%In the $k_2=\pi$, $R_s=\pm1$ blocks, $U(T,0)$ is then compatible with ch-COE($\nu_0,0$), where the imbalances are $\nu_0=\rm{Tr}(\widetilde{\mathcal C}_{\mathcal F})$ and $\nu_\pi=0$ (Table~\ref{tab:Floquet_stag_imbalances}).

In Model II, the two BDI sets are directly linked by the dynamical symmetry $G=T_1$ (see discussion around Eq.~\eqref{eq:symmG}), which exchanges the two half-period drives. We note that $G=T_1$ is not a symmetry of each half-period Hamiltonian separately. Rather, translation by one site reverses the staggered detuning term while leaving the PXP term invariant, so that $GH(t)G^{-1}=H(t+T/2)$, which gives $U_2=GU_1 G^{-1}$. Since $T_1^2=T_2=e^{ik_2}$ in a fixed momentum sector, the unitary operator
\begin{equation} \label{eq:modelII_W_k2}
 W_{k_2}(T,0)=e^{-ik_2/2}T_1 U_1
\end{equation}
satisfies $W_{k_2}(T,0)^2=U(T,0)$. $W_{k_2}(T,0)$ obeys the following relations 
\begin{equation} \label{eq:BDI_symm_W}
\begin{aligned}
\mathcal C_{\mathcal F,k_2}W_{k_2}(T,0) \mathcal C_{\mathcal F,k_2}^{-1}&=W_{k_2}(T,0)^{-1},\\
\mathcal T_{\mathcal F,k_2}W_{k_2}(T,0) \mathcal T_{\mathcal F,k_2}^{-1}&=W_{k_2}(T,0)^{-1},\\
\mathcal P_{\mathcal F,k_2}W_{k_2}(T,0) \mathcal P_{\mathcal F,k_2}^{-1}&=W_{k_2}(T,0) \ .
\end{aligned}
\end{equation}
Thus $W_{k_2}(T,0)$ is itself compatible with the BDI symmetries of $U(T,0)$ (Eq.~\eqref{eq:BDIFloquetmodelIIset1}) in reflection-unresolved fixed-$k_2$ momentum sectors and in $k_2=0,R_s=\pm1$ sectors. The operator $\widetilde{\mathcal C}_{\mathcal F}$ is also a chiral symmetry of $W_{k_2}(T,0)$, since $\widetilde{\mathcal C}_{\mathcal F} W_{k_2}(T,0) \widetilde{\mathcal C}_{\mathcal F}^{-1}=W_{k_2}(T,0)^{-1}$.

\begin{figure}
\centering
\begin{tikzpicture}[
    >={Stealth[length=2.4mm,width=2mm]},
    wlabel/.style={fill=white, inner sep=1pt, font=\small},
    tiplabel/.style={font=\footnotesize}
]
\def\Rt{3.35}
\def\rw{1.5}
\def\hw{1.55}
\def\dd{24}
\foreach \ang/\col in {180/red, 0/blue} {
    \begin{scope}[rotate=\ang]
        \path[fill=\col!35, fill opacity=0.65, draw=\col!70!black, thick]
            (0,0) .. controls (1.0,\hw) and (\Rt-1.2,\hw) .. (\Rt,0)
                  .. controls (\Rt-1.2,-\hw) and (1.0,-\hw) .. (0,0) -- cycle;
    \end{scope}
}
\node[align=center, font=\small] at (-2.3,0) {$\widetilde{\mathcal{C}}_{\mathcal{F}}=\Gamma$\\[2pt]$\widetilde{\mathcal{T}}_{\mathcal{F},k_2}$};
\node[align=center, font=\small] at (2.3,0) {$\mathcal{C}_{\mathcal{F},k_2}$\\[2pt]$\mathcal{T}_{\mathcal{F},k_2}$};
\node[tiplabel] at (-\Rt-0.45,0) {$a=0$};
\node[tiplabel] at (\Rt+0.45,0) {$a=1$};
\draw[->, thick] (180-\dd:\rw) arc[start angle=180-\dd, end angle=\dd, radius=\rw];
\node[wlabel] at (90:\rw) {$W_{k_2}$};
\draw[->, thick, dashed] (-\dd:\rw) arc[start angle=-\dd, end angle=-180+\dd, radius=\rw];
\node[wlabel] at (-90:\rw) {$W_{k_2}$};
\node[circle, draw=black, very thick, fill=white, fill opacity=0.8, text opacity=1, minimum size=1.15cm, inner sep=0pt, font=\large] at (0,0) {$\mathcal{P}_{\mathcal{F},k_2}$};
\end{tikzpicture}
\caption{The two sets of BDI operators of model II in a fixed momentum sector $k_2$ (petals), sharing the particle-hole operator $\mathcal{P}_{\mathcal{F},k_2}$ (circled intersection). The set $a=0$ is $(\widetilde{\mathcal{C}}_{\mathcal{F}},\widetilde{\mathcal{T}}_{\mathcal{F},k_2})$ of Eq.~\eqref{eq:BDIFloquetmodelIIset2} and the set $a=1$ is $(\mathcal{C}_{\mathcal{F},k_2},\mathcal{T}_{\mathcal{F},k_2})$ of Eq.~\eqref{eq:BDIFloquetmodelIIset1}. The solid arrow shows the action of $W_{k_2}(T,0)$ taking the set $a=0$ to the set $a=1$. Since $W_{k_2}(T,0)^2=U(T,0)$, the dashed arrow leads to $(U(T,0)\,\widetilde{\mathcal{C}}_{\mathcal{F}},\,U(T,0)\,\widetilde{\mathcal{T}}_{\mathcal{F},k_2})$, so that the sequence closes on the set $a=0$ only up to the factor $U(T,0)$.}
\label{fig_3}
\end{figure}
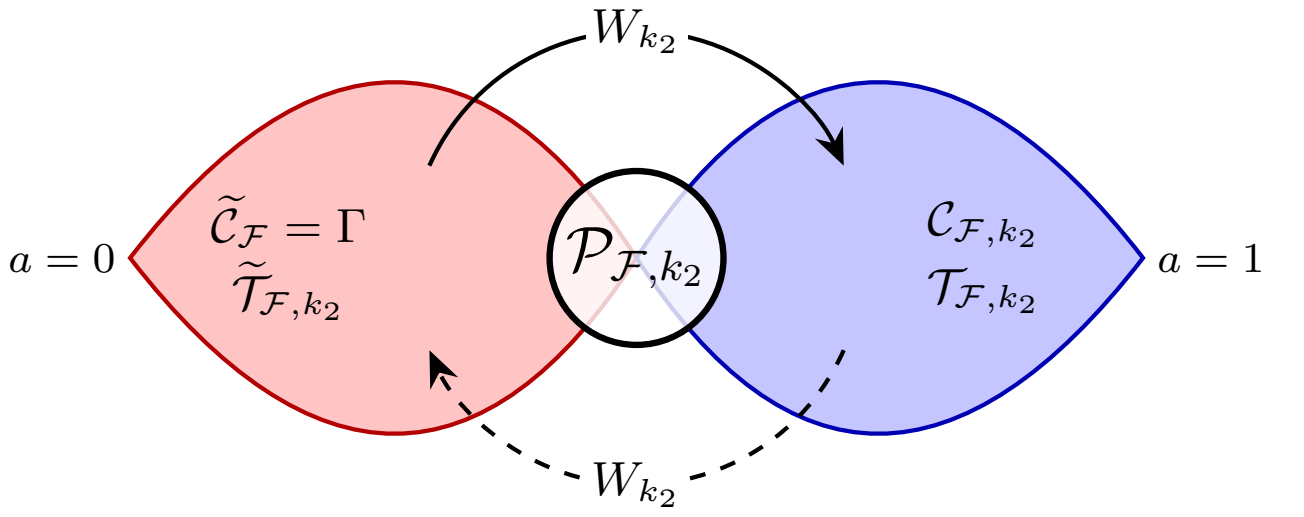

Importantly, $W_{k_2}(T,0)$ connects the two BDI realizations of Eq.~\eqref{eq:BDIFloquetmodelIIset2} and Eq.~\eqref{eq:BDIFloquetmodelIIset1} of the same Floquet unitary $U(T,0)$, which share the same particle-hole operator and are labeled by $a=0,1$ respectively, as shown schematically in Fig.~\ref{fig_3},
\begin{equation} \label{eq:modelII_W_sets}
 \begin{aligned}
     \mathcal C_{\mathcal F,k_2} &= W_{k_2}(T,0)\,\widetilde{\mathcal C}_{\mathcal F}, \quad
     \mathcal T_{\mathcal F,k_2} = W_{k_2}(T,0)\,\widetilde{\mathcal T}_{\mathcal F,k_2}, \\
     \mathcal P_{\mathcal F,k_2} &= \widetilde{\mathcal P}_{\mathcal F,k_2} \ .
 \end{aligned}
\end{equation}
These two BDI sets can be represented as the two petals of Fig.~\ref{fig_3}, with $\mathcal{P}_{\mathcal F,k_2}$ at their intersection, and $W_{k_2}(T,0)$ takes the petal $a=0$ to the petal $a=1$. Since $\widetilde{\mathcal C}_{\mathcal F}^2=\mathbbm{1}$, the two chiral operators are related by $\mathcal C_{\mathcal F,k_2}\widetilde{\mathcal C}_{\mathcal F}=W_{k_2}(T,0)$. A second application of $W_{k_2}(T,0)$ gives $W_{k_2}\mathcal C_{\mathcal F,k_2}=U(T,0)\,\widetilde{\mathcal C}_{\mathcal F}$ and $W_{k_2}\mathcal T_{\mathcal F,k_2}=U(T,0)\,\widetilde{\mathcal T}_{\mathcal F,k_2}$, so that the sequence of sets closes on the first one only up to the factor $U(T,0)$ (dashed arrow in Fig.~\ref{fig_3}). At $k_2=\pi$, $W_\pi(T,0)$ exchanges the two reflection sectors ($R_s=\pm1$) and therefore does not act within either reflection-resolved block. Only the petal $a=0$ is then available, 
%: the BDI construction of $U(T,0)$ in these individual blocks uses $\widetilde{\mathcal C}_{\mathcal F}=\Gamma$ and the dressed time-reversal operator $\widetilde{\mathcal T}_{\mathcal F,\pi}=U(T,0)^{-1/2}\mathcal T_{\mathcal F,\pi}$ of Eq.~\eqref{eq:BDIFloquetmodelIIset2}.

%----------------------------------------------------------------------------------------------------------------------------------------------------------------------------------------
%-----------------------------------------------------#  Fig. 3  #----------------------------------------------------------------------------------------------------------
%----------------------------------------------------------------------------------------------------------------------------------------------------------------------------------------

The two distinct chiral operators have a natural interpretation for the spectrum of $W_{k_2}(T,0)$. Since $W_{k_2}(T,0)$ does not approach $\mathbbm{1}$ as $T \rightarrow 0$, it follows that it can have two chiral imbalances in general, at $\phi=0$ and $\phi=\pi$, unlike $U(T,0)$. The chiral imbalances at eigenphases $\phi=0$ and $\phi=\pi$ are related to the trace of the two chiral operators $\widetilde{\mathcal C}_{\mathcal F}$ and $\mathcal C_{\mathcal F,k_2}=W_{k_2}(T,0)\widetilde{\mathcal C}_{\mathcal F}$
\begin{equation} \label{eq_Wk2_tr_chiral}
\begin{aligned}
    {\rm{Tr}}\left(W_{k_2}\widetilde{\mathcal C}_{\mathcal F}\right) &=\nu^W_{0,k_2}-\nu^W_{\pi,k_2} \ ,\\
    {\rm{Tr}}\left(\widetilde{\mathcal C}_{\mathcal F}\right) &=\nu^W_{0,k_2}+\nu^W_{\pi,k_2}\ , 
\end{aligned}
\end{equation}
and $W_{k_2}(T,0)$ is then expected to belong to a ``generalized" ch-COE($\nu^W_{0,k_2}, \nu^W_{\pi,k_2}$). 
%and hence
%\begin{equation}
 %   \nu_{0,\pi}=\frac12\left[\rm{Tr}\!\left(\widetilde{\mathcal C}_{\mathcal F}\right) 
  %  \pm\rm{Tr}\left(W_{k_2}(T,0)\, \widetilde{\mathcal C}_{\mathcal F}\right) \right].
%\end{equation}
%These expressions apply in symmetry sectors in which $(W_{k_2}(T,0)$ acts within the corresponding spaces. 
The values of $\nu^W_{0,k_2}$ and $\nu^W_{\pi,k_2}$ for all the symmetry sectors at $L=32$ are listed in Table~\ref{tab:Floquet_stag_imbalances}. For example, in the $k_2=0, R_s=+1$ sector, we obtain $\nu_0=342$ and $\nu_\pi=-152$. At $k_2=\pi$, however, $W_\pi(T,0)$ exchanges the two reflection sectors and therefore cannot be restricted to either reflection-resolved block. The imbalances listed for these individual blocks are instead obtained from the BDI construction in Eq.~\eqref{eq:BDIFloquetmodelIIset2}, for which $\nu_0=\rm{Tr}(\widetilde{\mathcal C}_{\mathcal F})$ and $\nu_\pi=0$.

%For momentum sectors without reflection resolution, the trace of $\mathcal C_{\mathcal F,k_2}$ can be evaluated using the projector onto the $T_2=e^{ik_2}$ sector. It turns out that $\mathcal C_{\mathcal F,k_2}$ and $\mathcal C_{s,k_2}$ have the same trace (Table~\ref{tab:Floquet_stag_imbalances} and Table~\ref{tab:static_stag_imbalances}) and using this we obtain
%\begin{equation}
%    \rm{Tr}(\mathcal C_{\mathcal F,k_2}) =\frac{e^{-ik_2/2}}{N} \sum_{n=0}^{N-1} e^{-ik_2 n} \, \rm{Tr}\!\left(T_1^{2n+1} \Gamma\right), \quad N=L/2.
%\end{equation}
%Only basis states invariant under the odd translations $(T_1^{2n+1}$ contribute. For $L=32$, the vacuum is the only such state, so each trace in the sum is unity and $\rm{Tr}(\mathcal C_{\mathcal {F},k_2})=\delta_{k_2,0}$ in the reflection-unresolved momentum sectors. At $k_2=\pi$, $\mathcal C_{\mathcal F,k_2}$ exchanges the two reflection sectors, and although it is defined on their direct sum, it is not an operator within either reflection-resolved block.

\begin{figure}
    \centering
    \includegraphics[width=\linewidth]{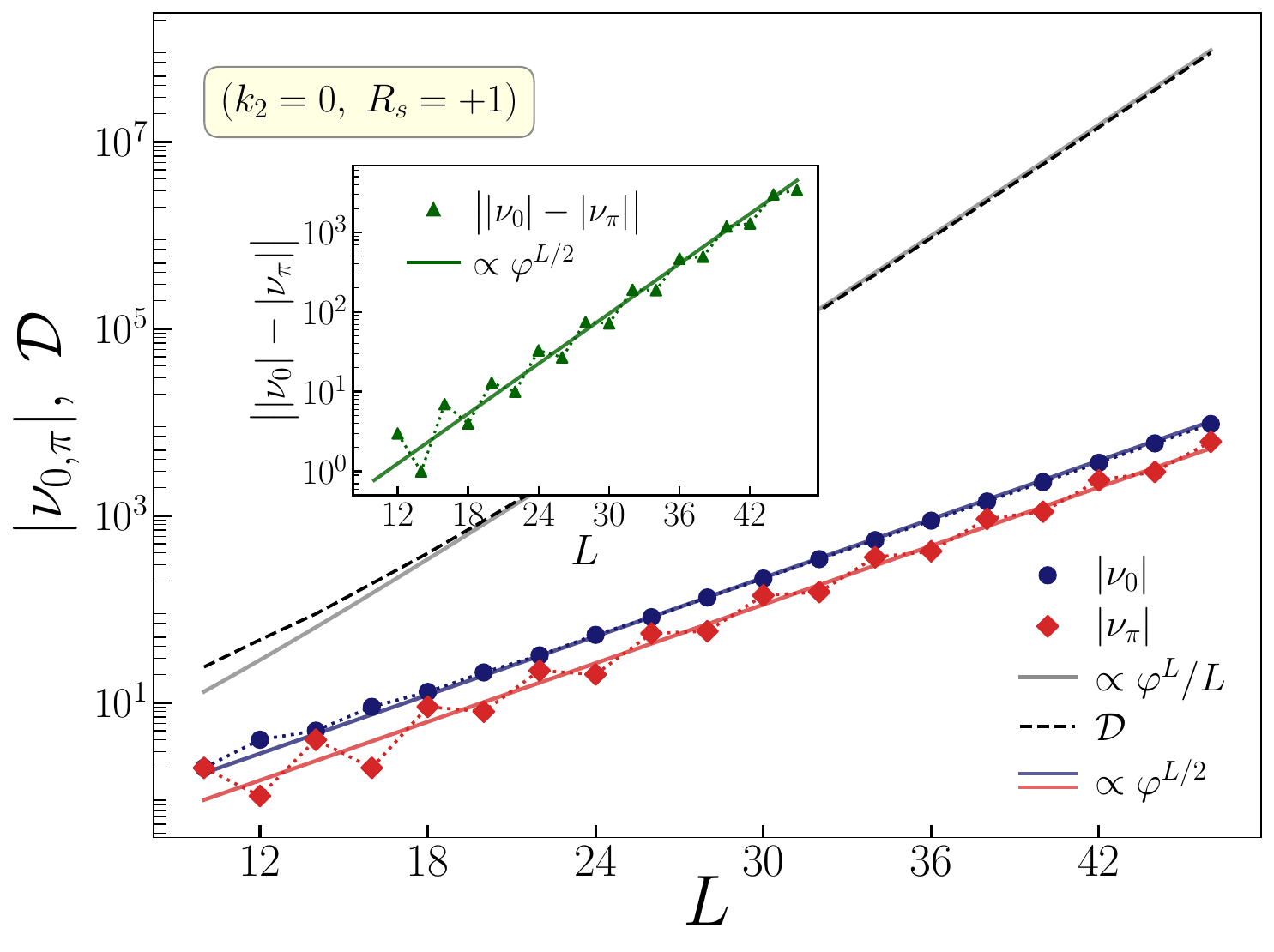}
    \caption{Chiral imbalances of the staggered driven PXP chain in the sector $k_2=0$, $R_s=+1$ as a function of the system size $L$. The imbalance $\nu_\pi$ is negative for all sizes shown, and we plot the absolute values. Here $\mathcal{D}$ is the sector dimension and the zero and $\pi$ imbalances, $\nu_{0}$, $\nu_{\pi}$ respectively are obtained from $\nu_{0,\pi}=(\mathrm{Tr}(\widetilde{\mathcal{C}}_{\mathcal{F}})\pm\mathrm{Tr}(W\widetilde{\mathcal{C}}_{\mathcal{F}})/2$ with $W=T_1U_1$. Inset shows the the scaling of $\big| |\nu_0| - |\nu_\pi|\big|$ with $L$ in the same sector. The data shown are for even $L$ ranging from $10$ to $46$ in both the main figure and the inset.}
    \label{fig_4}
\end{figure}

In the sector $k_2=0$, $R_s=+1$, $|\nu_0|$ and $|\nu_\pi|$ are shown as a function of the system size in Fig.~\ref{fig_4} for even $L$ between $10$ and $46$. Both $|\nu_0|$ and $|\nu_\pi|$ grow exponentially with $L$, approximately as $\varphi^{L/2}\approx 1.27^{L}$, whereas $\mathcal{D}$ grows asymptotically as $\varphi^{L}/L$, with $\varphi$ the golden ratio, so that the fraction $|\nu_{0,\pi}|/\mathcal{D}$ of eigenphases that are pinned to $0$ and $\pi$ by chirality decreases exponentially. The number of protected modes is nonetheless exponentially large in the system size. The imbalance $\nu_\pi$ is negative for all sizes shown, and $\nu_0$ and $\nu_\pi$ do not grow monotonically with $L$.

%-------------------------------------------------------------# Fig. 4 #----------------------------------------------------------------------------------------------------------------

%---------------------------------------------------------------------------------------------------------------------------------------------------------------------------------------

In sectors where $W_{k_2}(T,0)$ acts within the chosen sector (all sectors except $k_2=\pi, R_s=\pm 1$), the relation $U(T,0)=W_{k_2}(T,0)^2$ implies that their eigenphases are related by $\theta=2\phi\pmod{2\pi}$. Consequently, the eigenphase density of $U(T,0)$ is
\begin{equation} \label{eq_dos_form_U}
    \rho_U(\theta)=\frac12\left[\rho_{W_{k_2}}\!\left(\frac{\theta}{2}\right) +\rho_{W_{k_2}}\!\left(\frac{\theta}{2}+\pi\right) \right],
\end{equation}
where the arguments are understood modulo \(2\pi\). This simplifies to
\begin{equation} \label{eq_dos_sim_form_U}
\begin{aligned}
\rho_U(\theta) =& \frac{1}{2} \left. \left[\rho_{W_{k_2}}\left(\frac{\theta}{2}\right) + \rho_{W_{k_2}}\left(\frac{\theta}{2}+\pi\right) \right] \right|_{\theta_0 \leq |\theta| \leq \pi}\\ %_{\mathrm{bulk}}
&+ \frac{\left|\nu^W_{0,k_2}\right|+\left|\nu^W_{\pi,k_2}\right|}{\mathcal D}\delta(\theta)  \ ,
\end{aligned}
\end{equation}
with $\theta_0 = 2 \min(\phi_0, \pi-\phi_\pi)$. The modes of $W_{k_2}(T,0)$ at $\phi=0$ and $\phi=\pi$ both fold onto $\theta=0$. Thus, $U(T,0)$ has at least $\left|\nu^W_{0,k_2}\right|+\left|\nu^W_{\pi,k_2}\right|$ exact zero-quasienergy modes, while its protected chiral imbalance at $\theta=\pi$ vanishes. The hard edge of the folded continuous spectrum for $U(T,0)$ is $\theta_0=2\min(\phi_0,\pi-\phi_\pi),$ and a second square-root onset occurs at $\theta=2\max(\phi_0,\pi-\phi_\pi)$. On the other hand, $\rho_U(\theta)$ is expected to follow ch-COE($\nu_0,0$) (Eq.~\eqref{eq_circ_BDI_dos}) with $\nu_0=\rm{Tr}(\widetilde{\mathcal C}_{\mathcal F})$ for $k_2=\pi, R_s=\pm 1$ since only one chiral operator is defined in these two symmetry-resolved sectors.

\begin{figure*} 
    \centering
    \includegraphics[width=0.49\linewidth]{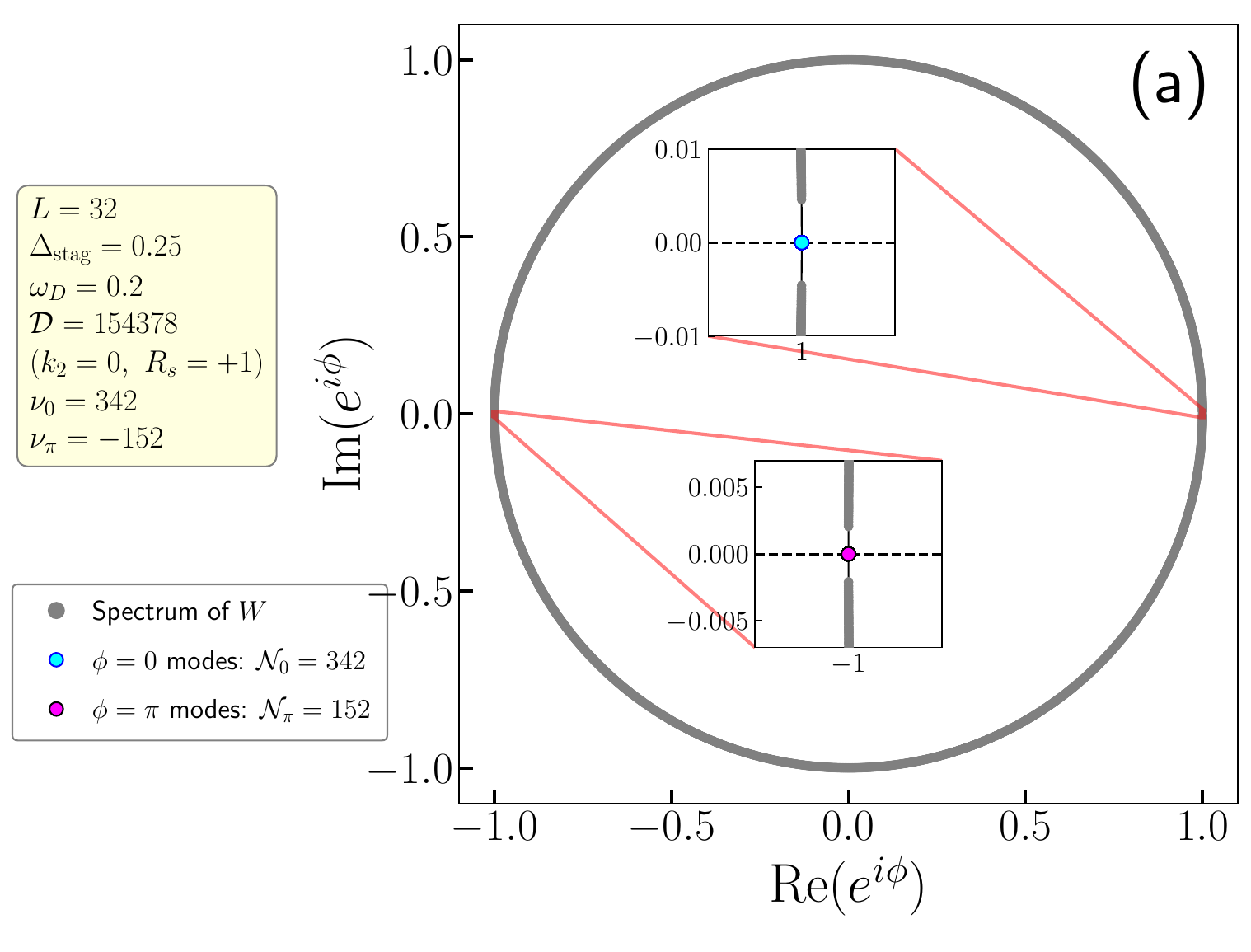}
    \includegraphics[width=0.49\linewidth]{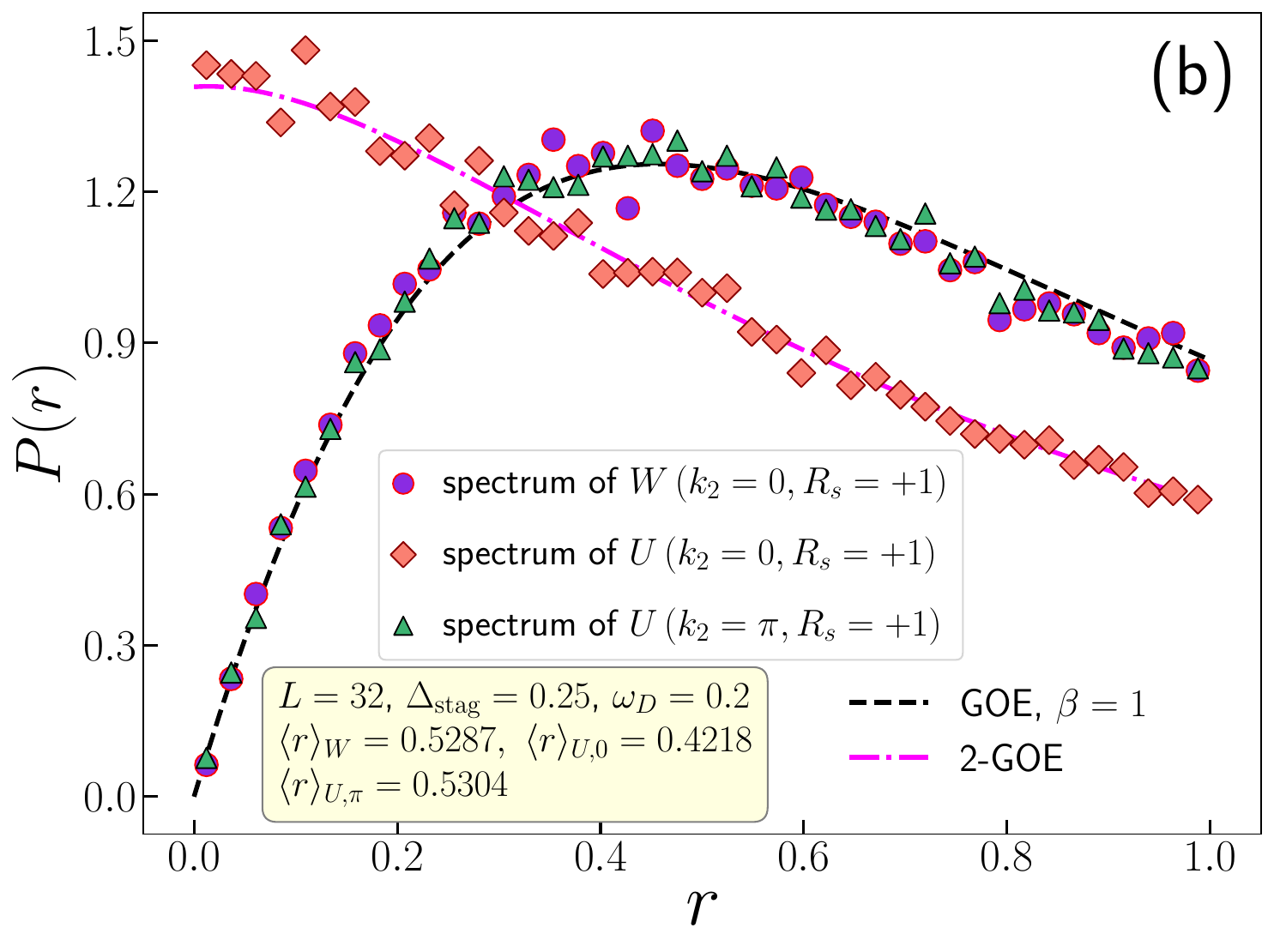}
    \includegraphics[width=0.49\linewidth]{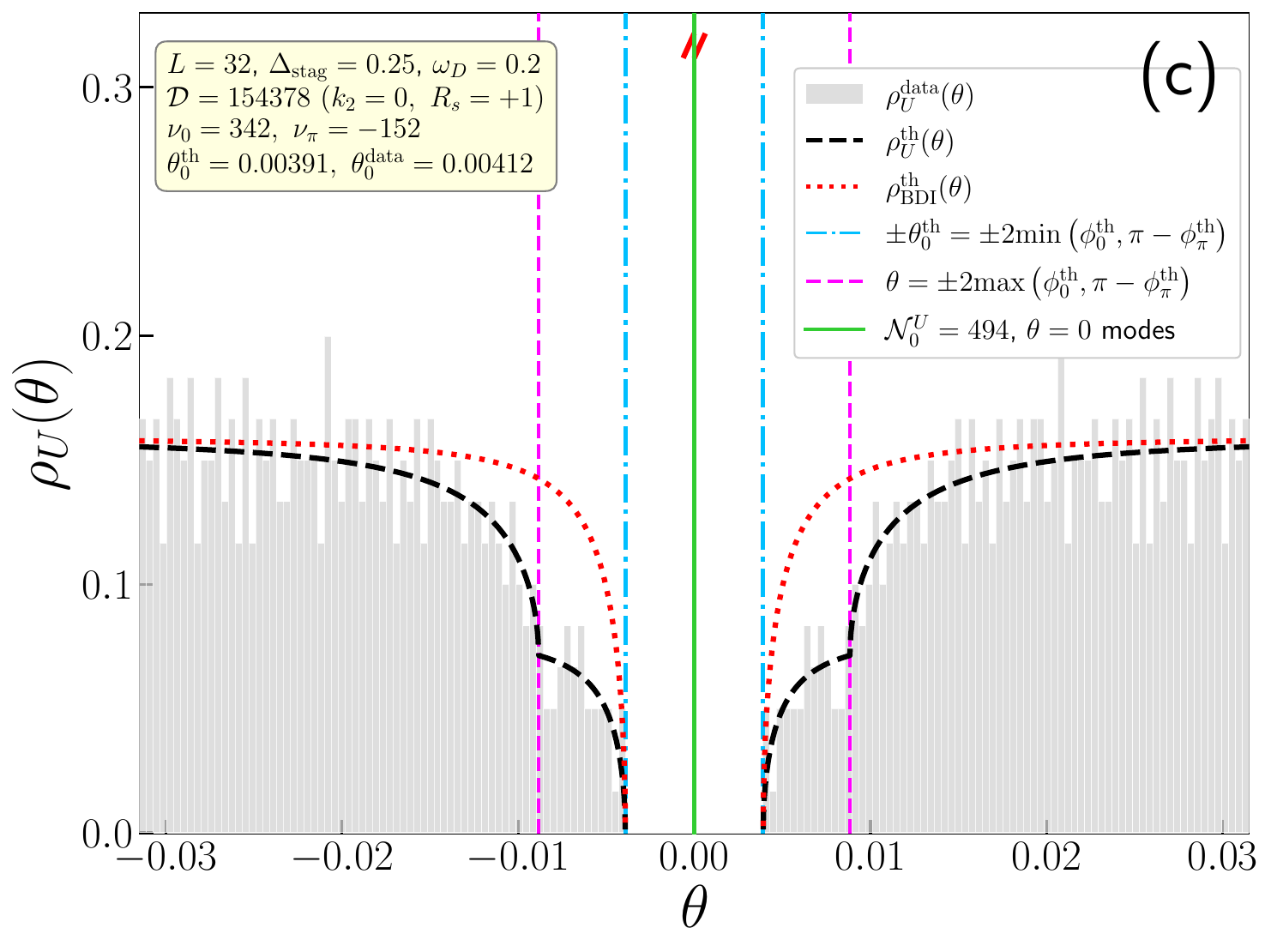}
    \includegraphics[width=0.49\linewidth]{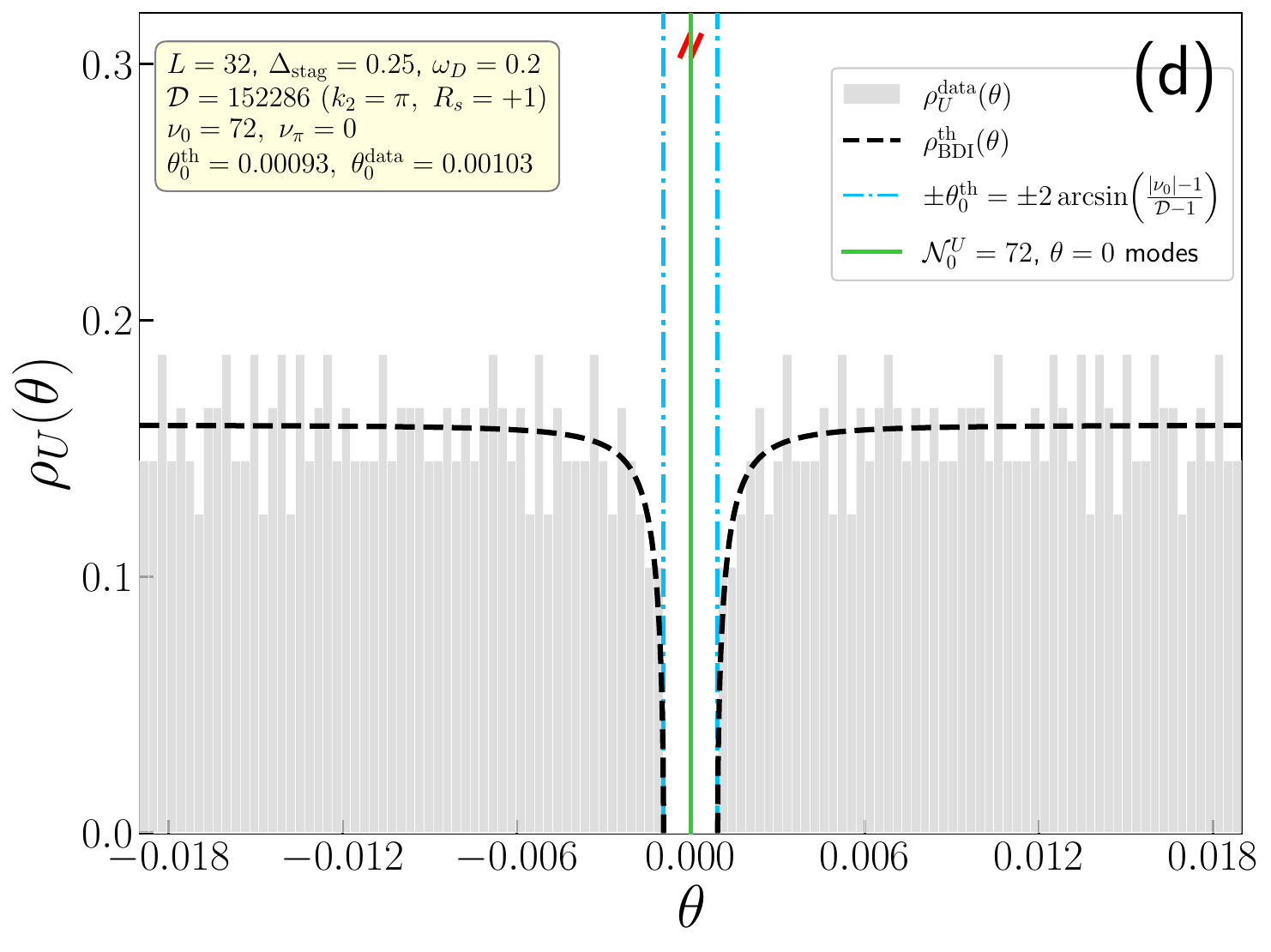}
    \caption{(a) Eigenvalues $\{e^{i\phi}\}$ of $W$ for the Floquet staggered PXP problem (model II) in the sector $k_2=0$, $R_s=+1$ ($\nu_0=342$, $\nu_\pi=-152$). The insets magnify the neighborhoods of $\phi=0$ and $\phi=\pi$ modulo $2\pi$: the exact modes at $\phi=0$ (cyan, $342$) and $\phi=\pi$ (magenta, $152$) are each surrounded by a gap. (b) Distribution of the ratios of consecutive level spacings for the positive-branch eigenphases in the bulk of $W$ and of $U(T,0)=W^2$ in the sector $k_2=0$, $R_s=+1$, and of $U(T,0)$ in the sector $k_2=\pi$, $R_s=+1$, together with the GOE and the $2$-GOE distributions. (c) Density of states of the eigenphases $\{\theta\}$ of $U(T,0)=W^2$ in the sector $k_2=0$, $R_s=+1$ in the vicinity of $\theta=0$; the black dashed curve is the analytical form $\rho_U^{\rm th}$, Eq.~\eqref{eq_dos_sim_form_U}. The $342$ modes at $\phi=0$ and the $152$ modes at $\phi=\pi$ of $W$ both fold to $\theta=0$, so that $U$ has $|\nu_0|+|\nu_\pi|=494$ exact zero modes (green line). The red dotted curve is the circular BDI RME density, Eq.~\eqref{eq_circ_BDI_dos}, whose hard edge matches $\theta_0^{\rm th}$. (d) Density of states of $U(T,0)$ in the vicinity of $\theta=0$ in the sector $k_2=\pi$, $R_s=+1$, where $W$ is not defined and $U(T,0)$ is an ordinary BDI unitary with $\nu_0=72$, $\nu_\pi=0$; the black dashed curve is Eq.~\eqref{eq_circ_BDI_dos}. In all the panels, the system size is $L=32$, and the staggering strength and the drive frequency are $\Delta_{\rm stag}=0.25$ and $\omega_D=0.2$, respectively.}
    \label{fig_5}
\end{figure*}

For numerical demonstration, we consider the largest symmetry sector $k_2=0$, $R_s=+1$ at $L=32$ ($\mathcal{D}=154378$), with $\Delta_{\rm stag}=0.25$ and $\omega_D=0.2$ which gives $\nu_0=342$ and $\nu_\pi=-152$ (Table~\ref{tab:Floquet_stag_imbalances}) for the corresponding $W(T,0)$. Fig.~\ref{fig_5}(a) shows the eigenvalues of $W(T,0)$ on the unit circle. The insets, which magnify the neighborhood of $e^{i \phi}=+1$ and $e^{i \phi}=-1$, show the $342$ degenerate modes at $\phi=0$ and the $152$ degenerate modes at $\phi=\pi$, each surrounded by a spectral gap. The saddle-point results for the edges of these gaps are given by the two-endpoint density of Eq.~\eqref{eq_W_dos}, which has no adjustable parameters: $\phi_0^{\rm th}=4.4178\times10^{-3}$ and $\pi-\phi_\pi^{\rm th}=1.9563\times10^{-3}$, whereas the smallest eigenphases of the data are $4.5726\times10^{-3}$ and $2.0605\times10^{-3}$, which are larger by $3.5\%$ and $5.3\%$, respectively. 
%The edges of the ordinary circular BDI density of Eq.~\eqref{eq_circ_BDI_dos} with a single imbalance ($\nu_\pi=0$ for the gap at $\phi=0$ and $\nu_0=0$ for the gap at $\phi=\pi$), $4.41776\times10^{-3}$ and $1.95625\times10^{-3}$, differ from the two-endpoint values by only about $10^{-5}$ in relative terms.
The form of $\rho_U(\theta)$ near $\theta=0$ for $k_2=0, R_s=+1$ is fundamentally different (Fig.~\ref{fig_5}(c)) from that of $\rho_U(\theta)$ for $k_2=\pi, R_s=+1$ (Fig.~\ref{fig_5}(d)). In the former case, the $\phi=0$ and $\phi=\pi$ modes of $W(T,0)$ both fold to $\theta=0$, so that $U$ has $|\nu_0|+|\nu_\pi|=494$ exact zero modes, and the two spectral gaps of $W(T,0)$ appear in $\rho_U(\theta)$ as two scales: the hard edge $\theta_0^{\rm th}=2\min(\phi_0^{\rm th},\pi-\phi_\pi^{\rm th})=3.91\times10^{-3}$ and, a second square-root onset at $\theta=2\max(\phi_0^{\rm th},\pi-\phi_\pi^{\rm th})=8.84\times10^{-3}$. The data in Fig.~\ref{fig_5}(c) follows the black dashed curve of Eq.~\eqref{eq_dos_sim_form_U}: the smallest positive eigenphase is $\theta_0^{\rm data}=4.12\times10^{-3}$, about $5\%$ above $\theta_0^{\rm th}$, and between the two scales the density is about half of the bulk value, with a second square-root feature that is clearly visible. For comparison, the red dotted curve shown in Fig.~\ref{fig_5}(c) is the circular BDI RME density of Eq.~\eqref{eq_circ_BDI_dos} whose hard edge is matched to $\theta_0^{\rm th}$, which corresponds to an effective imbalance $\nu_{\rm eff}=1+(\mathcal{D}-1)\sin(\theta_0^{\rm th}/2)\simeq303$ instead of $494$. It reproduces the hard edge, but it lacks another onset of a square-root feature, and it approaches the bulk value much faster than the black curve, overestimating the density between the two scales by about a factor of two. Hence, the two-endpoint structure of $W(T,0)$ cannot be replaced by a single-imbalance ensemble for $U (T,0)$, even if the imbalance is tuned to reproduce the gap. On the other hand, in the sector $k_2=\pi$, $R_s=+1$, $U(T,0)$ is itself an ordinary BDI unitary: $\widetilde{\mathcal{C}}_{\mathcal{F}}=\Gamma$ is its only chiral operator, the time-reversal symmetry is the dressed one, $U(T,0)^{-1/2}\mathcal{T}_{\mathcal{F}}$, and $\nu_\pi=0$, $\nu_0=\mathrm{Tr}(\widetilde{\mathcal{C}}_{\mathcal{F}})=72$ for $L=32$ ($\mathcal{D}=152286$, Table~\ref{tab:Floquet_stag_imbalances}). Its density of states is therefore given by the single-imbalance Eq.~\eqref{eq_circ_BDI_dos} and not by Eq.~\eqref{eq_dos_sim_form_U}: a single hard edge at $\theta_0^{\rm th}=2\arcsin[(|\nu_0|-1)/(\mathcal{D}-1)]=9.32\times10^{-4}$, 
%, and neither a plateau nor a kink. 
Fig.~\ref{fig_5}(d), at the same parameters $\Delta_{\rm stag}=0.25$ and $\omega_D=0.2$, shows that the histogram follows this curve without any adjustable parameter. The smallest positive eigenphase of the data, $\theta_0^{\rm data}=1.03\times10^{-3}$, lies $10.3\%$ above $\theta_0^{\rm th}$. We note that in all cases above, the deviation of the closest eigenphase to $\theta=0$ and $\phi=0,\pi$ from the corresponding saddle-point prediction is consistent with the scale predicted by Eq.~\eqref{eq:fluctuation_theta0} when the correct chiral imbalance is used.  

%------------------------------------------------------------#  Fig. 5  #----------------------------------------------------------------------------------------------------------------
%----------------------------------------------------------------------------------------------------------------------------------------------------------------------------------------

The distribution of the ratios of consecutive level spacings, $P(r)$, calculated for $W(T,0)$ and $U(T,0)$ are compared in Fig.~\ref{fig_5}(b). $P(r)$ for $W(T,0)$ follows the GOE distribution ($\langle r\rangle_W=0.5287$), whereas those of $U(T,0)=W(T,0)^2$ in the sector $k_2=0$, $R_s=+1$ are described by two superimposed GOE spectra ($\langle r\rangle_U=0.4218$), to be compared with $\langle r\rangle\simeq 0.4234$ calculated for the superposition of two independent GOE spectra~\cite{Alet_paper}. In contrast, the spectrum of $U(T,0)$ in the sector $k_2=\pi$, $R_s=+1$ has $\langle r\rangle_U=0.5304$, close to the single-GOE value $0.5359$. The $2$-GOE structure of $U(T,0)$ in $k_2=0$ is therefore a consequence of the quasienergy wrapping $\theta=2\phi$ mod $2\pi$, by which the spectrum of $W(T,0)$ is folded onto itself to generate the spectrum of $U(T,0)$. This structure is absent in the $k_2=\pi, R_s=+1$ sector, where $W_\pi(T,0)$ does not act within the reflection-resolved block.

\subsection{Model III: Three-spin Ising model under PBC}
\label{subsec_model_III}

We now consider $L$ spin-$1/2$ particles in one spatial dimension under PBC with three-spin Ising interactions in the presence of a transverse field,
\begin{align} \label{eq_three_spin_ham}
    H = \sum_{j=1}^{L} \sigma_{j}^z \sigma_{j+1}^z \sigma_{j+2}^z  + h \sum_{j=1}^L \sigma_j^x \ .
\end{align}
Unlike models I and II, model III is defined in an unconstrained Hilbert space. The ground state of $H$ has a phase transition at $h_c=1$, and $H$ has a duality between $h$ and $1/h$, with $h_c=1$ being the self-dual point~\cite{PhysRev.60.252, RevModPhys.52.453}. 
%Under open boundary conditions and for $L\to\infty$, $H$ also has a $\mathbb{Z}_2\times\mathbb{Z}_2$ symmetry~\cite{PhysRevB.108.214430}. 
Under PBC, $H$ commutes with the one-site translation $T_1=e^{ik_1}$ and the inversion $\mathcal{I}$, as in model I, and we work in the same sector with $k_1=0$, $\mathcal{I}=+1$.
%All the operators below are products of single-site operators over the entire chain and complex conjugation $\mathcal{K}$, so they preserve this sector.

%----------------------------------------------#  Fig. 6   #-----------------------------------------------------------------------------------------------------------------------------
\begin{figure*} 
    \centering
    \includegraphics[width=0.49\linewidth]{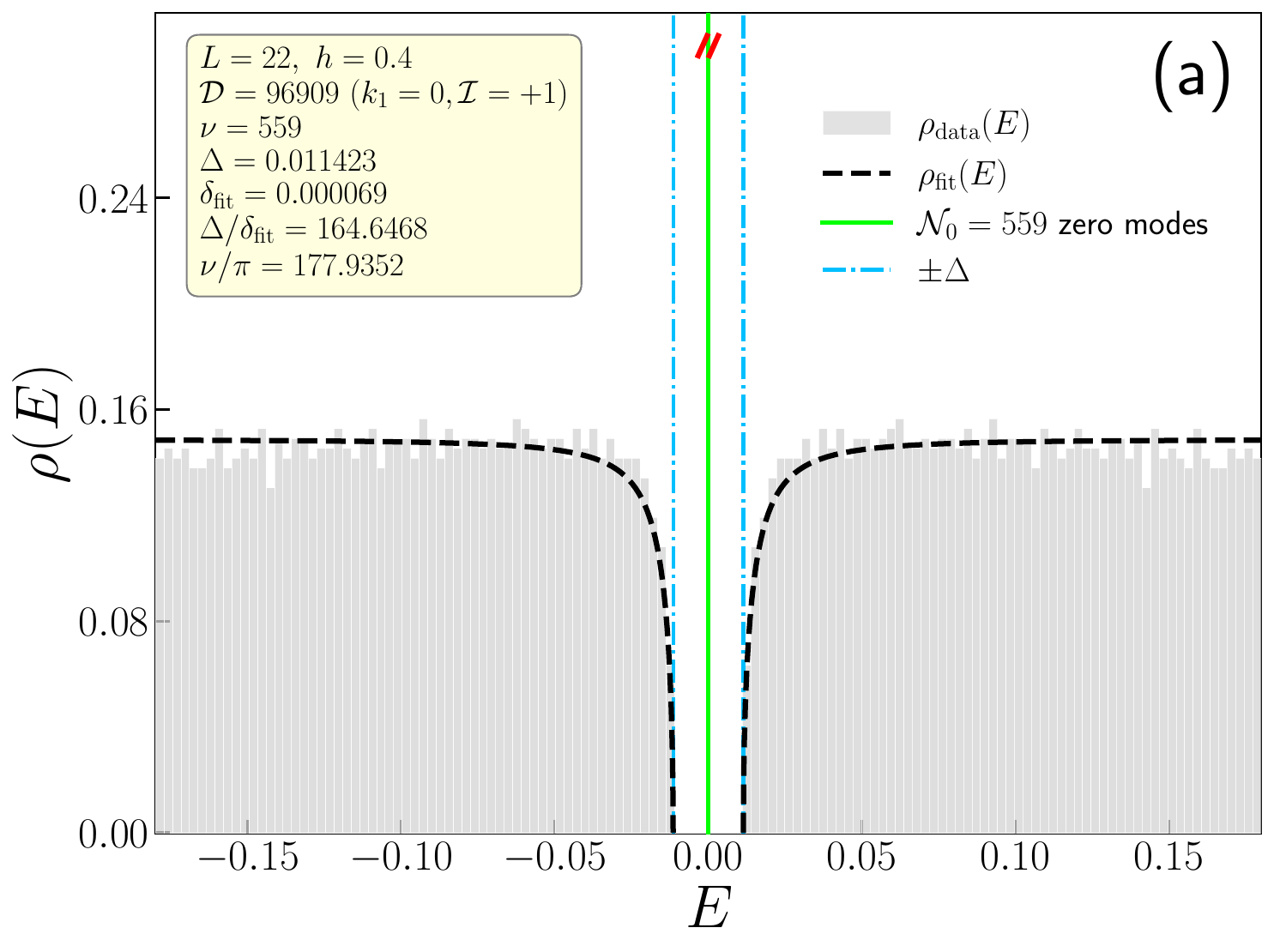}
    \includegraphics[width=0.49\linewidth]{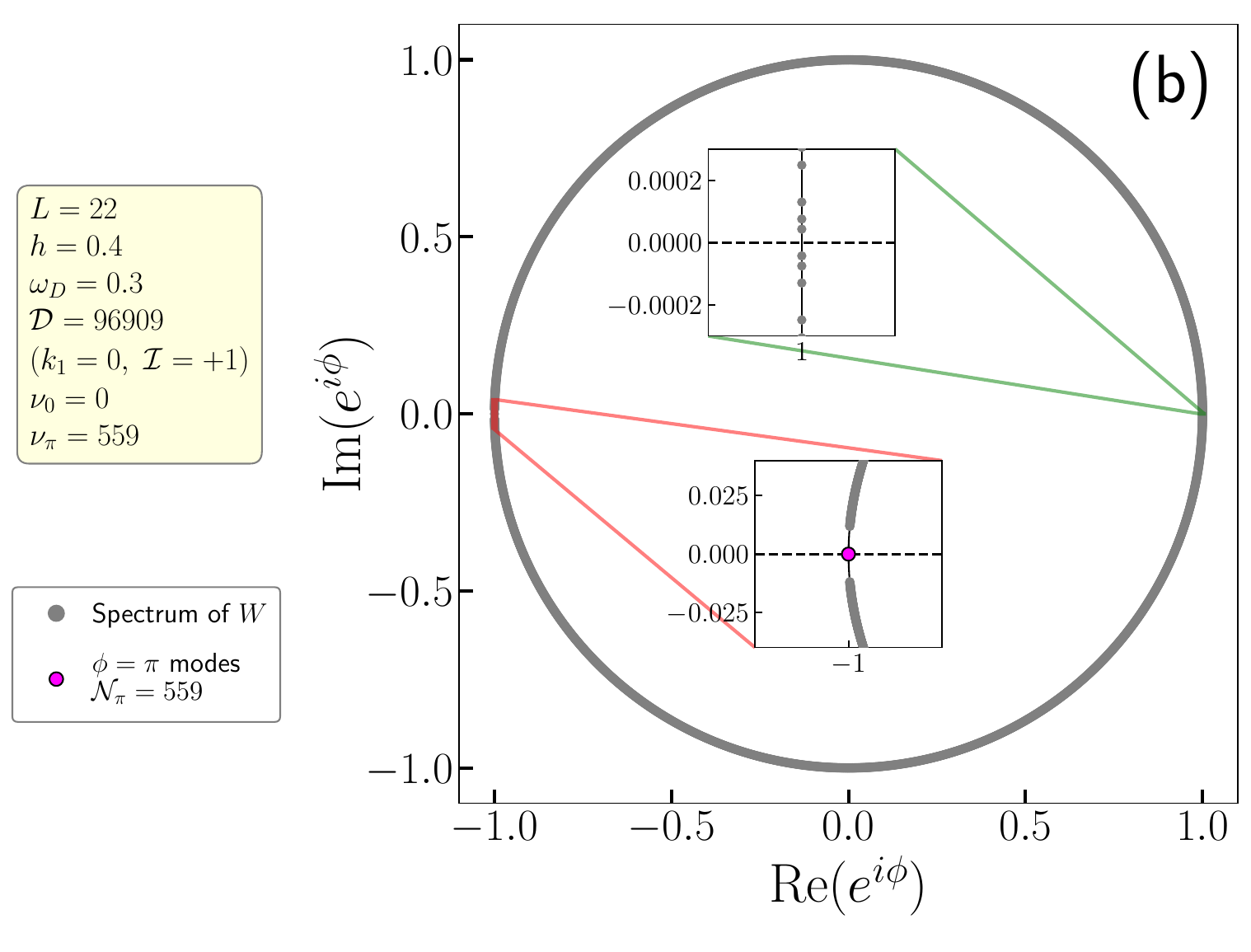}
    \includegraphics[width=0.49\linewidth]{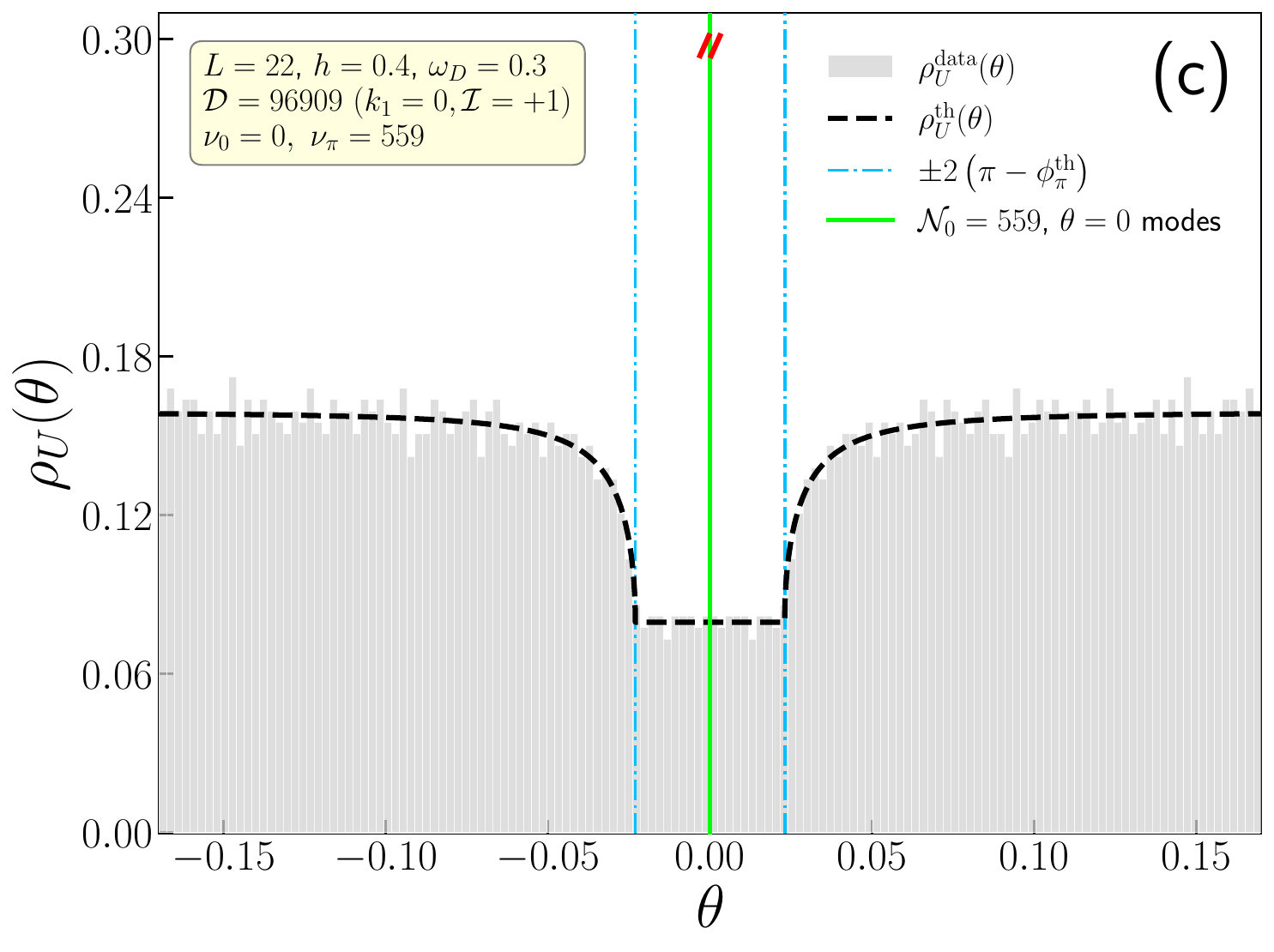}
    \includegraphics[width=0.49\linewidth]{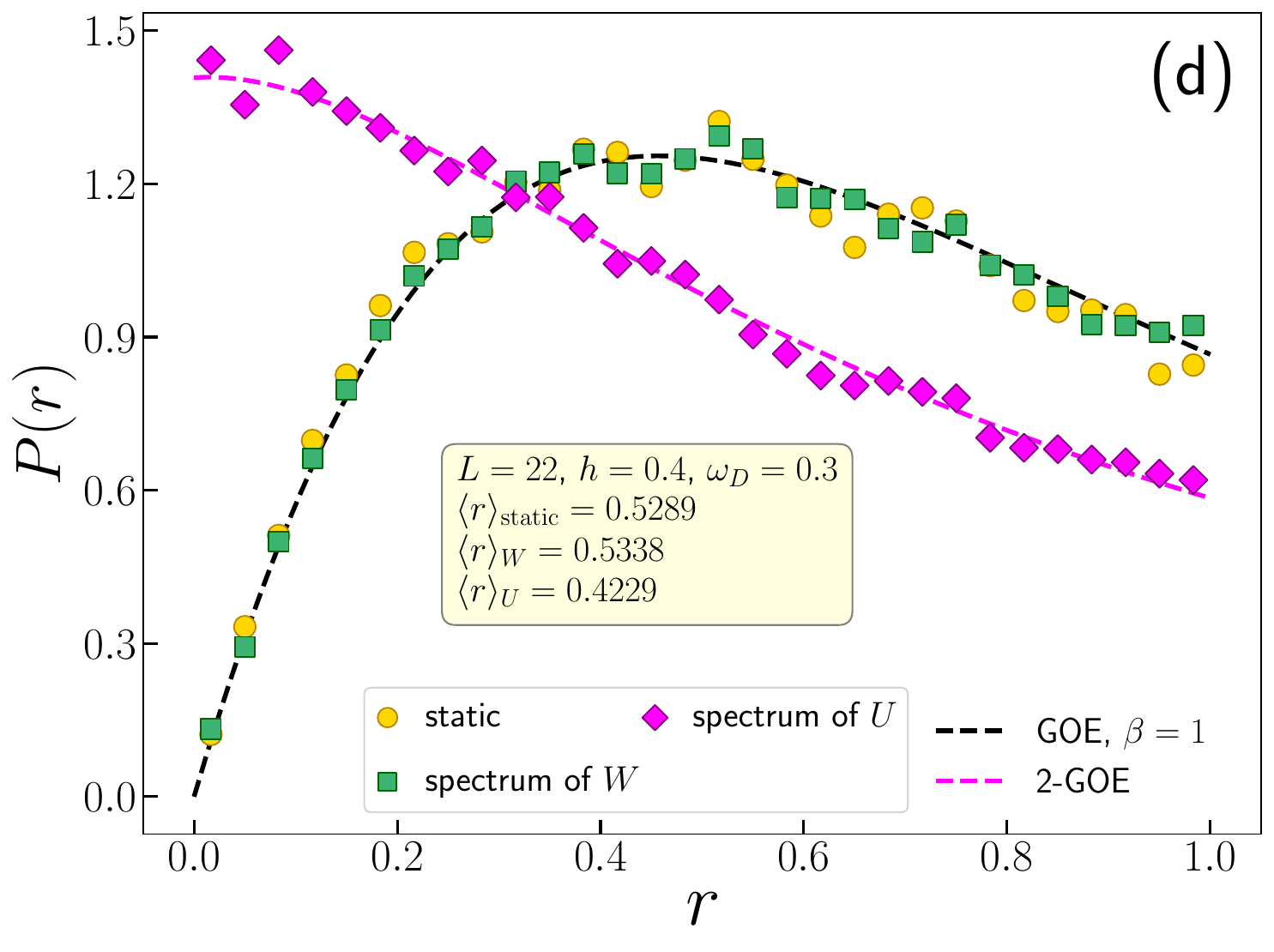}
    \caption{(a) Density of states of the static three-spin Hamiltonian (model III) with $h=0.4$ in the vicinity of the zero-mode subspace. (b) Eigenvalues $\{e^{i\phi}\}$ of $W$ for the Floquet three-spin problem. The insets magnify the neighborhoods of $\phi=0$ and $\phi=\pi$ (modulo $2\pi$): the $\nu_\pi=559$ exact modes at $\phi=\pi$ (magenta) are surrounded by a gap, while there is no gap at $\phi=0$ ($\nu_0=0$). (c) Density of states of the eigenphases $\{\theta\}$ of $U(T,0)=W^2$ in the vicinity of $\theta=0$; the black dashed curve is the analytical form $\rho_U^{\rm th}$, Eq.~\eqref{eq_dos_sim_form_U}, and the dash-dotted vertical lines mark $\pm2(\pi-\phi_\pi^{\rm th})$. The $559$ modes of $W$ at $\phi=\pi$ fold to $\theta=0$ (green line). (d) Distribution of the ratios of consecutive level spacings for the positive bulk eigenvalues of the static Hamiltonian and the positive bulk eigenphases of $W$ and $U(T,0)$, together with the GOE and the $2$-GOE distributions. In (b)--(d), the drive frequency is $\omega_D=0.3$, and the drive amplitude is $h=0.4$. In all the panels the system size is $L=22$ and the symmetry-resolved sector is $k_1=0$, $\mathcal{I}=+1$ ($\mathcal{D}=96909$).}
    \label{fig_6}
\end{figure*}
%-----------------------------------------------------------------------------------------------------------------------------------------------------------------------------------------

$H$ is real in the $\sigma^z$ basis, and $\prod_j\sigma_j^y$ flips both $\sigma^z_j$ and $\sigma^x_j$, so it anticommutes with each term of $H$. For $k_1=0$, $\mathcal{I}=+1$ and even $L$ the BDI operators of Eq.~\eqref{eq_Gaussian_BDI_action} are
\begin{equation} \label{eq_three_spin_static_BDI}
    \mathcal{C}_s = \prod_j \sigma_j^y\ , \qquad \mathcal{T}_s = \mathcal{K}\ , \qquad \mathcal{P}_s = \mathcal{C}_s\mathcal{K}\ ,
\end{equation}
where $\mathcal{K}\mathcal{C}_s\mathcal{K}=(-1)^L\mathcal{C}_s$ gives $\mathcal{P}_s^2=(-1)^L\mathbbm{1}$. Hence, $H$ belongs to ch-GOE for even $L$, which is the only case we consider; for odd $L$, $\mathcal{P}_s^2=-\mathbbm{1}$ and $H$ belongs to the CI class~\cite{Altland1997}.  In the sector $k_1=0$, $\mathcal{I}=+1$, $\mathrm{Tr}(\mathcal{C}_s)=74,\,143,\,284,\,559$ for $L=16,\,18,\,20,\,22$ ($\mathcal{D}=2250,\,7685,\,27012,\,96909$). It grows approximately as $2^{L/2}\approx1.41^{L}$, whereas $\mathcal{D}\simeq2^L/L$.

For $L=22$, where $\mathcal{D}=96909$ and $\nu=\mathrm{Tr}(\mathcal{C}_s)=559$, the density of states $\rho(E)$ shows a $\delta$-function at $E=0$ with the weight $\nu/\mathcal{D}$ predicted from the trace, followed by a spectral gap and then a square-root onset to a flat bulk density (Eq.~\eqref{eq_modified_MP}). The fitted form using Eq.~\eqref{eq_modified_MP} follows the histogram over the whole range $|E|\leq W_{\rm cut}$ (Fig.~\ref{fig_6}(a)). The measured ratio is $\Delta/\delta_{\rm fit}=164.6$, to be compared with the RMT estimate $\nu/\pi=177.9$, a relative deviation of $7.5\%$, much smaller than for the static versions of either model I or model II.

For the Floquet version of model III, we drive the transverse field periodically by a square-pulse protocol with drive frequency $\omega_D=2\pi/T$,
\begin{equation} \label{eq_h_pulse_three_spin}
    h(t) = \begin{cases}
        h, \ 0\leq t< \frac{T}{2}  \ , \\
        -h , \ \frac{T}{2} \leq t <T \ ,
    \end{cases}
\end{equation}
so that the Floquet unitary is
\begin{equation} \label{eq_time_evolution_op_three_spin}
\begin{aligned}
U(T,0) &= U_2U_1 \\ &= \exp\left(-i\frac{T}{2}H(-h)\right)\exp\left(-i\frac{T}{2}H(h)\right) \ .
\end{aligned}
\end{equation}
With $\mathcal{C}_x=\prod_j\sigma_j^x$ and $G=\prod_j\sigma_j^z$, the identities $\mathcal{C}_xH(\pm h)\mathcal{C}_x^{-1}=-H(\mp h)$ and $GH(\pm h)G^{-1}=H(\mp h)$ give
\begin{equation} \label{eq_three_spin_ingredients}
\begin{aligned}
    \mathcal{C}_x U_{1(2)}\mathcal{C}_x^{-1} = U_{2(1)}^{-1}\ , &\qquad
    \mathcal{P}_s U_{1(2)}\mathcal{P}_s^{-1} = U_{1(2)}\ ,\\
    \mathcal{K} U_{1(2)}\mathcal{K} = U_{1(2)}^{-1}\ , &\qquad
    G\, U_{1(2)}G^{-1} = U_{2(1)}\ .
\end{aligned}
\end{equation}
The first three relations give a set of BDI operators for $U(T,0)$ with the chiral operator $\mathcal{C}_x$ and the particle-hole operator of the static model,
\begin{equation} \label{eq_three_spin_floquet_set1}
\begin{aligned}
    \mathcal{C}_{\mathcal{F}} = \prod_j\sigma_j^x\ , \
    \mathcal{P}_{\mathcal{F}} = \mathcal{P}_s\ , \
    \mathcal{T}_{\mathcal{F}} = \mathcal{C}_{\mathcal{F}}\mathcal{P}_{\mathcal{F}} = i^L \prod_j\sigma_j^z\,\mathcal{K}\ ,
\end{aligned}
\end{equation}
which square to $\mathbbm{1}$ for even $L$ and satisfy Eq.~\eqref{eq_circular_BDI_action}. The static operators dressed by $U_2$ give a second set with the same particle-hole operator, as in model II,
\begin{equation} \label{eq_three_spin_floquet_set2}
\begin{aligned}
    \widetilde{\mathcal{C}}_{\mathcal{F}} = U_2\,\mathcal{C}_s\ , \qquad
    \widetilde{\mathcal{T}}_{\mathcal{F}} = U_2\,\mathcal{K}\ , \qquad
    \widetilde{\mathcal{P}}_{\mathcal{F}} = \widetilde{\mathcal{C}}_{\mathcal{F}}\widetilde{\mathcal{T}}_{\mathcal{F}} = \mathcal{P}_s\ ,
\end{aligned}
\end{equation}
which also square to $\mathbbm{1}$ and satisfy Eq.~\eqref{eq_circular_BDI_action}.

The last relation of Eq.~\eqref{eq_three_spin_ingredients} means that $G$ exchanges the two halves of the drive, $U_2=GU_1G^{-1}$. Since $G^2=\mathbbm{1}$, this dynamical symmetry implies that 
\begin{equation} \label{eq_three_spin_W}
    W(T,0) = G\,U_1 = U_2\,G\ , \qquad W(T,0)^2 = U(T,0)\ .
\end{equation}
$W(T,0)$ commutes with $T_1$ and $\mathcal{I}$, so it acts within the sector $k_1=0$, $\mathcal{I}=+1$. It connects the two sets of BDI operators,
\begin{equation} \label{eq_three_spin_W_connect}
    W(T,0) = (-1)^{L/2}\,\widetilde{\mathcal{C}}_{\mathcal{F}}\,\mathcal{C}_{\mathcal{F}} = (-1)^{L/2}\,\widetilde{\mathcal{T}}_{\mathcal{F}}\,\mathcal{T}_{\mathcal{F}}\ ,
\end{equation}
i.e., $W(T,0)$ is the product of the two chiral operators and of the two time-reversal operators, which share the particle-hole operator $\mathcal{P}_s$. Using Eq.~\eqref{eq_three_spin_W_connect} one finds that all four operators invert $W(T,0)$ and $\mathcal{P}_{\mathcal{F}}$ leaves it invariant,
\begin{equation} \label{eq_three_spin_W_BDI}
\begin{aligned}
    \mathcal{C}_{\mathcal{F}} W \mathcal{C}_{\mathcal{F}}^{-1} = \widetilde{\mathcal{C}}_{\mathcal{F}} W \widetilde{\mathcal{C}}_{\mathcal{F}}^{-1} &= \mathcal{T}_{\mathcal{F}} W \mathcal{T}_{\mathcal{F}}^{-1} = \widetilde{\mathcal{T}}_{\mathcal{F}} W \widetilde{\mathcal{T}}_{\mathcal{F}}^{-1} = W^{\dagger}\ , \\ \mathcal{P}_{\mathcal{F}} W \mathcal{P}_{\mathcal{F}}^{-1} &= W\ ,
\end{aligned}
\end{equation}
so $W(T,0)$ obeys the same relations as $U(T,0)$ and is BDI compatible (Eq.~\eqref{eq_circular_BDI_action}). The eigenphase density of $U(T,0)=W(T,0)^2$ is then given by the spectrum of $W(T,0)$ through Eq.~\eqref{eq_dos_form_U}.

The imbalances of $W(T,0)$ follow from the traces of its two chiral operators,
\begin{equation} \label{eq_three_spin_tr}
\begin{aligned}
    \mathrm{Tr}(\mathcal{C}_{\mathcal{F}}) &= \nu_0+\nu_\pi\ , \\
    \mathrm{Tr}(\mathcal{C}_{\mathcal{F}}W(T,0)) &= \nu_0-\nu_\pi = (-1)^{L/2}\,\mathrm{Tr}(\widetilde{\mathcal{C}}_{\mathcal{F}})\ ,
\end{aligned}
\end{equation}
where the last equality uses Eq.~\eqref{eq_three_spin_W_connect}. Moreover $\widetilde{\mathcal{C}}_{\mathcal{F}}=V\mathcal{C}_sV^{-1}$ with $V=U_2^{1/2}$, so $\mathrm{Tr}(\widetilde{\mathcal{C}}_{\mathcal{F}})=\mathrm{Tr}(\mathcal{C}_s)$. In the projector onto $k_1=0$, $\mathcal{I}=+1$ only basis states with equally many up and down spins contribute to the trace of an operator that flips all spins, and for these the matrix elements of $\prod_j\sigma^x_j$ and $\prod_j\sigma^y_j$ coincide, so that $\mathrm{Tr}(\mathcal{C}_{\mathcal{F}})=\mathrm{Tr}(\mathcal{C}_s)$. Hence
\begin{equation} \label{eq_three_spin_nu}
    \nu_0-\nu_\pi = (-1)^{L/2}\,(\nu_0+\nu_\pi)\ ,
\end{equation}
so that only one of the two imbalances is nonzero: $\nu_\pi=0$ and $\nu_0=\mathrm{Tr}(\mathcal{C}_s)$ for even $L/2$, whereas $\nu_0=0$ and $\nu_\pi=\mathrm{Tr}(\mathcal{C}_s)$ for odd $L/2$.

For $L=22$, $h=0.4$ and $\omega_D=0.3$ in the sector $k_1=0$, $\mathcal{I}=+1$ ($\mathcal{D}=96909$), $L/2$ is odd, so that $\nu_0=0$ and $\nu_\pi=559$. The eigenvalues of $W(T,0)$ are shown in Fig.~\ref{fig_6}(b). The $559$ exact modes at $\phi=\pi$ are surrounded by a spectral gap, whereas the spectrum lacks this feature at $\phi=0$. The gap edge agrees with the circular BDI density with $\nu_\pi=559$: the predicted edge is $\pi-\phi_\pi^{\rm th}=1.1516\times10^{-2}$ and the numerical one is $1.1867\times10^{-2}$, which is $3.0\%$ larger. The deviation of the closest eigenphase to $\phi=\pi$ from the corresponding saddle-point prediction is consistent with the scale predicted by Eq.~\eqref{eq:fluctuation_theta0} when using $\nu_\pi$.

Since $U(T,0)=W(T,0)^2$, all the protected modes of $W(T,0)$ at $\phi=\pi$ are mapped to $\theta=0$ for $U(T,0)$, and they appear as the $\delta$-function at $\theta=0$ with $559$ zero modes. As shown in Fig.~\ref{fig_6}(c), the density of $U(T,0)$ in the window $|\theta|<2(\pi-\phi_\pi^{\rm th})$ is half of the bulk value and rises to the full bulk value outside it, in agreement with Eq.~\eqref{eq_dos_sim_form_U}: inside the window only the branch $\rho_W(\theta/2)$ contributes, because $\rho_W(\theta/2+\pi)$ vanishes in the gap of $W(T,0)$ around $\phi=\pi$.

The level statistics from the ratios of consecutive gaps, $P(r)$, are given in Fig.~\ref{fig_6}(d). For the static model and for $W(T,0)$ we find the GOE ratio distribution ($\langle r\rangle=0.5289$ and $0.5338$, respectively), while the spectrum of $U(T,0)$, in which the two branches of the folded spectrum are superimposed, follows the $2$-GOE distribution with $\langle r\rangle_U=0.4229$.

For $L=6p\, ,\  p \in \mathbb{Z}^+$, the Hamiltonian has an additional $\mathbb{Z}_2\times\mathbb{Z}_2$ symmetry~\cite{Udupa2023}. Defining
\begin{equation} \label{eq_D_op_def}
\begin{aligned}
F_a&:=\prod_{j\equiv a\,(\mathrm{mod}\,3)}\sigma_j^x\ ,\\ D_1=F_2 F_3,\quad& D_2=F_1 F_3,\quad D_3=F_1F_2,
\end{aligned}
\end{equation}
we have $D_a^2=\mathbbm{1}$ and $D_1D_2D_3=\mathbbm{1}$, so only two of the three $D_a$ are independent. Since translation cyclically permutes $D_1, D_2, D_3$, we resolve the spectrum using the translation- and inversion-invariant operator $D=D_1+D_2+D_3$, whose $D=3$ sector corresponds to $D_1=D_2=D_3=+1$. All the operators of Eqs.~\eqref{eq_three_spin_static_BDI}, \eqref{eq_three_spin_floquet_set1}, \eqref{eq_three_spin_floquet_set2} and \eqref{eq_three_spin_W} commute with $D_a$, so the static and Floquet constructions above apply unchanged in the $D=3$ sector: $W(T,0)$ has the circular BDI symmetry as before, and the spectrum of $U(T,0)$ is obtained from that of $W$ by the phase-folding $\theta=2\phi$. We present a Floquet generalization of this model in Appendix~\ref {app_W_n}, where $U(T,0)=W(T,0)^n$ with $2 < n < \infty$, based on the same principle of dynamical symmetries. 

%----------------------------------------------------------------------------------------------------------------------------------------------------------------------------------------

\section{Conclusions and outlook}
\label{sec_discussions_conclusions}

We considered static and Floquet versions of three one-dimensional spin-$1/2$ models using square-pulse protocols such that the static Hamiltonian $H$ (Floquet unitary $U(T,0)$) has an exponentially large number (in system size) of zero (zero-quasienergy) modes in appropriate symmetry-resolved sectors at $E=0$ ($\theta=0$). Although these (zero) zero-quasienergy modes occupy a vanishing fraction of the entire Hilbert space, the exponentially growing chiral imbalance at $E=0$ ($\theta=0$) reorganizes the surrounding spectrum of $H$ ($U(T,0)$) over scales much larger than the mean level spacing. Our central result is that the symmetry class of a Floquet unitary does not, by itself, determine the structure of its eigenphase density or adjacent-gap-ratio distribution. Additional dynamical symmetry in the Floquet drive can produce new spectral features, even when the relevant unitary $U(T,0)$ belongs to the same circular BDI class.

The static counterparts provide a useful baseline. In the large symmetry-resolved sectors considered here, the extensive chiral imbalance produces an exact zero-energy manifold separated from the remaining states by a hard gap, with a square-root onset toward an approximately flat bulk density near $E=0$. The bulk level statistics match the Gaussian orthogonal ensemble (GOE), as expected for the chiral GOE class. Quantitative agreement with random matrix theory is strongest for the three-spin Ising model (model III) defined in an unconstrained Hilbert space, while the staggered PXP model (model II), with an $O(1)$ staggered detuning term,  gives a substantially better match than the clean PXP chain (model I), where the latter models are defined in a constrained Hilbert space using $L \leq 32$. In particular, the deviations from chiral GOE predictions are more significant near $E=0$ than for quantities like the mean level-spacing ratio, $\langle r \rangle$, which depend on the entire spectrum for models I and II. We do not have an explanation for this unusual finite-size effect in the PXP model or how it is remedied as the staggered detuning is progressively increased.

When the two halves of the Floquet drive are not related by a unitary symmetry, the Floquet operator $U(T,0)$ is described by the chiral circular orthogonal ensemble ch-COE $(\nu_0,0)$. Its eigenphase density $\rho(\theta)$ is fixed solely by the Hilbert-space dimension and the protected zero-mode
imbalance at $\theta=0$. In particular, the extensive imbalance separates the zero modes from the remaining spectrum by a hard edge, followed by a square-root onset toward the uniform bulk density. We find this behavior in the driven PXP chain (model I), where the match to random matrix theory predictions is significantly better compared to the static version. The level-spacing ratio of adjacent gaps follows GOE, as expected for the chiral circular orthogonal ensemble.

A different structure arises when a unitary operation $G$ exchanges the two halves of the Floquet drive. This leads to a time-glide-like dynamical symmetry in the Floquet problem, using which one can define a ``half-period" unitary $W(T,0)=GU_1=U_2G$, where $U_1=U(T/2,0)$ ($U_2=U(T, T/2)$), that satisfies $U(T,0)=W(T,0)^2$. In the symmetry sectors preserved by the half-period unitary $W(T,0)$, the Floquet operator can be written as $U(T,0) = W(T,0)^2$, with $W(T,0)$ itself having circular BDI
symmetry. Its protected modes at eigenphases $0$ and $\pi$ are both mapped to zero quasienergy of $U(T,0)$. The resulting eigenphase density of $U(T,0)$ then depends on which branches of $W(T,0)$ are gapped: it can exhibit a hard edge followed by a second square-root onset (Floquet version of model II), or a plateau
at half the bulk density without any spectral gap followed by a square-root onset to the bulk density (Floquet version of model III). Where the two folded branches overlap, the level statistics of $U(T,0)$ match a superposition of two GOE spectra. The staggered PXP and three-spin
Ising chains illustrate these possibilities. In sectors where $W(T,0)$ does not preserve the
symmetry block of $U(T,0)$, this folding does not occur within that block, and $U(T,0)$ retains the
ordinary single-spectrum ch-COE description. The Floquet version of the staggered PXP model clearly shows this effect: while $U(T,0)$ follows a ch-COE description in the sector $k_2=\pi, R_s=\pm 1$, it shows a hard edge followed by a second square-root onset in the sector $k_2=0, R_s=\pm 1$. In fact, in the sector $k_2=\pi, R_s=\pm 1$, the static version of model II only has time-reversal symmetry and belongs to class AI, whereas the Floquet drive supplies the chiral operator $\prod_j\sigma^z_j$, so that $U(T,0)$ is an ordinary BDI unitary with $\nu_\pi=0$ and $\nu_0=\mathrm{Tr}\left(\prod_j\sigma^z_j\right)$ protected modes.

More generally, our construction rests on a dynamical symmetry of the drive, i.e.\ a unitary $G$ with $H(t+T/n)=G\,H(t)\,G^{-1}$ whose $n$-th power is a complex phase in the resolved sector. Whether a model admits $G$ is a property of its Hamiltonian and its drive protocol. For model I, it fails because the spectrum of the uniform detuning $\sum_j\sigma^z_j$ in the constrained Hilbert space is one-sided, so no unitary can reverse its sign; for models II and III, however, $ G=T_1$ and $ G=\prod_j\sigma^z_j$ reverse the sign of the driven term, respectively. When $G$ exists, $W(T,0)$ is unique up to a sign for even $n$ and unique for odd $n$, once the BDI relations are imposed. The same principle gives $n$-step drives with $U(T,0)=W(T,0)^n$ in a concrete Floquet realization for model III (see Appendix.~\ref{app_W_n}). The density of $U(T,0)$ then shows a plateau at the fraction $(n-1)/n$ of the bulk value before a square-root approach to the bulk value, and the levels are those of $n$ superimposed GOE spectra, with $\langle r\rangle=0.4033$ for $n=3$ and the Poisson value as $n \rightarrow \infty$. This Poisson-like behavior at large $n$ does not signal integrability or hidden conservation laws, since the powers of $W(T,0)$ are generated by the drive and do not resolve $U(T,0)$ into smaller blocks.

Our work leaves some natural directions open for exploration. 
\begin{enumerate}
\item[(i)] The spectral edge of $W(T,0)$, $\phi_{0/\pi}$ (or $\theta_0$ for $U(T,0)$), where the exact zero/$\pi$ modes are separated from the bulk, is a soft edge of the Jacobi ensemble with $\beta=1$~\cite{johnstone2008multivariate}. We therefore expect the smallest positive eigenphase to follow the Tracy--Widom GOE law~\cite{Tracy1994, Tracy1996, johnstone2008multivariate}. The offsets of $1.6\%$--$5.3\%$ quoted when we compare the numerical data for the spectral edge to the saddle-point result are of the order of $\delta\theta/\theta_0^{\rm BDI}$ (Eq.~\eqref{eq:fluctuation_theta0}), which is a few percent for $|\nu_0|\sim10^2$, but a single realization of $W(T,0)$/ $U(T,0)$ spectrum cannot test the distribution of $\phi_{0/\pi}$ or $\theta_0$. This requires many realizations of these unitaries, for instance sampled over the drive parameters at different drive frequencies and drive amplitudes, together with a large Hilbert-space dimension so that $|\nu_0|\gg1$. The computational cost of such a sampling is large, and we leave this test of extreme-value statistics for the future.
\item[(ii)] Throughout this work, we have explored the regime  $1\ll|\nu_{0,\pi}|\ll\mathcal D$. It is interesting to ask what happens in the driven case when $|\nu_{0,\pi}|/\mathcal D$ is $\mathcal{O}(1)$ as $\mathcal D\to\infty$, in analogy with the static star-graph model of Ref.~\cite{jonay2026}, where a finite fraction of the spectrum is exactly degenerate. This is particularly interesting when both $\nu_0$ and $\nu_\pi$ are present (e.g., for $W(T,0)$): the protected modes at zero and at $\pi$ may then interact strongly through the common continuous spectrum, so that the two gaps and the spectral edges can no longer be treated independently. Their effect on the folded density of $U(T,0)=W(T,0)^2$ and on the level statistics remains to be studied.
\item[(iii)] The dynamical consequences of the spectral signatures in $U(T,0)$, with and without the dynamical symmetry, should be explored. The spectral form factor and operator growth are two quantities that may help answer this question. 
\item[(iv)] We have considered square-pulse protocols only. The time-glide-like construction itself carries over to smooth drives, for example $H(t)=H_0+\cos(\omega_Dt)\, V$ with $GH_0G^{-1}=H_0$ and $GVG^{-1}=-V$, for which $H(t+T/2)=G\,H(t)\,G^{-1}$. Whether the BDI symmetry relations hold for $U(T,0)$ and $W(T,0)$ in such protocols, and what the imbalances and spectral features near $\theta=0$ look like, remains for future work.
\end{enumerate}

%----------------------------------------------------------------------------------------------------------------------------------------------------------------------------------------------
\section{Acknowledgement}
\label{sec_acknowledgement}

S.R. acknowledges the support from UGC under the SRF scheme. During the preparation of this work, the authors used Claude (Anthropic, model Claude Sonnet 5.5) as an AI assistant for writing and testing numerical code (exact diagonalization in symmetry-resolved sectors and plotting scripts), cross-checking symmetry relations, and verifying the analytical steps. The authors also used ChatGPT (OpenAI), GPT-6.1 Sol, to help refine the presentation and articulation of scientific arguments. The authors checked and verified all physical results, derivations, numerical data, and the final text and take full responsibility for the content of this paper.

%----------------------------------------------------------------------------------------------------------------------------------------------------------------------------------------------------------

\begin{appendix}

%--------------------------------------------------------------------------------------------------------------------------------------------------------------------------------------------------------------

\begin{table*}
\centering
\begin{tabular}{|c|c|c|c|c|c|}
\hline
$L$ & $\mathcal{D}$ & $\Delta$ & $W_H$ & $\delta_{\mathrm{fit}}$ & $\delta_{\mathrm{k}}$ range \\
\hline \hline 
32 (Fig.~\ref{fig_1}(a)) & 77436 & $2.07031\times10^{-3}$ & $4.24112$ & $1.53087\times10^{-4}$ & $(1.420811$--$1.43767)\times10^{-4}$ \\
\hline
32 (Fig.~\ref{fig_1}(b)) & 154378 & $7.76123\times10^{-3}$ & $4.49917$ & $7.31442\times10^{-5}$ & $(7.57182$--$7.644071)\times10^{-5}$ \\
\hline \hline
30 (PXP) & 31836 & $1.51674\times10^{-3}$ & $4.12441$ & $3.05326\times10^{-4}$ & $(3.38066$--$3.43987)\times10^{-4}$ \\
\hline
30 (Staggered PXP) & 63320 & $9.26753\times10^{-3}$ & $4.17794$ & $1.73358\times10^{-4}$ & $(1.73051$--$1.75617)\times10^{-4}$ \\
\hline 
\end{tabular}
\caption{Comparison of the mean level spacing obtained from the hard-edge DOS curve fit and the level-counting procedure. For each spectrum, the quoted range is obtained from the five energy windows considered in the analysis. The curve-fit value is stable under variation of $W_{\mathrm{cut}}/W_H$ between $0.15$ and $0.20$, while five different $(a,b)$ windows test the stability of $\delta_{\rm k}$.}
\label{tab:app_spacing_comparison}
\end{table*}

\section{Numerical extraction of $\delta_{\rm fit}$ from DOS}
\label{app_delta_fit}
To assess the robustness of the mean level spacing extracted from the density of states near zero-modes, we compare two complementary procedures: (i) a nonlinear fit of the hard gapped spectral edge density of states form and (ii) the direct level-counting procedure introduced by Kawabata et al.~\cite{kawabata2026}. The comparison is performed for the two spectra corresponding to Fig.~\ref{fig_1}(a) and Fig.~\ref{fig_1}(b) in the main text.

For the curve-fit procedure, we fit the density of states to the form in Eq.~\eqref{eq_modified_MP}. The positive and negative parts of the spectrum are separately binned. 
For an energy bin centered at $E_i$ with width $\mathrm{d}E$, let $N_{+,i}$ and $N_{-,i}$ denote the number of positive and negative eigenvalues in the corresponding energy intervals. 
The normalized DOS is then
\begin{equation}
\rho_{+}(E_i)
=
\frac{N_{+,i}}{\mathcal{D}\,\mathrm{d}E},
\qquad
\rho_{-}(E_i)
=
\frac{N_{-,i}}{\mathcal{D}\,\mathrm{d}E}.
\label{eq:rho_pm}
\end{equation}

Using the spectral symmetry, we define the symmetrized DOS as

\begin{equation}
\rho_{\mathrm{sym}}(E_i)
=
\frac{1}{2}
\left[
\rho_{+}(E_i)+\rho_{-}(E_i)
\right].
\label{eq:rho_sym}
\end{equation}
The parameter $\delta_{\mathrm{fit}}$ is then obtained by minimizing the difference
\begin{equation*}
\begin{aligned}
 \underset{\delta>0}{\operatorname{arg\ min}} \sum_{E_i\in[\Delta,W_{\mathrm{cut}}]} \left[ \rho_{\mathrm{sym}}(E_i)
- \frac{1}{\mathcal{D}\,\delta} \sqrt{ 1-\left(\frac{\Delta}{E_i}\right)^2 } \right]^2 \ .
\end{aligned}
\end{equation*}

Next, we define the spectral width
\begin{equation}
W_H=\sqrt{\frac{1}{\mathcal D}\operatorname{Tr}(H^2)} =\sqrt{\frac{1}{\mathcal D}\sum_n E_n^2}\ .
\end{equation}
We use $W_H$ to set the energy scale for the level-counting procedure. Specifically, we consider five energy windows, whose boundaries are expressed in units of $W_H$. For a window $E_{\mathrm{low}}=a W_H, \ E_{\mathrm{high}}=bW_H$, with $b>a$, we count the number of positive-energy eigenvalues in that interval,
\begin{equation}
N_{\mathrm{window}} =\#\left\{E_n>0: aW_H\leq E_n<bW_H\right\}\ ,
\end{equation}
and estimate the mean level spacing as
\begin{equation}
\delta_{\mathrm{k}} =\frac{E_{\mathrm{high}}-E_{\mathrm{low}}} {N_{\mathrm{window}}} =\frac{(b-a)W_H}{N_{\mathrm{window}}}\ .
\end{equation}

Table~\ref{tab:app_spacing_comparison} summarizes the comparison between the two examples. The five chosen windows, $(a,b)=(0.15,0.25), (0.17,0.27), (0.12,0.28),(0.10,0.24)$, and $(0.08,0.20)$, provide five estimates of $\delta_{\mathrm{k}}$, whose minimum and maximum are reported in the table. For Fig.~\ref{fig_1}(b), the curve-fit procedure gives $\delta_{\mathrm{fit}}=7.31442\times10^{-5}$; whereas for $L=30$ in the same sector of the staggered PXP problem, it gives  $\delta_{\mathrm{fit}}=1.73358\times10^{-4}$. The Kawabata estimates obtained from different bulk windows lie in the range $7.57182\times10^{-5}$--$7.644071\times10^{-5}$ and $1.73051\times10^{-4}$--$1.75617\times10^{-4}$, respectively. For Fig.~\ref{fig_1}(a), the corresponding curve-fit value is $\delta_{\mathrm{fit}}=1.53087\times10^{-4}$, whereas the level-counting estimates lie in the range $1.420811\times10^{-4}$--$1.43767\times10^{-4}$. When it comes to $L=30$, we obtain $\delta_{\mathrm{fit}}=3.05326\times10^{-4}$ and $\delta_{\mathrm k}= 3.38066\times10^{-4}$--$3.43987\times10^{-4}$ from the two procedures. For the staggered PXP problem, the two procedures give very close estimates, with the $\delta_k$ values differing from the curve-fit result by only a few percent across the different windows. This agreement indicates that the extracted spacing does not strongly depend on the method used to characterize the bulk density of states. Nevertheless, the Kawabata estimates are mutually consistent across the different energy windows, providing a useful measure of the systematic variation associated with the choice of fitting interval.

For comparison with the random-matrix prediction, the expected hard-edge relation is
\begin{equation}
\frac{\Delta_{\rm RMT}}{\delta_{\mathrm{RMT}}} \approx \frac{\nu}{\pi}\ .
\end{equation}
For $L=30$ PXP, $\nu/\pi\simeq 67.4817$, whereas for $L=30$ staggered PXP $\nu/\pi\simeq 112.0451$. For Fig.~\ref{fig_1}(a), $\nu/\pi\simeq 108.8620$, while for Fig.~\ref{fig_1}(b), $\nu/\pi\simeq 157.2451$. Comparing these values with the corresponding numerical ratios $\Delta/\delta$ provides an additional diagnostic of how closely the spectrum near protected zero modes follows the expected sharp-edge random-matrix scaling.

To quantify the agreement with the hard-edge relation $\Delta/\delta\simeq\nu/\pi$, we consider the relative deviation $\epsilon=\left[\left|(\Delta/\delta_{\rm fit})-\nu/\pi\right|\right]/(\nu/\pi)$. For the PXP model, the numerical ratio is substantially smaller than the random-matrix estimate: for $L=32$ we find $\Delta/\delta_{\rm fit}=13.52$ against $\nu/\pi=108.86$, i.e.\ $\epsilon\simeq88\%$, and for $L=30$, $4.97$ against $67.48$, i.e.\ $\epsilon\simeq93\%$. The staggered PXP model is considerably closer to the prediction: $\Delta/\delta_{\rm fit}=106.1$ against $\nu/\pi=157.2$ for $L=32$ ($\epsilon\simeq33\%$), and $53.5$ against $112.0$ for $L=30$ ($\epsilon\simeq52\%$). For comparison, the three-spin model with periodic boundary conditions ($L=22$) deviates by only $\epsilon\simeq7.5\%$ (Fig.~\ref{fig_6}(a)). In all cases, the numerically obtained quantity $\Delta/\delta_{\rm fit}$ is smaller than the random-matrix estimate, and $\epsilon$ decreases as $L$ is increased.

%-------------------------------------------------------------------------------------------------------------------------------------------------------------------------------------

\section{Derivation of circular BDI density of states from JPD function}
\label{app_derivation}
We begin by considering the circular Jacobi distribution~\cite{Nishigaki_2003, annurev.nucl.50.1.343} in terms of the variables $y_i = \sin^2\left( \frac{\theta_i}{2}\right)$, where $\theta \in [0,\pi]$ are the positive eigenphases of a random BDI unitary matrix $U$,
\begin{equation} \label{app_eq_circ_BDI_JPDF}
\begin{aligned} 
    & P(y_1,y_2, \cdots , y_N) \\ & \quad \propto \left(\prod_i y_i^{(|\nu_0| -1)/2} \ (1-y_i)^{-1/2} \right) \ \prod_{i<j} |y_i - y_j|\ .
\end{aligned}
\end{equation}
Here we have taken $\mathcal{D}=2N+|\nu_0|$ to be the dimensionality of $U$ and $\nu_0$ is the imbalance in $\theta=0$ subspace. Taking the logarithm on both sides of Eq.~\eqref{app_eq_circ_BDI_JPDF} gives
\begin{equation}
    \begin{aligned}
        \log P &= \frac{|\nu_0| -1}{2} \sum_i \log y_i - \frac{1}{2} \sum_i \log(1-y_i) \\
        & \quad + \sum_{i <j} \log|y_i-y_j| +  \rm{const} \  ,
    \end{aligned}
\end{equation}
where the constant term arises from the proportionality constant in Eq.~\eqref{app_eq_circ_BDI_JPDF}. Define the density of states in the large-$N$ limit
\begin{align}
    \rho(y) := \frac{1}{N} \sum_{i=1}^N \delta(y - y_i) \ .
\end{align}
Then for $N\to \infty$, we have the action in the continuum limit
\begin{equation} \label{app_eq_action}
    \begin{aligned}
        & S[\rho] \\&= -\frac{|\nu_0| - 1}{2} \int_0^1 dy' \rho(y') \log y' + \frac{1}{2}\int_0^1 dy' \rho(y') \log(1-y') \\
        &  - \int_0^1 \int_0^1 dy \, dy'\rho(y) \rho(y') \log|y-y'| + \mu \left[ \int_0^1 dy' \rho(y') -1 \right] \ ,
    \end{aligned}
\end{equation}
where $\mu$ is a Lagrange multiplier assuring the normalization of DOS. An extremization of the action in Eq.~\eqref{app_eq_action}
\begin{align}
    \frac{\delta S[\rho]}{\delta \rho(y)} = 0 \ ,
\end{align}
leads to the saddle equation
\begin{equation} \label{app_eq_saddle}
    \mathrm{P.V.} \int_0^1 dy'\,  \frac{\rho(y')}{y-y'} = - \frac{|\nu_0 |-1}{2Ny} - \frac{1}{2N(1-y)} \ , 
\end{equation}
where `$\rm{P.V.}$' denotes the Cauchy principal value of the integral. 
%We consider the thermodynamic limit $N\to\infty$ with $\lim_{N\to\infty} (\nu_0/N) \geq 0 $ held fixed. The first term on the right-hand side of Eq.~\eqref{app_eq_saddle} then becomes $-\alpha/(2y)$ with $\alpha:=(\nu_0-1)/N$, while the second one is $\mathcal{O}(1/N)$ and hence can be dropped. The saddle equation then reduces to
We consider the limit $N\to\infty$ with $|\nu_0|\to\infty$; in the models studied in the main text $|\nu_0|/N\to0$. Both terms on the right-hand side of Eq.~\eqref{app_eq_saddle} are then small, and it is their ratio that matters: $\frac{1/[2N(1-y)]}{(|\nu_0|-1)/(2Ny)}=\frac{y}{(|\nu_0|-1)(1-y)}$ is of order $1/|\nu_0|$ except in around the boundary at $y=1$. For $|\nu_0|\gg1$ the second term can therefore be dropped, and with $\alpha:=(|\nu_0|-1)/N>0$ the saddle equation reduces to
\begin{equation} \label{app_eq_saddle_reduced}
    \mathrm{P.V.} \int_a^1 dy' \, \frac{\rho(y')}{y-y'} = V'(y) = -\frac{\alpha}{2y} \ ,
\end{equation} 
where the confining potential is $V(y) = -\frac{\alpha}{2} \log y$. We note that the point $\theta=\pi$ is a regular point of the unit circle, i.e., the density in $\theta$ is smooth across $\theta = \pi$. In the variable $y=\sin^2(\theta/2)$, it corresponds to the edge $b=1$, where the Jacobian $dy/d\theta = \sin(\theta/2)\cos(\theta/2)$ vanishes and $\rho(y)$ may diverge as $(1-y)^{-1/2}$. For $\alpha>0$ the $-\alpha/(2y)$ term repels the eigenphases from $\theta = 0$, so the lower edge $a>0$ is an edge with $\rho(a)=0$. Hence, we take the support as $y \in [a,1]$. The remaining task is to solve this saddle equation to extract the equilibrium DOS $\rho(y)$~\cite{Ramli_2012, Marino_2016, tricomi1985integral}.

Introduce the resolvent~\cite{brezin1978} (often called the Stieltjes transform~\cite{PhysRevLett.112.254101})
\begin{equation} 
    G(z) := \int_a^1 dy\  \frac{\rho(y)}{z - y}  \ .
\end{equation}
$G(z)$ is analytic in entire complex plane except on the cut $[a,1]$ and since $\int \rho(y) dy =1$, $G(z) = 1/z + \mathcal{O}(z^{-2})$ when $|z| \to \infty$. Call $G_{\pm} (y) := \lim_{\epsilon \to 0^+} G(y \pm i \epsilon)$. Then  Sokhotski–Plemelj formula~\cite{Muskhelishvili1953, ablowitz2003complex} gives
\begin{equation} \label{app_eq_SP_formula}
    G_{\pm}(y) = \mathrm{P.V.} \int_0^1 dy' \frac{\rho(y')}{y-y'} \mp i \pi \rho(y) \ ,
\end{equation}
which translates into the following equations.
\begin{align} 
    & G_+(y) + G_-(y) = 2 V'(y) = - \frac{\alpha}{y} \ , \\
    &\rho(y) = - \frac{1}{\pi} \mathrm{Im} \, G_+(y) \ . \label{app_eq_SP_formula_imply}
\end{align}
Next let
\begin{equation}
    S(z) := \sqrt{(z-a)(z-1)} \ , \quad w(y) := \sqrt{(y-a)(1-y)} \ ,
\end{equation}
where $S$ is analytic in $\mathbb{C}$ except on the cut $[a,1]$ with $S(z)\simeq z$ as $|z|\to\infty$. Its boundary values on the cut are $S_\pm(y) = \pm i w(y)$ so that $S_{+}(y)=-S_{-}(y)$ and $S(z)<0$ for real $z<a$. Now define
\begin{equation} \label{app_eq_F_G_rel}
    F(z) := S(z)\,G(z) \ .
\end{equation}
$F$ is analytic in $\mathbb{C}$ except on the cut $[a,1]$ and $F(z)\to 1$ as $|z|\to\infty$. Because $S_-=-S_+$, the jump of $F$ across the cut is determined by Eq.~\eqref{app_eq_SP_formula_imply}, as
\begin{equation}\label{app_eq_F_jump}
    \begin{aligned}
    & F_+(y)-F_-(y) = S_+(y)\left[G_+(y)+G_-(y)\right] \\ & =  -i \alpha \,\frac{w(y)}{y} =: J(y) \ .
    \end{aligned}
\end{equation}
We construct $F$ from its jump. Define
\begin{equation}
    F_{\rm cut}(z) := \frac{1}{2\pi i}\int_a^1  dt\ \frac{J(t)}{t-z} \ .
\end{equation}
This function (i) is analytic everywhere except on the segment $[a,1]$, (ii) vanishes as $|z|\to\infty$, (iii) jumps across the segment by exactly $J$,
$F_{{\rm cut},+}-F_{{\rm cut},-}=J$ (Sokhotski-Plemelj formula), and (iv) stays finite near $z=a$ and $z=1$, because $J$ is continuous and vanishes there. $F$ has the same jump, so $\chi(z):=F(z)-F_{\rm cut}(z)$ has none: its values just above and just below the segment coincide, and $\chi$ is analytic across it. In fact, $\chi$ is analytic on the whole complex plane. As $|z|\to\infty$ we have $F\to1$ and $F_{\rm cut}\to0$, so that $\chi\to1$. A function that is analytic on the whole plane and bounded is constant (Liouville's theorem~\cite{ablowitz2003complex}), so $\chi\equiv1$:
\begin{equation} \label{app_eq_F_decomp}
    F(z) = 1 + F_{\rm cut}(z) \ .
\end{equation}
To evaluate $F_{\rm cut}$ we use $\frac{1}{t(t-z)}=\frac{1}{z}\left[\frac{1}{t-z}-\frac{1}{t}\right]$, which gives
\begin{equation} \label{app_eq_Fcut_I}
    F_{\rm cut}(z) = \frac{\alpha}{2\pi z}\left[I_1 - I_2(z)\right] \ .
\end{equation}
The integrals can be evaluated using contour integration; however, we quote the results below, which have also been verified using Mathematica~\cite{Mathematica}.
\begin{align} 
    I_1 &:= \int_a^1  dt\,\frac{w(t)}{t} =  \frac{\pi}{2} \left(1 - \sqrt{a}\right)^2\ , \\ 
    I_2(z) &:= \int_a^1 dt\,\frac{w(t)}{t-z} \nonumber \\&= \pi \left[ \sqrt{(z-a)(z-1)} + \frac{1+a}{2} - z \right] \ .
\end{align}

Plugging these into Eq.~\eqref{app_eq_F_decomp}, we get
\begin{equation} \label{app_eq_F_final}
    F(z) = 1 + \frac{\alpha}{2} - \frac{\alpha}{2 z} \left(\sqrt{a} + S(z)\right) \ .
\end{equation}
Notice that for $z=a$, $S(a)=0$ while $G(a)$ is finite. So we get the condition for the edge $y=a$ such that $F(a)=0$, which implies from Eq.~\eqref{app_eq_F_final}
\begin{equation}
    a = \left(\frac{\alpha}{\alpha +2} \right)^2 = \left( \frac{|\nu_0 |-1}{\mathcal{D}-1}\right)^2\ .
 \end{equation}
Now using the form of $F$ in Eq.~\eqref{app_eq_F_final}, we get from Eq.~\eqref{app_eq_F_G_rel}
\begin{equation} \label{app_eq_G_intm}
    \begin{aligned}
        G(z) = -\frac{\alpha}{2z} + \frac{1}{S(z)} \left( 1 + \frac{\alpha}{2} - \frac{\alpha \sqrt{a}}{2z}\right) \ .
    \end{aligned}
\end{equation}
The second term in right-hand side of Eq.~\eqref{app_eq_F_G_rel} can be simplified to $\frac{1}{z}\sqrt{(z-a)/(z-1)}\, (1+\alpha/2)$, so that it can be rewritten as
\begin{align} \label{app_eq_G_simplified}
    G(z) = -\frac{\alpha}{2z} +  \frac{1}{z}\left(1+\frac{\alpha}{2}\right) \sqrt{\frac{(z-a)}{(z-1)}} \ .
\end{align}
Thus for $z=y+i0^+$ with $a<y<1$, 
\begin{align}
    G_+(y) = -\frac{\alpha}{2y} -  \frac{i}{y}\left(1+\frac{\alpha}{2}\right) \sqrt{\frac{(y-a)}{(1-y)}} \ .
\end{align}
Finally, we get the density of states using Eq.~\eqref{app_eq_SP_formula}
\begin{equation} \label{app_eq_dos_y}
    \begin{aligned} 
        \rho(y) &= \frac{\left(1+\frac{\alpha}{2}\right)}{\pi y} \sqrt{\frac{(y-a)}{(1-y)}} \ .
    \end{aligned}
\end{equation}
Returning to $\theta$ with $y=\sin^2(\theta/2)$, $ dy=\sin(\theta/2) \cos(\theta/2)\,d\theta$ and $\sqrt{1-y}=\cos(\theta/2)$, the normalized density for $
\theta \in[\theta_0,\pi]$ is given by 
\begin{equation}\label{app_eq_rho_theta}
\begin{aligned}
\rho(\theta) &=\rho(y)\ \frac{d y}{d\theta} \\
    &= \frac{1}{\pi}\left(\frac{\mathcal{D}- 1}{\mathcal{D}-\nu_0}\right)\sqrt{1-\frac{\sin^2(\theta_0/2)}{\sin^2(\theta/2)}}\ ,
\end{aligned}
\end{equation}
and $\rho(\theta)=0$ for $0\le\theta<\theta_0$, with the gap edge
\begin{equation} \label{app_eq_gap}
    \theta_0 = 2\arcsin\sqrt a = 2\arcsin\!\left(\frac{|\nu_0|-1}{\mathcal{D}-1}\right) \ .
\end{equation}
For the whole interval $(-\pi, \pi]$, the normalized DOS is given by Eq.~\eqref{eq_circ_BDI_dos} in the main text.

%---------------------------------------------------------------------------------------------------------------------------------------------------------------------------------------------------------
\section{Derivation of circular BDI density of states in presence of an additional chiral symmetry}
\label{app_derivation_additional}
As shown in the main text, for a BDI random matrix $W$ having two chiral symmetries $\mathcal{C}$ and $W\mathcal{C}$, we have imbalances $\nu_0$ and $\nu_\pi$ in the zero- and $\pi$-mode subspaces, where both of these can be macroscopically large. Then the joint probability distribution function in terms of $y_i=\sin^2(\theta_i/2)$ variables reads
\begin{equation} \label{app_add_eq_circ_BDI_JPDF}
\begin{aligned} 
    & P(y_1,y_2, \cdots , y_N) \\ & \quad \propto \left(\prod_i y_i^{(|\nu_0| -1)/2} \ (1-y_i)^{(|\nu_{\pi}| -1)/2} \right) \ \prod_{i<j} |y_i - y_j|\ .
\end{aligned}
\end{equation}
Here $\theta_i\in[0,\pi]$, $i=1,\dots,N$, are the positive eigenphases of $W$, while the remaining $\nu_0$ ($\nu_\pi$) eigenvalues sit exactly at $\theta=0$ ($\pi$); the dimension of $W$ is therefore $\mathcal{D}=2N+|\nu_0|+|\nu_\pi|$. Taking the logarithm on both sides of Eq.~\eqref{app_add_eq_circ_BDI_JPDF} gives
\begin{equation}
    \begin{aligned}
        \log P &= \frac{|\nu_0| -1}{2} \sum_i \log y_i + \frac{|\nu_\pi| -1}{2} \sum_i \log(1-y_i) \\
        & \quad + \sum_{i <j} \log|y_i-y_j| +  \mathrm{const} \  ,
    \end{aligned}
\end{equation}
where the constant term appears due to the proportionality constant in Eq.~\eqref{app_add_eq_circ_BDI_JPDF}. Define the density of states (DOS) in the large $N$ limit
\begin{align}
    \rho(y) := \frac{1}{N} \sum_{i=1}^N \delta(y - y_i) \ , \qquad \int_0^1 dy\, \rho(y) = 1 \ .
\end{align}
For $N\to\infty$ we replace
\begin{equation}
\begin{aligned}
    \sum_i f(y_i) &\to N\int dy\, \rho(y)\, f(y) \ , \\
    \sum_{i<j} g(y_i,y_j) &\to \frac{N^2}{2}\iint dy\, dy'\, \rho(y)\rho(y')\, g(y,y') \ .
\end{aligned}
\end{equation}
Writing $P\propto e^{-S[\rho]}$, we obtain the action in the continuum limit
\begin{equation} \label{app_add_eq_action}
    \begin{aligned}
        S[\rho] = &-N\,\left(\frac{|\nu_0 |- 1}{2}\right) \int_0^1 dy\, \rho(y) \log y \\
        & - N\, \left(\frac{|\nu_\pi| - 1}{2}\right)\int_0^1 dy\, \rho(y) \log(1-y) \\
        &  - \frac{N^2}{2} \int_0^1 \!\! \int_0^1 dy \, dy'\, \rho(y) \rho(y') \log|y-y'| \\
        & + \mu \left[ \int_0^1 dy\, \rho(y) -1 \right] \ ,
    \end{aligned}
\end{equation}
where $\mu$ is a Lagrange multiplier assuring the normalization of the DOS. Extremization
\begin{align}
    \frac{\delta S[\rho]}{\delta \rho(y)} = 0 \ ,
\end{align}
followed by a derivative with respect to $y$ leads to the saddle equation
\begin{equation} \label{app_add_eq_saddle}
     \mathrm{P.V.} \int_0^1 dy'\,  \frac{\rho(y')}{y-y'} = - \frac{|\nu_0| -1}{2Ny} + \frac{|\nu_{\pi} |-1}{2N(1-y)} \ ,
\end{equation}
where $\mathrm{P.V.}$ denotes the Cauchy principal value of the integral. We consider the limit $N\to\infty$ with both $|\nu_{0, \pi}|\gg1$ and $|\nu_{0,\pi}|/N \to 0$, and define
\begin{equation}
    \alpha := \frac{|\nu_0 |- 1}{N} \ , \qquad \beta := \frac{|\nu_\pi| - 1}{N} \ , \qquad \alpha,\beta>0 \ .
\end{equation}
Then the saddle equation (Eq.~\eqref{app_add_eq_saddle}) takes the form
\begin{equation} \label{app_add_eq_saddle_reduced}
\begin{aligned}
\mathrm{P.V.} \int_a^b dy' \, \frac{\rho(y')}{y-y'} &= V'(y) = -\frac{\alpha}{2y} + \frac{\beta}{2(1-y)} \ , \\
V(y) &= -\frac{\alpha}{2}\log y - \frac{\beta}{2}\log(1-y) \ ,
\end{aligned}
\end{equation}
where $V$ is the confining potential. In contrast to the single chiral imbalance case in Appendix~\ref{app_derivation}, no term can be dropped here: both one-body terms are of the same order as the repulsion. Since $V(y)\to+\infty$ for $y\to0^+$ and for $y\to1^-$, the eigenvalues are repelled from both ends of the interval $[0,1]$, so that the support of $\rho$ is $[a,b]$ with $0<a<b<1$. In terms of $\theta$, this means that gaps open around both $\theta=0$ and $\theta=\pi$. The remaining task is to solve Eq.~\eqref{app_add_eq_saddle_reduced} to extract the equilibrium DOS $\rho(y)$ and the edges $a$ and $b$~\cite{Ramli_2012, Marino_2016, tricomi1985integral}.

Introduce the resolvent~\cite{brezin1978}
\begin{equation}
    G(z) := \int_a^b dy\  \frac{\rho(y)}{z - y}  \ .
\end{equation}
$G(z)$ is analytic in the entire complex plane except on the cut $[a,b]$, and since $\int \rho(y)\, dy =1$, $G(z) = 1/z + \mathcal{O}(z^{-2})$ when $|z| \to \infty$. Call $G_{\pm} (y) := \lim_{\epsilon \to 0^+} G(y \pm i \epsilon)$. Then the Sokhotski--Plemelj formula~\cite{Muskhelishvili1953, ablowitz2003complex} gives 
\begin{equation}
    G_{\pm}(y) = \mathrm{P.V.} \int dy' \frac{\rho(y')}{(y-y')} \mp i \pi \rho(y) \ ,
\end{equation}
which together with Eq.~\eqref{app_add_eq_saddle_reduced} translates into
\begin{align}
    G_+(y) + G_-(y) &= 2 V'(y) = -\frac{\alpha}{y} + \frac{\beta}{1-y} \ ,\\
    \rho(y) &= - \frac{1}{\pi} \mathrm{Im} \, G_+(y) \ ,  \label{app_add_eq_SP}
\end{align}
for $y\in(a,b)$. Next let
\begin{equation}
    S(z) := \sqrt{(z-a)(z-b)} \ , \quad w(y) := \sqrt{(y-a)(b-y)} \ ,
\end{equation}
where $S$ is analytic in $\mathbb{C}$ except on the cut $[a,b]$ with $S(z)\simeq z$ as $|z|\to\infty$. Its boundary values on the cut are $S_\pm(y) = \pm i\, w(y)$, so that $S_{+}(y)=-S_{-}(y)$. Moreover, $S(z)<0$ for real $z<a$ and $S(z)>0$ for real $z>b$, hence
\begin{equation} \label{app_add_eq_S_01}
    S(0) = -\sqrt{ab} \ , \qquad S(1) = +\sqrt{(1-a)(1-b)} \ .
\end{equation}
Now define
\begin{equation} \label{app_add_eq_F_G_rel}
    F(z) := S(z)\,G(z) \ .
\end{equation}
$F$ is analytic in $\mathbb{C}$ except on the cut $[a,b]$ and $F(z)\to 1$ as $|z|\to\infty$. Because $S_-=-S_+$, the jump of $F$ across the cut is determined by Eq.~\eqref{app_add_eq_SP}, as
\begin{equation}\label{app_add_eq_F_jump}
    \begin{aligned}
    F_+(y)-F_-(y) &= S_+(y)\left[G_+(y)+G_-(y)\right] \\ & =  i\, w(y)\left[-\frac{\alpha}{y}+\frac{\beta}{1-y}\right] =: J(y) \ .
    \end{aligned}
\end{equation}
We construct $F$ from its jump. Define
\begin{equation}
    F_{\rm cut}(z) := \frac{1}{2\pi i}\int_a^b  dt\ \frac{J(t)}{t-z} \ .
\end{equation}
This function (i) is analytic everywhere except on the segment $[a,b]$, (ii) vanishes as $|z|\to\infty$, (iii) jumps across the segment by exactly $J$ as $F_{{\rm cut},+}-F_{{\rm cut},-}=J$ (Sokhotski--Plemelj formula), and (iv) stays finite near $z=a$ and $z=b$, because $J$ is continuous and vanishes there. $F$ has the same jump, so $\chi(z):=F(z)-F_{\rm cut}(z)$ has none and as before by Liouville's theorem~\cite{ablowitz2003complex}, $\chi\equiv1$:
\begin{equation} \label{app_add_eq_F_decomp}
    F(z) = 1 + F_{\rm cut}(z) \ .
\end{equation}
To evaluate $F_{\rm cut}$ we use $\frac{1}{t(t-z)}=\frac{1}{z}\left[\frac{1}{t-z}-\frac{1}{t}\right]$ and $\frac{1}{(1-t)(t-z)}=\frac{1}{1-z}\left[\frac{1}{1-t}+\frac{1}{t-z}\right]$, which gives
\begin{equation} \label{app_add_eq_Fcut_I}
    F_{\rm cut}(z) = \frac{\alpha}{2\pi z}\left[I_1 - I_2(z)\right] + \frac{\beta}{2\pi (1-z)}\left[I_3 + I_2(z)\right] \ ,
\end{equation}
where
\begin{align*}
    I_1 &:= \int_a^b dt\,\frac{w(t)}{t} \ , \qquad I_2(z) := \int_a^b dt\,\frac{w(t)}{t-z}\ , \\ I_3 &:= \int_a^b dt\,\frac{w(t)}{1-t} \ .
\end{align*}
These integrals can be evaluated using contour integration, yielding the following results.
\begin{align}
    I_1 &= \frac{\pi}{2} \left(\sqrt{b} - \sqrt{a}\right)^2\ , \\
    I_2(z) &= \pi \left[ \sqrt{(z-a)(z-b)} + \frac{a+b}{2} - z \right] \ ,\\
    I_3 &= \frac{\pi}{2} \left(\sqrt{1-a} - \sqrt{1-b}\right)^2\ . 
\end{align}
Plugging these into Eq.~\eqref{app_add_eq_Fcut_I} and then into Eq.~\eqref{app_add_eq_F_decomp}, we obtain
\begin{equation} \label{app_add_eq_F_final}
    \begin{aligned}
        F(z) &= \eta - \frac{\alpha}{2 z} \left(\sqrt{ab} + S(z)\right)\\ &\quad + \frac{\beta}{2(1-z)}\left(S(z) - \sqrt{(1-a)(1-b)}\right) \ , \\
        \eta &:= 1+\frac{\alpha+\beta}{2} = \frac{\mathcal{D}-2}{2N} \ .
    \end{aligned}
\end{equation}
Notice that for $z=a$ and $z=b$, $S(z)=0$ while $G(a)$ and $G(b)$ are finite, since $a$ and $b$ are the edges at which $\rho$ vanishes. So we get two conditions for the edges, $F(a)=0$ and $F(b)=0$, which imply from Eq.~\eqref{app_add_eq_F_final}
\begin{equation} \label{app_add_eq_edge_two}
    \eta = \frac{\alpha}{2}\sqrt{\frac{b}{a}} + \frac{\beta}{2}\sqrt{\frac{1-b}{1-a}} \ , \quad
    \eta = \frac{\alpha}{2}\sqrt{\frac{a}{b}} + \frac{\beta}{2}\sqrt{\frac{1-a}{1-b}} \ .
\end{equation}
The difference of the two equations in Eq.~\eqref{app_add_eq_edge_two} and the inequality $a \neq b$ then gives
\begin{equation}
   \frac{\alpha}{\sqrt{ab}}=\frac{\beta}{\sqrt{(1-a)(1-b)}} = 2\eta \ ,
\end{equation}
Therefore
\begin{equation} \label{app_add_eq_edge_conditions}
\begin{aligned}
    \sqrt{ab} &= \frac{\alpha}{2\eta} = \frac{\alpha}{2+\alpha+\beta} = \frac{|\nu_0|-1}{\mathcal{D}-2} \ , \\ 
    \sqrt{(1-a)(1-b)} &= \frac{\beta}{2\eta} = \frac{\beta}{2+\alpha+\beta} = \frac{|\nu_\pi|-1}{\mathcal{D}-2} \ ,
\end{aligned}
\end{equation}
Writing $ab=A^2$ and $(1-a)(1-b)=B^2$, we get $a+b=1+A^2-B^2$, so that $a$ and $b$ are the roots of the equation $x^2-(1+A^2-B^2)\,x + A^2=0$:
\begin{equation} \label{app_add_eq_edges}
    \begin{aligned}
        a (b)  &= \frac{1}{2}\left[ 1+A^2-B^2 -(+) \mathcal{R} \right] \ , \\
        \mathcal{R} &:= \sqrt{\left[(1-A)^2-B^2\right] \left[(1+A)^2-B^2\right]} \ .
    \end{aligned}
\end{equation}
It is convenient to express the gaps in terms of $\theta$ variables, which, after a few algebraic steps, leads to the following results.
\begin{equation} \label{app_add_eq_theta_edges}
    \begin{aligned}
        \theta_0 &= \arcsin\!\left(\frac{|\nu_0|+|\nu_\pi|-2}{\mathcal{D}-2}\right) + \arcsin\!\left(\frac{|\nu_0|-|\nu_\pi|}{\mathcal{D}-2}\right) , \\
        \pi-\theta_{\pi} &= \arcsin\!\left(\frac{|\nu_0|+|\nu_\pi|-2}{\mathcal{D}-2}\right) - \arcsin\!\left(\frac{|\nu_0|-|\nu_\pi|}{\mathcal{D}-2}\right)  \ .
    \end{aligned}
\end{equation}

Now using the form of $F$ in Eq.~\eqref{app_add_eq_F_final}, we get from Eq.~\eqref{app_add_eq_F_G_rel}
\begin{equation} \label{app_add_eq_G_intm}
\begin{aligned}
G(z) &= -\frac{\alpha}{2z} + \frac{\beta}{2(1-z)} \\
& \quad + \frac{1}{S(z)}\left[\eta - \frac{\alpha\sqrt{ab}}{2z} - \frac{\beta\sqrt{(1-a)(1-b)}}{2(1-z)}\right] \ .
\end{aligned}
\end{equation}
The square bracket in Eq.~\eqref{app_add_eq_G_intm} can be simplified to
\begin{equation}
    \eta\left[1-\frac{ab}{z}-\frac{(1-a)(1-b)}{1-z}\right] = \eta\,\frac{(z-a)(z-b)}{z(z-1)} \ ,
\end{equation}
and since $(z-a)(z-b)/S(z)=S(z)$,
\begin{equation} \label{app_add_eq_G_simplified}
    G(z) = -\frac{\alpha}{2z} + \frac{\beta}{2(1-z)} + \eta\,\frac{\sqrt{(z-a)(z-b)}}{z(z-1)} \ .
\end{equation}
Thus for $z=y+i0^+$ with $a<y<b$,
\begin{align}
    G_+(y) = -\frac{\alpha}{2y} + \frac{\beta}{2(1-y)} - i\,\eta\,\frac{w(y)}{y(1-y)} \ .
\end{align}
Finally, we get the density of states using Eq.~\eqref{app_add_eq_SP}
\begin{equation} \label{app_add_eq_dos_y}
    \rho(y) = \frac{\eta}{\pi}\,\frac{\sqrt{(y-a)(b-y)}}{y(1-y)} \ , \qquad a\le y\le b \ ,
\end{equation}
and $\rho(y)=0$ otherwise. The density of states (Eq.~\eqref{app_add_eq_dos_y}) within the specified interval is normalized to unity.

Returning to $\theta$ with $y=\sin^2(\theta/2)$, $dy=\sin(\theta/2)\cos(\theta/2)\,d\theta$ and $\sqrt{1-y}=\cos(\theta/2)$, the normalized density of the positive eigenphases for $\theta\in[\theta_0,\theta_{\pi}]$ is given by
\begin{equation}\label{app_add_eq_rho_theta}
\begin{aligned}
\rho(\theta) &=\rho(y)\ \frac{d y}{d\theta}  \\
    &= \left(\frac{\mathcal{D}-2}{2\pi N}\right)\ \frac{\sqrt{\left(\cos\theta_0-\cos\theta\right)\left(\cos\theta-\cos\theta_{\pi}\right)}}{\sin\theta} \ .
\end{aligned}
\end{equation}
Finally, we extend the DOS to the whole interval $(-\pi,\pi]$. The $2N$ nonzero eigenphases come in pairs $\pm\theta_i$, so that their number density at $\theta$ is $N\rho(|\theta|)$, in addition to $\nu_0$ eigenvalues at $\theta=0$ and $\nu_\pi$ eigenvalues at $\theta=\pi$. Hence the normalized DOS is
\begin{equation} \label{app_add_eq_dos_full}
    \begin{aligned}
        \rho(\theta) ={}& \frac{|\nu_0|}{\mathcal{D}}\,\delta(\theta) + \frac{|\nu_\pi|}{\mathcal{D}}\,\delta(\theta-\pi) \\
        & + \left(\frac{\mathcal{D}-2}{2\pi\mathcal{D}}\right)\ \frac{\sqrt{\left(\cos\theta_0-\cos\theta\right)\left(\cos\theta-\cos\theta_{\pi}\right)}}{|\sin\theta|}\\
        & \qquad \times \Theta(|\theta|-\theta_0)\ \Theta(\theta_{\pi}-|\theta|) \ ,
    \end{aligned}
\end{equation}
with $\theta_0$ and $\theta_{\pi}$ given in Eq.~\eqref{app_add_eq_theta_edges}. 
 
%----------------------------------------------------------------------------------------------------------------------------------------------------------------------------------------------

\section{Technical remarks on exact diagonalization}
\label{app_tech_remarks}

In a given symmetry sector of dimension $\mathcal{D}$, chiral symmetry is used to avoid diagonalizing a dense $\mathcal{D}\times\mathcal{D}$ matrix, both for the static Hamiltonian and the Floquet operator. The only inputs are the sparse matrix for the problem in the sector and a unitary involution $\mathcal{C}$ that implements the chiral symmetry. This involution maps each basis state to another basis state up to a sign, so its eigenvectors are known explicitly. We denote the dimensions of its $+1$ and $-1$ eigenspaces by $p$ and $q$. Then $p-q=\mathrm{Tr}(\mathcal{C})$, obtained exactly by counting fixed points and requiring no diagonalization.

For a real symmetric chiral Hamiltonian (BDI), $\mathcal{C}_s H\mathcal{C}_s=-H$ implies that $H$ has no matrix elements within an eigenspace of $\mathcal{C}_s$, so that in the eigenbasis of $\mathcal{C}_s$
\begin{equation}
H=\begin{pmatrix}0&M\\M^{T}&0\end{pmatrix},\qquad M\in\mathbb{R}^{p\times q} \ .
\end{equation}
The matrix $M$ is real because $H$ is real symmetric in the basis. The spectrum of $H$ consists of $\pm\sigma_k$, where $\sigma_k$ ($k=1,\dots,\min(p,q)$) are the singular values of $M$, and of $|p-q|=|\mathrm{Tr}(\mathcal{C}_s)|$ exact zero modes. It is therefore enough to compute the singular values of $M$, a dense problem of size about $\mathcal{D}/2$ instead of $\mathcal{D}$, and the number of protected zero modes is known beforehand from the trace.

For a unitary circular BDI operator, let $V$ be the unitary whose spectrum is wanted ($V=W$, or $V=U$ when a square root $W$ is not available in the sector) and let $\mathcal{C_F}$ be an involution with $\mathcal{C_F}V\mathcal{C_F}=V^{\dagger}$. In the eigenbasis of $\mathcal{C_F}$
\begin{equation}
V=\begin{pmatrix}A&iB\\iB^{\rm T}&F\end{pmatrix},\qquad A=A^{\rm T},\quad F=F^{\rm T},
\end{equation}
so that $\frac{1}{2} (V+V^{\dagger})=A\oplus F$ is block diagonal, with eigenvalues $\cos\phi_n$ where $e^{i\phi_n}$ are the eigenvalues of $V$. Here $A\in \mathbb{R}^{p\times p}$, $F \in \mathbb{R}^{q\times q}$, and $B \in \mathbb{R}^{p\times q}$. Since $\mathcal{C_F}$ maps the eigenspace with phase $\phi$ onto the one with $-\phi$, every pair $\pm\phi$ gives the same $\cos\phi$ once in $A$ and once in $F$, whereas an eigenphase $\phi=0,\pi$ is its own partner and is unpaired. Only the larger block (here $A$ since we take $p\geq q$) has to be diagonalized. Notice that the variables $y_n:=\sin^2(\phi_n/2)=(1-\cos\phi_n)/2$ used in Appendices~\ref{app_derivation} and~\ref{app_derivation_additional} are the eigenvalues of $(\mathbbm{1}-A)/2$.

The unpaired phases are counted exactly. Paired eigenspaces do not contribute to the trace of $\mathcal{C_F}$, so that $\mathrm{Tr}(\mathcal{C_F}) =\nu_0+\nu_\pi$ and $\mathrm{Tr}(V\mathcal{C_F})=\nu_0-\nu_\pi$, where $\nu_{0,\pi}$ are the traces of $\mathcal{C_F}$ on the degenerate eigenspaces of $V$ with eigenvalue $+1$ and $-1$. Both traces are evaluated from the basis states. In the larger block (the one with $\mathcal{C_F}=+1$ when all fixed points have this value), the $|\mathrm{Tr}(\mathcal{C_F})|$ eigenvalues with $|\cos\phi|$ closest to one are the unpaired phases, and the sign of $\cos\phi$ separates $\phi=0$ from $\phi=\pi$. The remaining eigenvalues give the paired phases $\pm\phi_n=\pm\arccos(\cos\phi_n)$. The full set of eigenphases is the union of the $\pm\phi_n$ and the unpaired phases, and its size must equal $\mathcal{D}$, which serves as a check. When $V=W$ the phases of $U=W^2$ follow as $\theta_n=2\phi_n \ \rm{mod} \  2\pi$.

The matrix elements $\langle a|V|b\rangle$ of the relevant block $A$ are obtained by applying $V$ to the sparse basis vectors $|b\rangle$ and projecting onto the corresponding sparse vectors $\langle a|$, so that the full $\mathcal{D}\times\mathcal{D}$ matrix of $V$ is never formed~\cite{TalEzer1984}. We evaluate the exponential of each Hermitian generator $H$ entering the Floquet operator using a Chebyshev expansion. After rescaling $H$ as
\begin{equation}
X=\frac{H}{a}, \qquad a=\frac{E_{\max}}{2},
\end{equation}
where $E_{\max}$ can be obtained using Lanczos method, the spectrum of $X$ lies in $[-1,1]$ and
\begin{equation}
e^{-i\tau X} = J_0(x)\mathbbm{1} + 2\sum_{m=1}^{\infty}(-i)^mJ_m(x)T_m(X) \ ,
\end{equation}
where $x=\tau a$, $T_m$ denotes the Chebyshev polynomial of the first kind and $J_m$ is the Bessel function of the first kind. The expansion acts on a sparse vector and is generated recursively from $T_{m+1}(X)=2XT_m(X)-T_{m-1}(X)$, so no dense representation of the operator is required. The expansion is truncated at order $m_{\rm max}$, where the neglected Bessel tail is below the prescribed numerical tolerance. Equivalently, $m_{\rm max}$ is chosen such that the contribution of all terms with $m>m_{\rm max}$ is numerically negligible over the full spectral interval of the rescaled operator. We then use the same cutoff consistently when constructing the Floquet action. We explicitly checked convergence for small system sizes by increasing $m_{\rm max}$ until the resulting eigenphases and derived quantities were unchanged within the target numerical precision. This procedure also avoids introducing an uncontrolled approximation associated with a fixed polynomial order since the cutoff is determined by the magnitude of the Bessel coefficients and the required numerical accuracy.

In both static and Floquet cases, the dense step involves a matrix of size $\simeq\mathcal{D}/2$, which needs about $\mathcal{D}^2/4$ real numbers instead of $\mathcal{D}^2$ complex ones, and a real symmetric or singular value problem instead of a general one. For small sizes, we verified the procedure against the dense diagonalization of $H$ and of $V$ and found agreement of all eigenvalues and eigenphases to about $10^{-13}$.

%---------------------------------------------------------------------------------------------------------------------------------------------------------------------------------------
%-----------------------------------------------------#  Fig. 7  #----------------------------------------------------------------------------------------------------------------------
\begin{figure}
\centering
\begin{tikzpicture}[
    >={Stealth[length=2.4mm,width=2mm]},
    wlabel/.style={fill=white, inner sep=1pt, font=\small},
    tiplabel/.style={font=\footnotesize}
]
\def\rc{0.55}
\def\Rt{3.0}
\def\rw{1.45}
\def\rl{2.1}
\def\hw{0.95}
\def\dd{7}
\pgfmathsetmacro{\step}{360/7}
\foreach \i/\kk/\col in {0/0/red, 1/1/blue, 2/2/green, 3/3/orange, 5/{n-2}/violet, 6/{n-1}/cyan} {
    \pgfmathsetmacro{\ang}{90-\i*\step}
    \begin{scope}[rotate=\ang]
        \path[fill=\col!35, fill opacity=0.65, draw=\col!70!black, thick]
            (0,0) .. controls (1.0,\hw) and (\Rt-1.2,\hw) .. (\Rt,0)
                  .. controls (\Rt-1.2,-\hw) and (1.0,-\hw) .. (0,0) -- cycle;
    \end{scope}
    \node[align=center, font=\small] at (\ang:\rl) {$\mathcal{C}_{\kk}$\\[1pt]$\mathcal{T}_{\kk}$};
    \node[tiplabel] at ([yshift=-1mm]\ang:{\Rt+0.5}) {$k=\kk$};
}
\pgfmathsetmacro{\angd}{90-4*\step}
\begin{scope}[rotate=\angd]
    \path[draw=gray!70, dashed, thick]
        (0,0) .. controls (1.0,\hw) and (\Rt-1.2,\hw) .. (\Rt,0)
              .. controls (\Rt-1.2,-\hw) and (1.0,-\hw) .. (0,0) -- cycle;
\end{scope}
\foreach \a in {-5,0,5} {
    \node[font=\large] at ({\angd+\a}:{\rl+0.18*cos(\a)}) {$\cdot$};
}
\node[tiplabel] at (\angd:{\Rt+0.5}) {$k=4,\dots,n-3$};
\foreach \i in {0,1,2,5} {
    \pgfmathsetmacro{\aa}{90-\i*\step-\dd}
    \pgfmathsetmacro{\ab}{90-(\i+1)*\step+\dd}
    \pgfmathsetmacro{\am}{90-\i*\step-\step/2}
    \draw[->, thick] (\aa:\rw) arc[start angle=\aa, end angle=\ab, radius=\rw];
    \node[wlabel] at (\am:\rw) {$W$};
}
\pgfmathsetmacro{\aa}{90-3*\step-\dd}
\pgfmathsetmacro{\ab}{90-5*\step+\dd}
\draw[->, thick, dotted] (\aa:\rw) arc[start angle=\aa, end angle=\ab, radius=\rw];
\node[wlabel] at (\angd:\rw) {$W$};
\pgfmathsetmacro{\aa}{90-6*\step-\dd}
\pgfmathsetmacro{\ab}{90-7*\step+\dd}
\pgfmathsetmacro{\am}{90-6*\step-\step/2}
\draw[->, thick, dashed] (\aa:\rw) arc[start angle=\aa+3.5, end angle=\ab-3.5, radius=\rw+0.12];
\node[wlabel] at (\am:\rw) {$W$};
\node[circle, draw=black, very thick, fill=white, fill opacity=0.8, text opacity=1, minimum size=0.95cm, inner sep=0pt, font=\large] at (0,0) {$\mathcal{P}_s$};
\end{tikzpicture}
\caption{$n$ sets of BDI operators $(\mathcal{C}_k,\mathcal{T}_k)=(W^k\mathcal{C}_0,\,W^k\mathcal{T}_0)$ with $k=0,1,\dots,n-1$ (petals), all sharing the particle-hole operator $\mathcal{P}_s=\mathcal{P}_0$ (circled intersection of the petals). The arrows inside the flower show the action of $W$, which takes the set $k$ to the set $k+1$. Since $W^n=U(T,0)$, the last (dashed) arrow leads to $W^n\mathcal{C}_0=U(T,0)\,\mathcal{C}_0$, so that the sequence closes on the first set only up to the factor $U(T,0)$.}
\label{fig_7}
\end{figure}
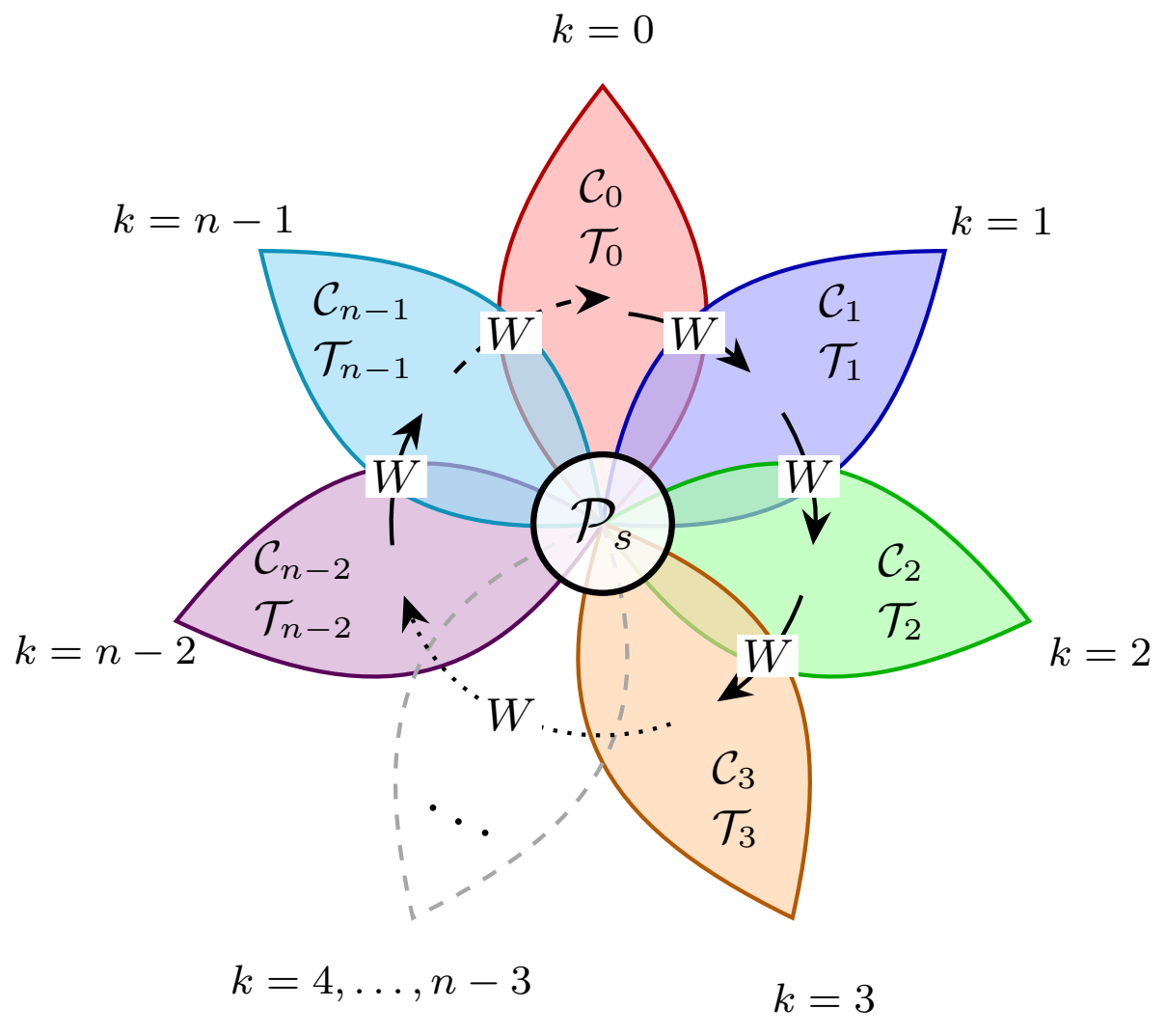
%--------------------------------------------------------------------------------------------------------------------------------------------------------------------------------

%---------------------------------------------------#  Fig.  8  #----------------------------------------------------------------------------------------------------------------------
\begin{figure*} 
    \centering
    \includegraphics[width=0.49\linewidth]{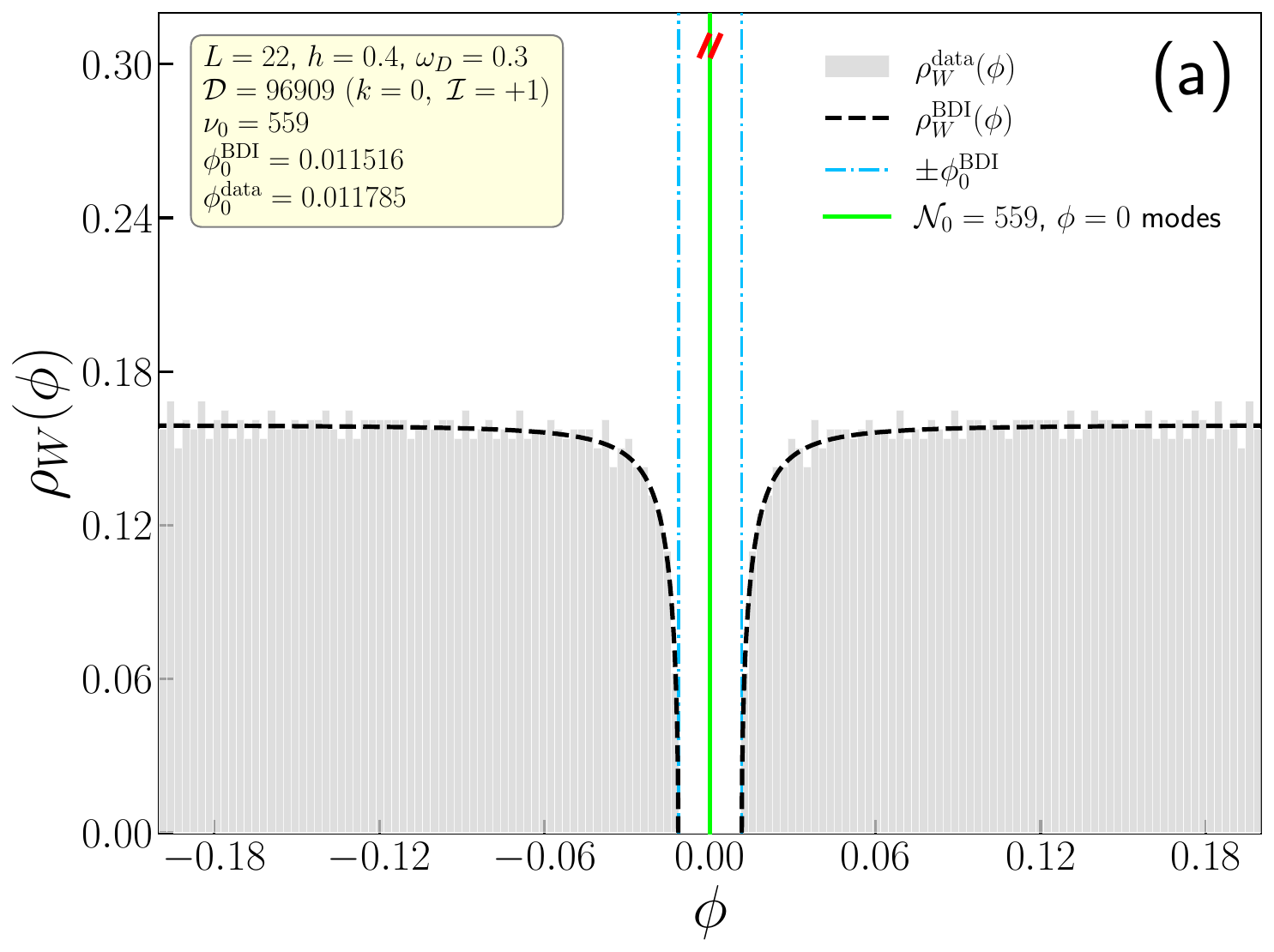}
    \includegraphics[width=0.49\linewidth]{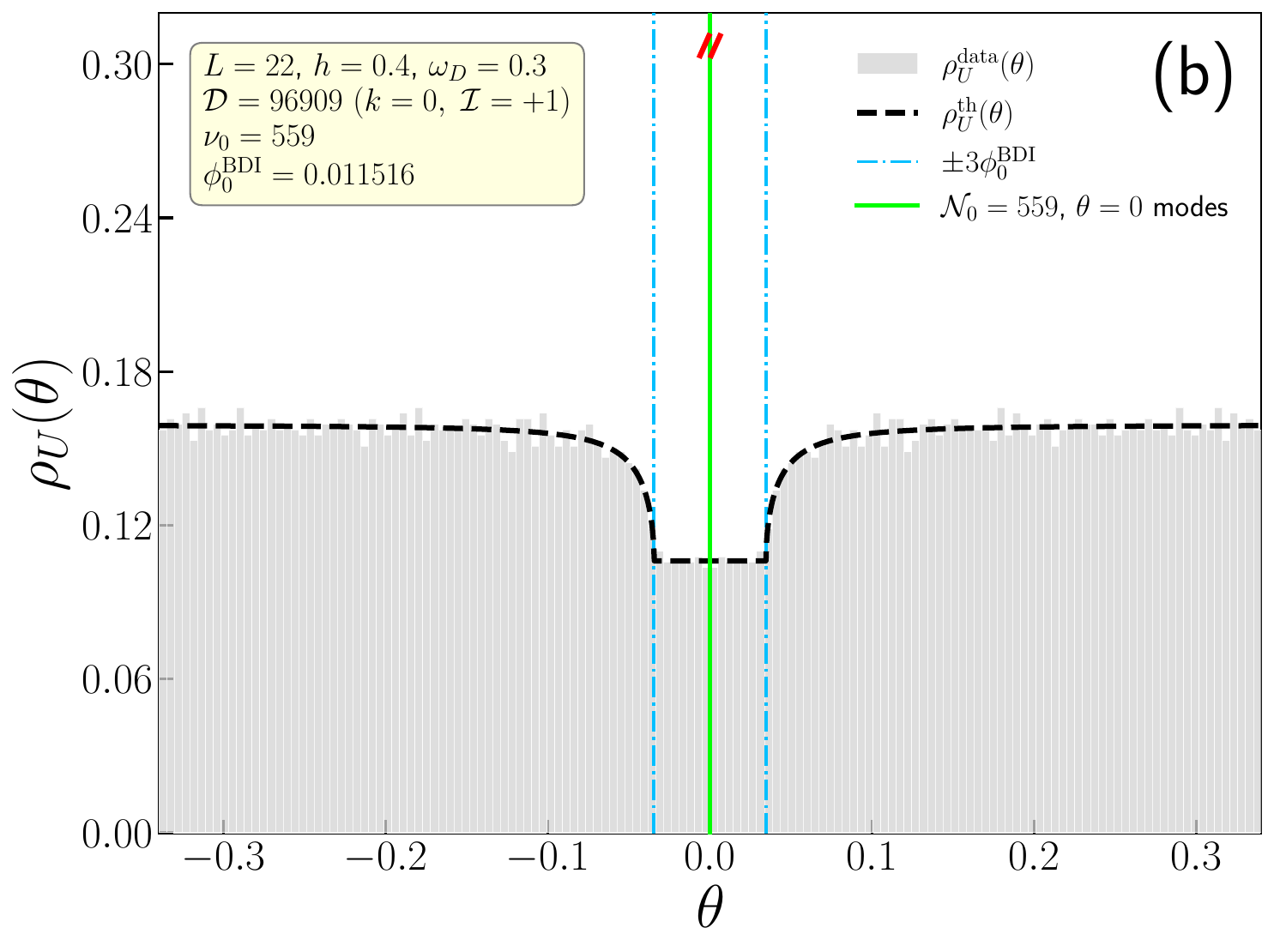}
    \caption{(a) Density of states of the eigenphases $\{\phi\}$ of $W(T,0)$ and (b) $\{\theta\}$ of $U(T,0)=W(T,0)^n$ for the three-spin three-step driven problem in the vicinity of $\phi=0$ and $\theta=0$ respectively. The black dashed curve is the analytical form $\rho_W^{\rm BDI}$ and $\rho_U^{\rm th}$ in panel (a) and (b), respectively. The $559$ modes of $W(T,0)$ at $\phi=0$ fold to $\theta=0$ (green line). The drive frequency is $\omega_D=0.3$, the drive amplitude is $h=0.4$ and the symmetry-resolved sector is $k_1=0$, $\mathcal{I}=+1$ ($\mathcal{D}=96909$) for the system size $L=22$.}
    \label{fig_8}
\end{figure*}
%-----------------------------------------------------------------------------------------------------------------------------------------------------------------------------------------

%-------------------------------------------#  Fig. 9  #----------------------------------------------------------------------------------------------------------------------------------
\begin{figure}
    \centering
    \includegraphics[width=\linewidth]{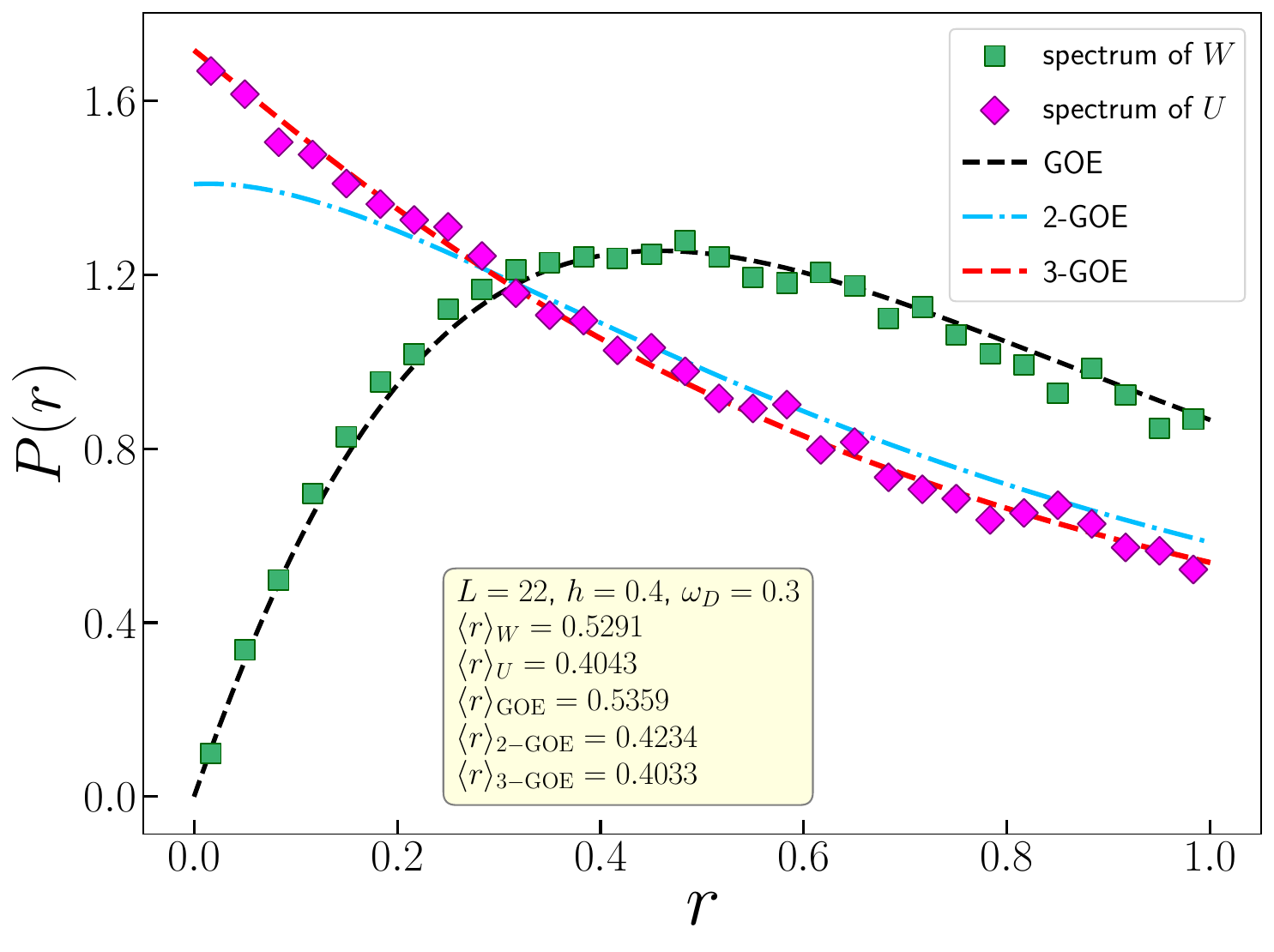}
    \caption{Distribution of the ratios of consecutive level spacings in the three-step Floquet three-spin model for the positive bulk eigenphases of $W(T,0)$ and $U(T,0)$, together with the GOE, $2$-GOE and the $3$-GOE distributions. The drive frequency is $\omega_D=0.3$, the drive amplitude is $h=0.4$, and the system size is $L=22$. The symmetry-resolved sector considered is $k_1=0$, $\mathcal{I}=+1$ ($\mathcal{D}=96909$).}
    \label{fig_9}
\end{figure}
%-----------------------------------------------------------------------------------------------------------------------------------------------------------------------------------------

%-----------------------------------------------------------------------------------------------------------------------------------------------------------------------------------------------------------

\section{Floquet unitary with $U(T,0)=W(T,0)^n$ in model III}
\label{app_W_n}

In the main text, the dynamical symmetry $G=\prod_j\sigma^z_j$ of model III gives $U(T,0)=W(T,0)^2$. Here we show that the same Hamiltonian admits $n$-step drives with $U(T,0)=W(T,0)^n$ for any integer $n\geq2$, where $W(T,0)$ is a circular BDI unitary with a single protected imbalance, and then discuss $n=3$ explicitly.

\subsection{Construction for general $n$}
Let $H_0$ be the Hamiltonian of Eq.~\eqref{eq_three_spin_ham} and
\begin{equation} \label{eq_Gn_def}
    G_n = \exp\left(-i\frac{\pi}{n}\sum_{j=1}^L\sigma_j^z\right) .
\end{equation}
Since $e^{-i\pi\sigma^z}=-\mathbbm{1}$, $G_n^n=(-1)^L\mathbbm{1}=\mathbbm{1}$ for even $L$. $G_n$ rotates the transverse field about the $z$-axis by $2\pi/n$ and leaves the three-spin term invariant, so that
\begin{equation} \label{eq_Ha_def}
\begin{aligned}
& H_a = G_n^{a} H_0 G_n^{-a} \\&= \sum_{j=1}^L\sigma_j^z\sigma_{j+1}^z\sigma_{j+2}^z + h\sum_{j=1}^L\left[\cos\left(\tfrac{2\pi a}{n}\right)\sigma_j^x+\sin\left(\tfrac{2\pi a}{n}\right)\sigma_j^y\right]\ , 
\end{aligned}
\end{equation}
with $a=0,\dots,n-1$. We drive the system with period $T=2\pi/\omega_D=n\tau$ by applying $H_a$ for a time $\tau$ in the order $a=0,1,\dots,n-1$,
\begin{equation} \label{eq_n_step_protocol}
H(t)=H_a\ , \ a\tau\leq t<(a+1)\tau\ ;\ a=0,\dots,n-1\ .
\end{equation}
With
\begin{equation}
A_a=e^{-i\tau H_a}=G_n^aA_0G_n^{-a}\ ,
\end{equation}
the Floquet unitary can be written as 
\begin{equation} \label{eq_U_nstep}
U(T,0)=A_{n-1}\cdots A_1 A_0 \ .
\end{equation}
Define
\begin{equation} \label{eq_Wn_def}
W(T,0) = G_n^{-1}A_0 \ .
\end{equation}
Using the relations $(G_n)^n=\mathbbm{1}$ and $A_{a-n}=A_a$ where the indices are to be understood modulo $n$, we get
\begin{equation}
\begin{aligned} 
W^n &= (G_n^{-1}A_0G_n)(G_n^{-2}A_0G_n^{2})\cdots(G_n^{-n}A_0G_n^{n})\,G_n^{-n} \\
&= A_{-1}A_{-2}\cdots A_{-n} = A_{n-1}\cdots A_1A_0 = U(T,0)\ .
\end{aligned}
\end{equation}
For $n=2$, Eq.~\eqref{eq_n_step_protocol} is the square-pulse of Eq.~\eqref{eq_h_pulse_three_spin} with the identifications $A_0=U_1$ and $A_1=U_2$, and $G_2=G=(-1)^{L/2}\prod_j\sigma^z_j$, so that $W(T,0)=(-1)^{L/2}\left(\prod_j\sigma_j^z\right)U_1$. The factor $(-1)^{L/2}$ is the sign ambiguity of $W$, and the convention of Eq.~\eqref{eq_Wn_def} places the protected modes at $\phi=0$ for any even $L$, which we show later.

We use the particle-hole operator $\mathcal{P}_s$ of the static problem (Eq.~\eqref{eq_three_spin_static_BDI}) and dress the chiral and time-reversal operators of the static problem, $\mathcal{C}_s=\prod_j\sigma^y_j$ and $\mathcal{T}_s= \mathcal{K}$, with the first pulse $A_0$ as the following.
\begin{equation} \label{eq_Wn_BDI_ops}
\mathcal{C}_0 = A_0^{-1}\mathcal{C}_s\ , \ \mathcal{T}_0 = A_0^{-1}\,\mathcal{K}\ , \
\mathcal{P}_0=\mathcal{C}_0\mathcal{T}_0 = \mathcal{C}_s\mathcal{K} =\mathcal{P}_s\ .
\end{equation}
They follow from
\begin{equation}
\begin{aligned}
    \mathcal{C}_s H_a \mathcal{C}_s^{-1}=-H_{-a}\ ,& \quad \mathcal{C}_s G_n \mathcal{C}_s^{-1}=G_n^{-1}\ , \\
    \mathcal{K}H_a\mathcal{K}=H_{-a}\ ,& \quad \mathcal{K}G_n\mathcal{K}=G_n^{-1}\ ,
\end{aligned}
\end{equation}
with $a$ modulo $n$, which give $\mathcal{C}_s A_a \mathcal{C}_s^{-1}=A_{-a}^{-1}$ and $\mathcal{K}A_a\mathcal{K}=A_{-a}^{-1}$. In particular $\mathcal{C}_s A_0 \mathcal{C}_s^{-1}=A_0^{-1}$, so that $\mathcal{C}_0^2=\mathcal{T}_0^2=\mathbbm{1}$ for even $L$, $\mathcal{C}_0$ and $\mathcal{T}_0$ commute, and
\begin{equation} \label{eq_Wn_BDI}
    \mathcal{C}_0W\mathcal{C}_0^{-1}=W^\dagger\ , \ \mathcal{T}_0W\mathcal{T}_0^{-1}=W^\dagger\ , \ \mathcal{P}_0W\mathcal{P}_0^{-1}=W\ .
\end{equation}
For $U(T,0)=W(T,0)^n$, the same operators satisfy the relations of Eq.~\eqref{eq_circular_BDI_action}. The Hamiltonians $H_a$ with $a\neq0$ contain $\sigma^y$ and are not individually time-reversal symmetric under $\mathcal{K}$; the relations hold for the unitary operators $W(T,0)$ and $U(T,0)$. Both $G_n$ and all $H_a$ commute with $T_1$ and $\mathcal{I}$, so $W(T,0)$ and $U(T,0)$ act within the sector $k_1=0$, $\mathcal{I}=+1$, which is also preserved by $\mathcal{C}_0$, $\mathcal{T}_0$ and $\mathcal{P}_0$.

\iffalse
The operator $W$ connects different sets of BDI operators sharing the particle-hole operator,
\begin{equation} \label{eq_Wn_BDI_sets}
\begin{aligned}
\mathcal{C}_k = W^k\mathcal{C}_0\ , \ \mathcal{T}_k &= W^k\mathcal{T}_0\ , \
\mathcal{P}_k=\mathcal{C}_k\mathcal{T}_k=\mathcal{P}_0\ ; \\ k&=0,1,\dots,n-1\ ,
\end{aligned}
\end{equation}
each of which satisfies Eq.~\eqref{eq_Wn_BDI} and squares to $\mathbbm{1}$ since $\mathcal{C}_0W^k\mathcal{C}_0^{-1}=\mathcal{T}_0W^k\mathcal{T}_0^{-1}=W^{-k}$. The schematic in Fig.~\ref{fig_7} illustrates the $n$ BDI symmetry sets $(\mathcal{C}_k,\mathcal{T}_k)$, arranged cyclically and connected by the action of $W(T,0)$. All sets share the same particle-hole symmetry $\mathcal P_s$, shown at the center of the diagram. For $n=2$, $A_1 \mathcal{C}_s=U(T,0)\, \mathcal{C}_0$, which is $U_2\mathcal{C}_s$ of the main text, so that these are the two sets of model III.
\fi

The operator $W$ connects different sets of BDI operators sharing the particle-hole operator,
\begin{equation} \label{eq_Wn_BDI_sets}
\begin{aligned}
\mathcal{C}_k = W^k\mathcal{C}_0\ , \ \mathcal{T}_k &= W^k\mathcal{T}_0\ , \
\mathcal{P}_k=\mathcal{C}_k\mathcal{T}_k=\mathcal{P}_0\ ; \\ k&=0,1,\dots,n-1\ .
\end{aligned}
\end{equation}
Since $\mathcal{C}_0W^k\mathcal{C}_0^{-1}=\mathcal{T}_0W^k\mathcal{T}_0^{-1}=W^{-k}$, each set satisfies Eq.~\eqref{eq_Wn_BDI}, and its three operators commute mutually and each squares to $\mathbbm{1}$,
\begin{equation} \label{eq_petal_commute}
\begin{aligned}
\mathcal{C}_k\mathcal{T}_k=\mathcal{T}_k\mathcal{C}_k=\mathcal{P}_s\ &, \quad
\mathcal{C}_k^2=\mathcal{T}_k^2=\mathcal{P}_s^2=\mathbbm{1}\ , \\
[\mathcal{C}_k,\mathcal{P}_s]&=[\mathcal{T}_k,\mathcal{P}_s]=0\ .
\end{aligned}
\end{equation}
Operators of different sets, however, do not commute: for any $X_k\in\{\mathcal{C}_k,\mathcal{T}_k\}$ and $X_\ell \in\{\mathcal{C}_\ell,\mathcal{T}_\ell\}$ with $k\neq \ell$,
\begin{equation} \label{eq_petal_commutator}
X_kX_\ell X_k^{-1}X_\ell^{-1}=W^{2(k-\ell)}\ ,
\end{equation}
which equals $U(T,0)^{\pm1}=W^{\pm n}$ only for even $n$ and $|k-l|=n/2$. The schematic in Fig.~\ref{fig_7} illustrates the $n$ BDI symmetry sets $(\mathcal{C}_k,\mathcal{T}_k)$ as petals, with the operators of each petal mutually commuting (Eq.~\eqref{eq_petal_commute}), all sharing the particle-hole operator $\mathcal{P}_s$ at the center, and the arrows inside show the action of $W(T,0)$ taking the set $k$ to the set $k+1$. For $n=2$, $A_1\mathcal{C}_s=U(T,0)\,\mathcal{C}_0$, which is $U_2\mathcal{C}_s$ of the main text, so that these are the two sets of model III, and Eq.~\eqref{eq_petal_commutator} gives $X_1X_0X_1^{-1}X_0^{-1}=U(T,0)$.

Let $\mathrm{Tr}$ denotes the trace over the sector $k_1=0$, $\mathcal{I}=+1$. The imbalances of $W$(T,0) are $\nu_{0,\pi}=\frac12\left[\mathrm{Tr}(\mathcal{C}_0)\pm\mathrm{Tr}(\mathcal{C}_0 W(T,0))\right]$ as defined in the Introduction, and do not depend on $\tau$ or any other parameters of the problem. For $\tau\to0$, $A_0\to\mathbbm{1}$, $\mathcal{C}_0\to \mathcal{C}_s$ and $W(T,0) \to G_n^{-1}$. Since $\mathcal{C}_s$ flips every spin, only basis states with $\sum_j\sigma^z_j=0$ contribute to $\mathrm{Tr}(\mathcal{C}_s G_n^{-1})$ and $\mathrm{Tr}(\mathcal{C}_s)$, and on these states $G_n^{-1}=\mathbbm{1}$. Hence $\mathrm{Tr}(\mathcal{C}_0 W(T,0))=\mathrm{Tr}(\mathcal{C}_0)=\mathrm{Tr}(\mathcal{C}_s)$, i.e.
\begin{equation} \label{eq_Wn_imbalance}
    \nu_\pi=0\ , \qquad \nu_0=\mathrm{Tr}(\mathcal{C}_s)=\nu \quad \text{for every } n\, ,
\end{equation}
where $\nu$ is the imbalance of the static model in the same sector. Thus $W(T,0)$ belongs to ch-COE$(\nu_0,0)$ and has the density $\rho_W(\phi)$ of Eq.~\eqref{eq_circ_BDI_dos}, with the hard edge spectral gap $\phi_0=2\arcsin[(\nu_0-1)/(\mathcal{D}-1)]$ separating $\phi=0$. Since $U(T,0)\to\mathbbm{1}$ for $\tau\to 0$, $U(T,0)$ also has $\nu_0^U=\nu$ and $\nu_\pi^U=0$.

An eigenvector of $W(T,0)$ with eigenphase $\phi$ has the eigenphase $\theta=n\phi$ mod $2\pi$ of $U(T,0)$, so that, unlike the squaring of the main text, the density is the exact $ n$-folded
\begin{equation} \label{eq_rho_U_n}
\begin{aligned}
\rho_U^{\rm th}(\theta)&=\frac{1}{n}\sum_{m=0}^{n-1}\rho_W\!\left(\frac{\theta+2\pi m}{n}\right) \\
&= \frac{\nu_0}{\mathcal{D}}\,\delta(\theta)+\frac1n\sum_{m=0}^{n-1}\rho_W^{\rm bulk}\!\left(\frac{\theta+2\pi m}{n}\right),
\end{aligned}
\end{equation}
with the arguments understood modulo $2\pi$. The $\delta$-function of $W(T,0)$ at $\phi=0$ gives the $\nu_0$ protected zero modes of $U(T,0)$. The other preimages of $\theta=0$, namely $\phi=2\pi m/n$ with $m\neq0$, lie in the bulk of $W(T,0)$ and are not pinned by the symmetry. For $|\theta|<n\phi_0$ the $m=0$ branch vanishes inside the gap of $W$, and the remaining $n-1$ branches contribute the flat bulk value $1/2\pi$, so that
\begin{equation}
    \rho_U(\theta)\simeq\left(\frac{n-1}{n}\right)\,\frac{1}{2\pi}\qquad (0<|\theta|<n\phi_0)\ ,
\end{equation}
a plateau at the fraction $(n-1)/n$ of the bulk value instead of a spectral gap. For $n=2$, this is the plateau at half the bulk value of the main text. As $\theta$ increases from zero, the $m=0$ branch, corresponding to $\phi=\theta/n$, moves from $\phi=0$ through the gap of $W(T,0)$. It reaches the hard edge $\phi=\phi_0$ at $\theta=n\phi_0$, where it begins to contribute to $\rho_U(\theta)$ with a square-root onset:
\begin{equation}
\Delta\rho_U\simeq\frac{1}{2\pi n}\sqrt{\frac{\epsilon}{nA}}\ , \quad \theta=n\phi_0+\epsilon,\ \  1\gg \epsilon >0,
\end{equation}
where $A=(\nu_0-1)/(\mathcal D-2)$. This additional contribution grows from zero, bringing the density from the plateau towards the bulk value $1/(2\pi)$.

The $n$ arcs of $W(T,0)$ that are mapped onto the whole circle by $\theta=n\phi$ are separated in $\phi$. Provided $W(T,0)$ has GOE statistics in the bulk, the positive-branch levels of $U(T,0)$ are a superposition of $n$ independent GOE spectra, with $\langle r\rangle=0.5359$ for one, $0.4234$ for two, and $0.3863$ (Poisson) for $2\ll n \ll\mathcal{D}$~\cite{Alet_paper}. The Poisson-like behavior does not indicate integrability or the emergence of additional conservation laws in the chosen symmetry sector. Although $W(T,0), W(T,0)^2, \dots , W(T,0)^{n-1}$ commute with $U(T,0)$, they are the powers of the same unitary operator and do not provide any block structure of $U(T,0)$ to diagonalize. The resulting spectral superposition arises from the folding of the $n$ distinct spectral arcs of $W(T,0)$, rather than from an underlying integrable structure, and the bulk spectrum of $W(T,0)$ itself retains GOE statistics. In this work, we have numerically verified this for $n=2,3$, and we conjecture that it holds for any $n$ in the limit $L\to\infty$.

We finally note the model-independent uniqueness of the ``glide'' ($G_n$) defined Floquet root $W(T,0)$. The root $W(T,0)$ is not, in general, determined by $U(T,0)$ alone. Here the drive specifies $G_n$, which implements the cyclic shift $H_a\mapsto H_{a+1}$, i.e., $G_nH(t)G_n^{-1}=H(t+\tau)$, and thereby singles out $W(T,0)=G_n^{-1}A_0$. Let $G_n'$ be another operator implementing the same shift. Then $G_n'^{-1}G_n$ commutes with all $H_a$, and if the common unitary commutant of the $H_a$ consists only of scalars in the fully resolved symmetry sector, Schur's lemma~\cite{serre1977linear} gives $G_n'^{-1}G_n=e^{i\alpha}\mathbbm{1}$, so that $W'(T,0)=e^{i\alpha}W(T,0)$. The condition $G_n'^{\,n}=G_n^n=\mathbbm{1}$ requires $e^{in\alpha}=1$. The particle-hole relation $\mathcal P_0W'(T,0)\mathcal P_0^{-1}=W'(T,0)$ with the same antiunitary operator $\mathcal P_0$ further requires $e^{2i\alpha}=1$. Combining the two conditions leaves only $e^{i\alpha}=1$ for odd $n$ and $e^{i\alpha}=\pm1$ for even $n$. Hence the root is unique for odd $n$, and for even $n$ it is unique up to $W\to-W$, which shifts all eigenphases of $W$ by $\pi$, exchanges $\nu_0\leftrightarrow\nu_\pi$ and leaves $U(T,0)$ unchanged. For model III the definition $G_n=\exp[-i(\pi/n)\sum_j\sigma_j^z]$ selects the root with $\nu_\pi=0$; for $n=2$ this is the convention associated with the factor $(-1)^{L/2}$ in $G_2$.

%---------------------------------------------------------------------------------------------------------------------------------------------------------------------------------------------------------
\subsection{Construction for $n=3$ and numerical evidences}
For $n=3$, $G_3=\exp\left(-i\frac{\pi}{3}\sum_{j=1}^L\sigma^z_j\right)$ rotates the transverse field about the $z$-axis by $120^\circ$ (Eq.~\eqref{eq_Ha_def}),
\begin{equation} \label{eq_Ha_n3}
\begin{aligned}
    H_0 &= \sum_{j=1}^L\sigma^z_j\sigma^z_{j+1}\sigma^z_{j+2}+h\sum_{j=1}^L\sigma^x_j\ , \\
    H_{1(2)} &= \sum_{j=1}^L\sigma^z_j\sigma^z_{j+1}\sigma^z_{j+2}+h\sum_{j=1}^L\left(-\tfrac12\sigma^x_j\pm\tfrac{\sqrt3}{2}\sigma^y_j\right) ,
\end{aligned}
\end{equation}
and the drive of Eq.~\eqref{eq_n_step_protocol} applies $H_0$, $H_1$ and $H_2$ successively for a time $\tau=T/3$ each, with $T=2\pi/\omega_D$. The Floquet unitary and its cube root are
\begin{equation} \label{eq_cubic_UW}
\begin{aligned}
U(T,0)&=e^{-i\tau H_2}\,e^{-i\tau H_1}\,e^{-i\tau H_0}=W(T,0)^3\ , \\
W(T,0)&=G_3^{-1}\,e^{-i\tau H_0}\ .
\end{aligned}
\end{equation}
The BDI operators of Eq.~\eqref{eq_Wn_BDI_ops} are 
\begin{equation}
\mathcal{C}_0=e^{i\tau H_0}\prod_j\sigma^y_j\ , \ \mathcal{T}_0=e^{i\tau H_0}\mathcal{K}\ , \ \mathcal{P}_s=\prod_j\sigma^y_j\,\mathcal{K}\ ,
\end{equation}
and $W(T,0)$ generates the three sets of Eq.~\eqref{eq_Wn_BDI_sets}
\begin{equation}
(\mathcal{C}_k,\mathcal{T}_k)=(W^k\mathcal{C}_0,\,W^k\mathcal{T}_0) \quad \text{with } k=0,1,2 \ ,
\end{equation} 
which share $\mathcal{P}_s$. We verified Eq.~\eqref{eq_cubic_UW} and the relations of Eqs.~\eqref{eq_Wn_BDI} and \eqref{eq_Wn_BDI_sets} with dense matrices at small system sizes.

Since $\sum_j\sigma_j^z$ takes even integer values, the eigenvalues of $G_3^{-1}$ are the cube roots of unity, and none is $-1$. This agrees with Eq.~\eqref{eq_Wn_imbalance}: $W(T,0)$ has $\nu_\pi=0$, and $\nu_0=\mathrm{Tr}(\mathcal{C}_s)=\nu$ in the sector $k_1=0$, $\mathcal{I}=+1$. For $L=22$ ($\mathcal{D}=96909$), $\nu_0=559$, the same as the static imbalance of Sec.~\ref{subsec_model_III}. Cubing maps $\phi=\pi$ to $\theta=\pi$, so neither $W(T,0)$ nor $U(T,0)$ has protected modes at $\pi$; exact eigenphases at $\pi$ are not excluded by the imbalance, but they would be unprotected and are not generic. For $n=3$, Eq.~\eqref{eq_rho_U_n} reads
\begin{equation} \label{eq_rho_U_3}
\begin{aligned}
\rho_U^{\rm th}(\theta)=\frac{\nu_0}{\mathcal{D}}\delta(\theta)+\frac13\Big[&\rho_W^{\rm bulk}\!\left(\frac\theta3\right)+\rho_W^{\rm bulk}\!\left(\frac{\theta+2\pi}{3}\right) \\
&+\rho_W^{\rm bulk}\!\left(\frac{\theta-2\pi}{3}\right)\Big],\quad \theta\in(-\pi,\pi]\ ,
\end{aligned}
\end{equation}
where the three arguments lie in $(-\pi/3,\pi/3)$, $(\pi/3,\pi)$ and $(-\pi,-\pi/3)$, respectively. For $0<|\theta|<3\phi_0$, only the last two terms are nonzero, so the density has a plateau at $2/3$ of the bulk value, not a gap, followed by a square-root onset at $\theta=3\phi_0$. The positive-branch levels of $U(T,0)$ are expected to follow the statistics of three superposed GOE spectra.

Fig.~\ref{fig_8} shows the density of states of $W(T,0)$ and $U(T,0)$ for $L=22$, $h=0.4$ and $\omega_D=0.3$ in the sector $k_1=0$, $\mathcal{I}=+1$. In panel (a), the $559$ modes of $W(T,0)$ at $\phi=0$ are separated from the bulk by a gap. The numerically obtained edge $\phi_0^{\rm data}=0.011785$ is $2.3\%$ larger than the prediction $\phi_0^{\rm BDI}=2\arcsin[(\nu_0-1)/(\mathcal{D}-1)]=0.011516$, and the histogram follows the circular BDI density $\rho_W^{\rm BDI}(\phi)$ (black dashed curve) with a flat bulk value $\approx1/2\pi$. In panel (b), the $559$ modes at $\phi=0$ appear as the $\delta$-function of $U(T,0)$ at $\theta=0$. The density of $U(T,0)$ in the window $0<|\theta|<3\phi_0^{\rm BDI}=0.0345$ (vertical dash-dotted lines) is the plateau $\approx0.106=(2/3)\times 1/(2\pi)$, and there is no gap around $\theta=0$, in agreement with Eq.~\eqref{eq_rho_U_3} (black dashed curve), which has no fitting parameters. The density rises to the bulk value outside this window. The density of $U(T,0)$ near $\theta=0$ is therefore not that of ch-COE$(\nu_0,0)$ with a hard gap, although $U(T,0)$ has the same BDI symmetry and the same imbalance as $W(T,0)$.

Fig.~\ref{fig_9} shows the distribution of ratios of consecutive level spacings for the positive bulk eigenphases. The levels of $W(T,0)$ follow the GOE distribution, with $\langle r\rangle_W=0.5291$ compared with $0.5359$ for the GOE. For $U(T,0)$, $\langle r\rangle_U=0.4043$ agrees with $0.4033$ for three superposed GOE spectra and is clearly distinct from $0.4234$ for two superposed GOE spectra and from the Poisson value $0.3863$, and the histogram is closest to the 3-GOE curve. This is the expected result of the folding $\theta=3\phi$, in which three separated arcs of the spectrum of $W(T,0)$, each with GOE statistics, are superimposed to form the spectrum of $U(T,0)$. For comparison, the main-text model III has $U(T,0)=W(T,0)^2$, leading to a superposition of two GOE spectra with $\langle r\rangle_U=0.4229$, in contrast to the present case with three superposed GOE spectra. Nevertheless, both cases support the correspondence between the power $n$ in $U(T,0)=W(T,0)^n$ and the number of superposed GOE spectra.

\end{appendix}

%-----------------------------------------------------------------------------------------------------------------------------------------------------------------------------------------------------------
\bibliography{citations}
\end{document}